\documentclass[aps,prb,twocolumn,showpacs,longbibliography]{revtex4-2}
\usepackage{graphicx}

\usepackage[colorlinks=true, 
            linkcolor=blue, 
            urlcolor=blue,
            citecolor=blue]{hyperref}
\usepackage{color}
\usepackage{bm}
\usepackage{braket,soul}
\usepackage{mathbbol,amsmath,amssymb}
\usepackage{siunitx}
\usepackage{dcolumn}
\DeclareSymbolFontAlphabet{\amsmathbb}{AMSb}%

\begin{document}


\title{ Derivation of low-energy Hamiltonians for heavy-fermion Materials}
\author{E. A. Ghioldi$^{1}$, Zhentao Wang$^2$, L. M. Chinellato$^1$, Jian-Xin Zhu$^3$, Yusuke~Nomura$^4$, Ryotaro Arita$^{5,6}$, W. Simeth$^{7,8}$, M. Janoschek$^{8,9,10}$,  F. Ronning$^{10}$, C. D. Batista$^{1,11}$} 

\affiliation{
$^1$Department of Physics, The University of Tennessee, Knoxville, Tennessee 37996, USA \\
$^2$Center for Correlated Matter and School of Physics, Zhejiang University, Hangzhou 310058, China \\
$^3$Theoretical Division, Los Alamos National Laboratory; Los Alamos, New Mexico, 87545, USA\\
$^4$Institute for Materials Research, Tohoku University, Katahira, Aoba-ku, Sendai 980-8577, Japan \\
$^5$Department of Physics, The University of Tokyo, Hongo, Bunkyo-ku, Tokyo 113-0033, Japan \\
$^{6}$RIKEN Center for Emergent Matter Science, 2-1 Hirosawa, Wako 351-0198, Japan\\
$^7$Laboratory for Neutron and Muon Instrumentation, PSI Center for Neutron and Muon Sciences CNM, Paul Scherrer Institute, 5232 Villigen-PSI, Switzerland\\
$^8$Physik-Institut, Universit\"{a}t Z\"{u}rich, Winterthurerstrasse 190, CH-8057 Z\"{u}rich, Switzerland\\
$^9$Los Alamos National Laboratory, Los Alamos, New Mexico 87545, USA\\
$^{10}$Institute for Materials Science, Los Alamos National Laboratory, New Mexico, 87545, USA\\
$^{11}$Quantum Condensed Matter Division and Shull-Wollan Center, Oak Ridge National Laboratory, Oak Ridge, Tennessee 37831, USA 
}

\date{\today}
\begin{abstract}
By utilizing a multi-orbital periodic Anderson model with parameters obtained from \textit{ab initio} band structure calculations, combined with degenerate perturbation theory, we derive effective Kondo-Heisenberg and spin Hamiltonians that capture the interaction among the effective magnetic moments. This derivation encompasses fluctuations via both nonmagnetic $4f^0$ and magnetic $4f^2$ virtual states, and its accuracy is confirmed through comparison with experimental data obtained from CeIn$_3$. The significant agreement observed between experimental results and theoretical predictions underscores the potential of deriving minimal models from first-principles calculations for achieving a quantitative description of $4f$ materials. Moreover, our microscopic derivation unveils the underlying origin of anisotropy in the exchange interaction between Kramers doublets, shedding light on the conditions under which this anisotropy may be weak compared to the isotropic contribution.

\end{abstract}

\maketitle

\section{Introduction} 

Heavy-fermion materials serve as exemplary strongly correlated electron systems, exhibiting a wide range of novel phenomena~\cite{AndresK1975,SteglichF1979, OttHR1983, StewartGR1984_UPt3, FisherRA1989, BrulsG1990, AdenwallaS1990, AeppliG1992, JaccardD1992, MovshovichR1996, MathurND1998, SchroderA2000, PaglioneJ2003, BianchiA2003, PaschenS2004, ShishidoH2005, DzeroM2010, RonningF2017}. While simplified minimal model Hamiltonians inspired by these materials have offered valuable insights and novel conceptual ideas~\cite{ColemanP2007}, exploring a broader range of realistic models is essential for gaining deeper insights into the spectrum of states of matter that these materials can host.

In the realm of $d$-electron materials, the shift from oversimplified effective models to more realistic Hamiltonians that incorporate spin-orbit coupling has been pivotal in uncovering novel states of matter~\cite{JackeliG2009,Witczak-KrempaW2014_review}. Furthermore, microscopic insights into $d$-electron materials have provided essential guiding principles for understanding the sign of effective magnetic exchange interactions~\cite{GoodenoughJB1955,KanamoriJ1957,GoodenoughJB1958}, along with exploring potential origins of exchange anisotropy~\cite{MoriyaT1960b,JackeliG2009}. Developing similar rules for $4f$-electron materials could significantly impact the search for novel quantum states of matter, such as quantum spin liquids or skyrmion crystals. However, obtaining insights from first-principles effective models remains challenging for $f$-electron systems, primarily due to the significant disparity in relevant energy scales.


The initial step in deriving low-energy models involves identifying a hierarchy of energy scales. For $f$-electron ions, the relevant single-ion scales include the Coulomb interaction, spin-orbit coupling (SOC), and crystal field (CF) potential. The hierarchy 
$E_{\rm int} \gg E_{\rm SOC} \gg E_\text{CF}$, as assumed in the Russell-Saunders coupling scheme~\cite{RussellHN1925}, is commonly used to derive effective spin models~\cite{JangSH2019,JangSH2020}. Note, however, that this hierarchy may not be applicable to every $f$-ion. For instance, recent studies have shown that in the tetravalent Pr$^{4+}$ ion, the CF splitting and spin-orbit coupling are of comparable magnitude, i.e., $E_{\rm SOC} \sim E_\text{CF}$~\cite{DaumMJ2021,RamanathanA2023a,RamanathanA2023b}.

The absence of universal rules and the potential for significant exchange anisotropies, arising from the interplay of strong spin-orbit coupling and CF splitting, highlight the need for a microscopic approach in the field. Here we present a systematic derivation of effective low-energy models, such as the Kondo-Heisenberg lattice  model (KHLM) and effective spin Hamiltonians, from a many-body treatment of the periodic Anderson model (PAM) with parameters extracted from first-principles calculations. This integrated approach can provide valuable insights into the complex behavior of highly correlated $f$-electron systems.

As a starting point, we will focus on $4f$ electron ions such as Ce$^{3+}$, where the lowest-energy electronic configuration is $4f^{1}$, indicating one $f$ electron. These ions fluctuate between this configuration and higher-energy states $4f^0$ and $4f^2$. By particle-hole symmetry, this approach is equally valid for Yb$^{3+}$ with a $4f^{13}$ configuration and a single hole fluctuating to nonmagnetic $4f^{14}$ and magnetic $4f^{12}$ states.
We will also consider that CF splitting results in a low-energy Kramers doublet. Our aim is to derive an effective spin-$1/2$ Hamiltonian that captures the interaction between these low-energy doublets mediated by the conduction electrons. Via direct comparison against inelastic neutron scattering (INS) data, this effective spin model can be used to indirectly validate the PAM and the KHLM derived from our first-principles calculations.

Recently, we proposed a material-specific low-energy effective Hamiltonian for the heavy-fermion antiferromagnet CeIn$_3$~\cite{SimethW2023_CeIn3}. Beginning with a multiorbital PAM (MO-PAM) with parameters obtained from first-principles calculations, we employed degenerate perturbation theory, keeping only nonmagnetic $4f^0$ virtual states, to deduce a quantitative yet relatively straightforward effective (KHLM) Hamiltonian. This model includes only a narrow energy range (\SI{1}{eV}) of conduction band states and incorporates the rest of the band states via an effective ``superexchange'' Heisenberg interaction. Validation of this low-energy model was achieved through comparison with zero-field INS data, revealing the low-energy magnon dispersion originating from the low-energy $\Gamma_7$ doublet.

In this study, we aim to generalize the methodology outlined in Ref.~\cite{SimethW2023_CeIn3} to comprehend the origin of exchange anisotropy in the effective exchange interaction between Kramers doublets. As we will see below, only fluctuations via $4f^2$ virtual states can induce an effective exchange anisotropy for centrosymmetric bonds. Furthermore, we will ``dissect'' the $4f^2$ virtual processes to separate the ones that contribute to  the isotropic part of the effective exchange interaction from those that generate exchange anisotropy. This analysis will allow us to identify conditions under which the exchange anisotropy should be weak in comparison to the isotropic contribution. 


Additionally, we will demonstrate that the incorporation of $4f^2$ virtual states in deriving an effective spin Hamiltonian for CeIn$_3$ results in an energy difference between the Fermi level $E_F$ and the $4f$ energy level $\varepsilon_f$ of $E_F-\varepsilon_f=\SI{0.65}{eV}$.  While this value is closer to  photoemission experiments ($E_F-\varepsilon_f=1.5$ - \SI{2.0}{eV})~\cite{KimHD1997, ZhangY2016} in comparison to the value $E_F-\varepsilon_f=\SI{0.512}{eV}$ obtained in Ref.~\cite{SimethW2023_CeIn3}, which only considered fluctuations via the nonmagnetic $4f^0$ configuration, it is not enough to reproduce the photoemission data. As we will see below, this discrepancy suggests that the first-principles calculations underestimate the $f$-$c$ hybridization amplitudes by a factor $\approx 1.22$.

This manuscript is structured as follows. In Sec.~\ref{sec:TB-Ham}, we present the tight-binding Hamiltonian constructed from density functional theory (DFT). In Sec.~\ref{sec:PAM}, we outline the derivation of a MO-PAM based on first-principles calculations. In Sec.~\ref{sec:KLM}, we implement degenerate perturbation theory to derive the Kondo lattice model (KLM) from the PAM. Section~\ref{sec:KHM} elaborates on the derivation of the KHLM  using degenerate second-order perturbation theory in the hybridization term between the $4f$ orbitals and the conduction electron bands, encompassing fluctuations via $4f^0$ and $4f^2$ virtual states. In Sec.~\ref{sec:SPIN}, we derive an effective spin Hamiltonian employing fourth-order degenerate perturbation theory in the hybridization term, assuming the $f$ electrons are localized. In Sec.~\ref{sec:exan}, we  analyze the origin of exchange anisotropy and the relative contributions of processes occurring in the particle-hole and particle-particle channels, as well as via $4f^0/4f^2$ virtual states.  Section~\ref{sec:VAL} presents an experimental validation of the effective spin Hamiltonian through comparison with experimental data obtained from CeIn$_3$. Finally, Sec.~\ref{sec:CONC} discusses the conclusions and implications of this work.


\section{Tight-binding Hamiltonian} \label{sec:TB-Ham}

To extract the magnetic exchange interactions from first principles, we follow the methodology outlined in Ref.~\cite{SimethW2023_CeIn3}. This approach provides the dispersion of the conduction bands and their hybridization with the $f$ orbitals, forming a material-specific PAM. 
We note that the energy $\varepsilon_f$ of the $f$ level and the Coulomb integrals $F^0$, $F^2$, $F^4$, and $F^6$ associated with the double occupancy of $f$ orbitals must be determined by other means. 


We start with the reasonable assumption that the PAM serves as an accurate effective description for elucidating the low-energy physics of most, if not all, heavy-fermion systems based on $4f$ electrons. 
To exploit translational invariance, we identify the ions in the crystallographic unit cell by their type and coordinates ${\bm R}^{\alpha}_{j_{\alpha}}$, where $1 \leq j_{\alpha} \leq N^{\alpha}_s$ and $N^{\alpha}_s$ is the number of ions of type $\alpha$ included in the cell. The corresponding creation and annihilation operators are denoted by $a^{\dagger}_{ {\bm R}^{\alpha}_{j_{\alpha}}, n_{\alpha}}$ and $a^{\;}_{ {\bm R}^{\alpha}_{j_{\alpha}}, n_{\alpha}}$, respectively. 
The index  $1 \leq n_{\alpha} \leq N_{\alpha}$ runs over the different non-$f$ Wannier orbitals of a given $\alpha$ ion. 
For example, $\alpha=0,1$ for CeIn$_3$ because there are two types of ions. The $\alpha=0$ ion is chosen to be the one that contains the $f$ orbitals, meaning that $\alpha=0$ denotes the Ce ions for CeIn$_3$. 
For simplicity of notation, we will assume that there is only one $f$ ion in each crystallographic unit cell, i.e., $N_s^0=1$,  and that it is located at the origin: ${\bm R}^0_1 = {\bm 0}$.

Because of the Kramers degeneracy of each Wannier orbital, it is convenient to introduce $2 \times 2$ hopping matrices,
\begin{equation}
\mathbb{t}_{n_{\alpha} n_{\beta}}({\bm r}_{\alpha} - {\bm r}_{\beta}) = t_{n_{\alpha} n_{\beta}}({\bm r}_{\alpha} - {\bm r}_{\beta})
\mathbb{U}_{n_{\alpha} n_{\beta}}({\bm r}_{\alpha} - {\bm r}_{\beta}),
\label{eq:hopmat}
\end{equation}
where the scalar amplitudes  $t_{n_{\alpha} n_{\beta}}$ can be assumed to be real (for an adequate gauge choice) because of time-reversal invariance and $\mathbb{U}_{n_{\alpha} n_{\beta}}({\bm r}_{\alpha} - {\bm r}_{\beta})$  is a $2 \times 2$ matrix with determinant equal to one, i.e., 
$\mathbb{U}_{n_{\alpha} n_{\beta}}({\bm r}_{\alpha} - {\bm r}_{\beta}) \in$ SU(2). We note that in the absence of spin-orbit coupling, the 
$\mathbb{U}_{n_{\alpha} n_{\beta}}({\bm r}_{\alpha} - {\bm r}_{\beta})$ matrices become identity matrices for a particular gauge choice, i.e., for a particular choice of the local spin reference frame of  each orbital (Kramers doublet).

The tight-binding Hamiltonian for the conduction bands can then be expressed as:
\begin{equation}
{\cal H}_{\rm TB} = \!\!\! \sum_{{\bm R}, {\bm R}^\prime } \!
{\bm a}^{\dagger}_{ {\bm R}^{\alpha}_{j_{\alpha}}\!+{\bm R}, n_{\alpha}}  \mathbb{t}^{\alpha \beta}_{n_{\alpha} n_{\beta}}(\Delta {\bm R}^{\alpha \beta}_{j_{\alpha}, j_{\beta}} \!+ \Delta_{{\bm R}, {\bm R}^{\prime}})  {\bm a}^{\;}_{ {\bm R}^{\beta}_{j_{\beta}}\!+{\bm R}^{\prime}, n_{\beta}},
\end{equation}
where $ \Delta {\bm R}^{\alpha \beta}_{j_{\alpha}, j_{\beta}} \equiv {\bm R}^{\alpha}_{j_\alpha} - {\bm R}^{\beta}_{j_\beta}$,
$ \Delta_{{\bm R}, {\bm R}^{\prime}} \equiv {\bm R}- {\bm R}^\prime$,
and $\{ {\bm R} \}$ denotes the set of translations that leaves the system invariant. Unless specified otherwise, here and throughout this work, we adhere to the convention of summation over repeated indices,  except for the unit cell coordinates  (${\bm R}$) and wavevectors ($\bm{k}$ or $\bm q$) or for cases where we make the sum explicit for clarity of presentation. We  have also introduced the spinor notation, 
\begin{eqnarray}
{\bm a}^{\;}_{{\bm R}^{\alpha}_{j_{\alpha}}+{\bm R}, n_{\alpha}} &\equiv& ({a}^{\;}_{{\bm R}^{\alpha}_{j_{\alpha}}+{\bm R}, n_{\alpha}, \uparrow},  {a}^{\;}_{{\bm R}^{\alpha}_{j_{\alpha}}+{\bm R}, n_{\alpha}, \downarrow})^T,
\end{eqnarray}
Note that $\mathbb{t}^{\alpha \beta}_{n_{\alpha} n_{\beta}}({\bm r}_{\alpha} - {\bm r}_{\beta}) = [\mathbb{t}^{\beta \alpha}_{n_{\beta} n_{\alpha}}({\bm r}_{\beta} - {\bm r}_{\alpha})]^{\dagger} $ because ${\cal H}_{\rm TB} $ is Hermitian.

After introducing similar two-component spinor notation for the $f$ orbitals with fixed projection of the orbital angular momentum $l_z=m$ 
($-3 \leq m \leq 3$),
\begin{equation}
{\bm f}^{\;}_{{\bm R},m} \equiv ( f^{\;}_{{\bm R},m, \uparrow}, f^{\;}_{{\bm R},m, \downarrow}  )^T ,
\end{equation}
we can write the hybridization term between the $f$-Wannier orbitals and the rest of the Wannier orbitals,
\begin{equation}
{\cal H}_{\rm V} = \sum_{{\bm R}, {\bm R}^{\prime}} 
 {\bm f}^{\dagger}_{{\bm R},m}  \mathbb{V}^{\alpha}_{m j_{\alpha} n_{\alpha}}(\Delta_{{\bm R}, {\bm R}^{\prime}}- {\bm R}^{\alpha}_{j_\alpha})  {\bm a}^{\;}_{{\bm R}^{\alpha}_{j_{\alpha}}+{\bm R}^{\prime},n_{\alpha}} 
+ {\rm H. c. },
\end{equation}
where $\mathbb{V}^{\alpha}_{m j_{\alpha} n_{\alpha}}(\Delta_{{\bm R}, {\bm R}^{\prime}}- {\bm R}^{\alpha}_{j_\alpha})$ is the $2\times 2$ matrix that hybridizes the $l_z=m$ $f$ doublet   centered at the ion ${\bm R}$ with the spin doublet of the Wannier  orbital $n_{\alpha}$ centered at the ion ${\bm R}^{\alpha}_{j_\alpha} + {\bm R}^{\prime}$.

To exploit translational invariance, we introduce the  Fourier transform of the creation and annihilation  operators, 
\begin{eqnarray}
{\tilde {\bm a}}^{\dagger}_{{\bm k}, \alpha, {j_{\alpha}}, n_{\alpha}} &=& \frac{1}{\sqrt{N_u}} \sum_{\bm R} e^{i {\bm k} \cdot ({\bm R}+ {\bm R}_{j_{\alpha}} )} \bm a^{\dag}_{{\bm R}^{\alpha}_{j_{\alpha}}+{\bm R}, n_{\alpha}},
\nonumber \\
{\tilde  {\bm a}}^{\;}_{{\bm k}, \alpha, {j_{\alpha}}, n_{\alpha}} &=& \frac{1}{\sqrt{N_u}} \sum_{\bm R} e^{-i {\bm k} \cdot ({\bm R}+ {\bm R}_{j_{\alpha}} )} \bm a^{\;}_{{\bm R}^{\alpha}_{j_{\alpha}}+{\bm R}, n_{\alpha}},
\label{eq:FT}
\end{eqnarray}
where $N_u$ is the number of unit cells. After rewriting ${\cal H}_{\rm TB}$ in momentum space, we obtain
\begin{equation}
{\cal H}_{\rm TB} = \sum_{ {\bm k} }
\tilde{\bm a}^{\dagger}_{{\bm k}, \alpha, j_{\alpha}, n_{\alpha}}  \tilde{\mathbb{t}}^{\alpha \beta}_{ {j_{\alpha} j_{\beta},} n_{\alpha} n_{\beta}}({\bm k})  \tilde{\bm a}^{\;}_{{\bm k}, \beta, j_{\beta}, n_{\beta}},
\end{equation}
with

\begin{equation}
\tilde{\mathbb{t}}^{\alpha \beta}_{ {j_{\alpha} j_{\beta},} n_{\alpha} n_{\beta}}({\bm k}) = \! \sum_{\bm R} e^{-i {\bm k}\cdot (\Delta {\bm R}^{\alpha \beta}_{j_{\alpha}, j_{\beta}}+ {\bm R})} \; \mathbb{t}^{\alpha \beta}_{n_{\alpha} n_{\beta} }(\Delta {\bm R}^{\alpha \beta}_{j_{\alpha}, j_{\beta}}+ {\bm R}) .
\label{eq:mome}
\end{equation}
Note that if the system has spatial inversion symmetry, 
\begin{equation}
\mathbb{t}^{\alpha \beta}_{n_{\alpha} n_{\beta} }(\Delta {\bm R}^{\alpha \beta}_{j_{\alpha}, j_{\beta}} + {\bm R}) = \mathbb{t}^{\alpha \beta}_{n_{\alpha} n_{\beta} }(-\Delta {\bm R}^{\alpha \beta}_{j_{\alpha}, j_{\beta}} - {\bm R}),
\end{equation}
we can rewrite Eq.~\eqref{eq:mome} as 
\begin{widetext}
\begin{eqnarray}
\tilde{\mathbb{t}}^{\alpha \beta}_{j_{\alpha} j_{\beta}, n_{\alpha} n_{\beta} }({\bm k}) = 
 \sum_{\bm R} \cos{[{\bm k}\cdot  ({\bm R}+ \Delta {\bm R}^{\alpha \beta}_{j_{\alpha}, j_{\beta}}) ]} \; \mathbb{t}^{\alpha \beta}_{n_{\alpha} n_{\beta} }(\Delta {\bm R}^{\alpha \beta}_{j_{\alpha}, j_{\beta}} + {\bm R}).
\label{eq:mome2}
\end{eqnarray}
\end{widetext}

By using Eq.~\eqref{eq:hopmat} and the fact that the amplitudes $t_{n_{\alpha} n_{\beta}}^{\alpha\beta}$ are real, we conclude that, in the presence of spatial inversion symmetry,  the $2 \times 2$ matrix $\tilde{\mathbb{t}}^{\alpha \beta}_{ j_{\alpha} j_{\beta},n_{\alpha} n_{\beta}}({\bm k})$
can be expressed as 
\begin{equation}
\tilde{\mathbb{t}}^{\alpha \beta}_{ j_{\alpha} j_{\beta}, n_{\alpha} n_{\beta}}({\bm k}) = \tilde{t}^{\alpha \beta}_{j_{\alpha} j_{\beta}, n_{\alpha} n_{\beta} }({\bm k}) \tilde{\mathbb{U}}^{\alpha \beta}_{j_{\alpha} j_{\beta}, n_{\alpha} n_{\beta}}({\bm k}),
\label{eq:real}
\end{equation}
where $\tilde{t}^{\alpha \beta}_{j_{\alpha} j_{\beta}, n_{\alpha} n_{\beta}}({\bm k})$ is a real number and $\tilde{\mathbb{U}}^{\alpha \beta}_{j_{\alpha} j_{\beta}, n_{\alpha} n_{\beta}}({\bm k})  \in$ SU(2).

The diagonal form of ${\cal H}_{\rm TB}$ 
\begin{eqnarray}
{\cal H}_{\rm TB} = \sum_{{\bm k}} \varepsilon_{{\bm k}, \mu} {\bm c}^{\dagger}_{{\bm k}, \mu} {\bm c}^{\;}_{{\bm k}, \mu} 
\end{eqnarray}
can be finally obtained by implementing a unitary transformation:
\begin{eqnarray}
{\bm c}^{\;}_{{\bm k}, \mu} \!\!  &=& \!\!\! \sum_{\alpha, {j_{\alpha}, n_{\alpha}}} \!\!\! b_{\alpha, j_{\alpha}, n_{\alpha}}({\bm k}, \mu) \mathbb{U}_{\alpha, j_{\alpha}, n_{\alpha}}({\bm k}, \mu) \tilde{\bm a}_{{\bm k}, \alpha, {j_{\alpha}, n_{\alpha}}},
\nonumber \\
\tilde{\bm a}_{{\bm k}, \alpha, {j_{\alpha}, n_{\alpha}}} \!\! &=& \sum_{\mu} b^*_{\alpha, j_{\alpha}, n_{\alpha}}({\bm k}, \mu) 
\mathbb{U}^{\dagger}_{\alpha, j_{\alpha}, n_{\alpha}}({\bm k}, \mu) {\bm c}^{\;}_{{\bm k}, \mu},
\label{eq:uni}
\end{eqnarray}
where repeated indices are not implicitly summed.
The coefficients $b_{\alpha, j_{\alpha},n_\alpha}({\bm k}, \mu)$ are scalars that satisfy the orthonormality condition,
\begin{equation}
\sum_{\alpha, j_{\alpha},n_\alpha}  b^*_{\alpha, j_{\alpha},n_\alpha}({\bm k}, \mu) b_{\alpha, j_{\alpha},n_\alpha}({\bm k}, \mu') = \delta_{\mu \mu'},
\end{equation}
and $\mathbb{U}_{\alpha, j_{\alpha},n_\alpha}({\bm k}, \mu)$ are unitary $2 \times 2$ matrices with determinant equal to one. 
The dispersion of the conduction bands,  $\varepsilon_{\bm{k},\mu }$, does not depend on the spin index because of Kramers' theorem.
Equation~\eqref{eq:real} implies that 
$b^*_{\alpha, j_{\alpha},n_\alpha}({\bm k}, \mu) = b_{\alpha, j_{\alpha},n_\alpha}({\bm k}, \mu)$, i.e., the coefficients of the unitary transformation are real, in the presence of spatial inversion symmetry. 

Similarly, 
after introducing the Fourier transform of the ${\bm f}^{\;}_{ {\bm R},m}$ operators
\begin{eqnarray}
{\bm f}^{\;}_{{\bm k},m} &=& \frac{1}{\sqrt{N_u}} \sum_{\bm k} e^{-i {\bm k} \cdot {\bm R}} {\bm f}^{\;}_{ {\bm R},m},
\nonumber \\
{\bm f}^{\;}_{ {\bm R},m} &=& \frac{1}{\sqrt{N_u}} \sum_{\bm R} e^{i {\bm k} \cdot {\bm R}} {\bm f}^{\;}_{{\bm k},m},
\end{eqnarray}
we can rewrite ${\cal H}_{\rm V}$ as
\begin{equation}
{\cal H}_{\rm V} = \sum_{ {\bm k}} [{\bm f}^{\dagger}_{{\bm k}, m}
{\mathbb{V}}^{\alpha}_{m j_{\alpha} n_{\alpha}}({\bm k}) \tilde{\bm a}_{{\bm k}, \alpha, j_{\alpha}, n_{\alpha}} + {\rm H. c.}],
\end{equation}
with
\begin{equation}
{\mathbb{V}}^{\alpha}_{m j_{\alpha} n_{\alpha}}({\bm k})  = \sum_{\bm R} e^{i {\bm k} \cdot ({\bm R}-{\bm R}^{\alpha}_{j_{\alpha}})} \mathbb{V}^{\alpha}_{m j_{\alpha} n_{\alpha}}(  {\bm R} - {\bm R}^{\alpha}_{j_{\alpha}}).
\end{equation}

Finally, after applying the unitary transformation \eqref{eq:uni} that diagonalizes the tight-binding Hamiltonian, we get
\begin{equation}
{\cal H}_{\rm V} = \sum_{{\bm k} } [{\bm f}^{\dagger}_{{\bm k}, m}
{\mathbb{V}}_{{\bm k} m \mu}  {\bm c}_{{\bm k}, \mu} + {\rm H. c.}]
\label{eq:hyb3}
\end{equation}
with 
\begin{equation}
{\mathbb{V}}_{{\bm k} m \mu } =  b^*_{\alpha, j_{\alpha}, n_{\alpha}} ({\bm k}, \mu) {\mathbb{V}}_{m j_{\alpha} n_{\alpha}}^{\alpha}({\bm k}) \mathbb{U}^{\dagger}_{\alpha, j_{\alpha}, n_{\alpha}}({\bm k}, \mu). 
\label{eq:unitary}
\end{equation}
Equation~\eqref{eq:hyb3} can be further simplified to
\begin{equation}
{\cal H}_{\rm V} = \sum_{{\bm k}, \mu } [{\bm f}^{\dagger}_{{\bm k}}
{\mathbb{V}}_{{\bm k}\mu} {\bm c}_{{\bm k}, \mu} + {\rm H. c.}]
\end{equation}
by introducing the $14$-component Nambu spinor, 
\begin{equation}
\bm f^{\;}_{\bm k} = (f_{{\bm k},-3\uparrow},\dots;f_{{\bm k},3\uparrow},f_{{\bm k},-3\downarrow},\dots,f_{{\bm k},3\downarrow})^T,
\end{equation}
where the $14 \times 2$ hybridization matrix ${\mathbb{V}}_{{\bm k} \mu }$ is obtained from the seven $2 \times 2$ matrices ${\mathbb{V}}_{{\bm k} m \mu }$.

\section{Periodic Anderson Model} \label{sec:PAM}

The MO-PAM can be expressed as 
\begin{equation}\label{eq:AndersonModel}
    \mathcal{H}_{\textrm{MO-PAM}} = \mathcal{H}_0 + \mathcal{V},
\end{equation}
with
\begin{eqnarray}
    \mathcal{H}_0 \!\!\! &=& \!\!\!\sum_{\bm{k}} \varepsilon_{\bm{k},\mu }^{} {\bm c}^{\dagger}_{\bm{k},\mu }    {\bm c}_{\bm{k},\mu} +  \sum_{\bm R} \bm f_{\bm R}^\dag ( \mathbb{\Lambda} + \varepsilon_f ) \bm f_{\bm R} + \mathcal{H}_U , \ \  \\
    \mathcal{V} \!\!\! &=& \!\!\! \sum_{\bm R} \bm f_{\bm R}^\dag  \mathbb{C} \bm f_{\bm R} + \!\! \sum_{\bm{k}} \! \left( \bm f^{\dagger}_{\bm k} \mathbb{V}_{\bm{k}\mu}^{} \bm c_{\bm{k},\mu}^{} + \bm c_{\bm{k},\mu}^{\dag} \mathbb{V}_{\bm{k} \mu}^{\dag} \bm f_{\bm k}^{}  \right)\!\!, \ \ \label{eq:hyb}
\end{eqnarray}
where $\varepsilon_f$ is the energy of the $f$ orbital. The intra-atomic spin-orbit coupling is described by the matrix $[\mathbb{\Lambda}]_{m\sigma,m'\sigma'} = \bm L_{m,m'} \cdot \bm S_{\sigma,\sigma'} $, where $\bm L$ and $\bm S$ are the vector of angular momentum and spin, respectively. The matrix $\mathbb{C}$ represents the crystal-field
and the term $\mathcal{H}_U$ is the Coulomb repulsion in the ion that is determined by the four Slater integrals
$F^0$, $F^2$, $F^4$, and $F^6$(see Appendix \ref{app:Ce-ion}).

In the context of CeIn$_3$ and related materials, it is widely recognized that the $f^2$ state is considerably higher in energy compared to the $f^0$ excited state~\cite{SundermannM2016, BraicovichL2007}. In a previous study~\cite{SimethW2023_CeIn3}, we assumed an infinitely large on-site repulsive interaction between $f$ electrons, effectively excluding $f^2$ configurations from the Hilbert space.
However, as we will elucidate in this work, this assumption results in isotropic effective spin-spin interactions up to fourth order in the hybridization $\mathbb{V}_{\bm{k}\mu }$ whenever the center of the bond under consideration is a center of spatial inversion symmetry.
In other words, it becomes necessary to incorporate processes via the magnetic $f^2$ virtual states to account for the origin of exchange anisotropy in materials such as CeIn$_3$, where the center of each Ce-Ce bond is a center of inversion. 

The dispersion of the conduction bands, denoted as $\varepsilon_{\bm{k},\mu}$, and the hybridization between the conduction bands and the $f$ orbitals, represented as $\mathbb{V}_{\bm{k}\mu}$, are determined through the tight-binding fit to the first-principles DFT calculation.  These parameters  are identical to the ones used in Ref.~\cite{SimethW2023_CeIn3}. 
We project out the excited $J=7/2$ spin-orbit split states from the $f$ manifold of states, as they lie roughly 300 meV higher in energy than the $J=5/2$ states. Generically, the CF splits the $J=5/2$ states into three doublets. For CeIn$_3$ the cubic CF splits the $J=5/2$ states into a doublet $\Gamma_7$ and a quartet $\Gamma_8$. We set the energy of the $\Gamma_7$ doublet $\varepsilon_{f}^{}$ to  \SI{0.65}{eV} below the chemical potential, which was determined by the overall magnon bandwidth of \SI{2.75}{meV} measured by INS. We further fix the energy of the $\Gamma_8$ quartet $\varepsilon_{\Gamma_8}^{f}$ to lie $\simeq$ \SI{12}{meV} above the $\Gamma_7$ ground state doublet, as determined by INS measurements of the CF levels~\cite{LawrenceJM1980,KnafoW2003}. Thus, {\it all parameters of our reduced PAM can be determined from a combination of ab initio-based first-principles calculations and from spectroscopic measurements of the $f$ electrons}.

\section{Effective Low Energy models \label{sec:LE}}

\subsection{Kondo Lattice Model}\label{sec:KLM}

\begin{figure}[h!]
    \centering
    \includegraphics[width=0.48\textwidth]{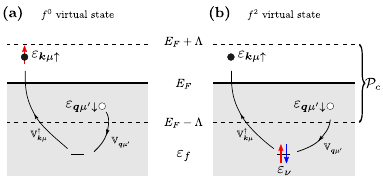}
    \caption{Schematic representation of the virtual processes of the  second-order perturbation theory that leads to the KLM. The virtual state consist of a particle-hole pair. The particle (hole) is created in a band state with energy $\varepsilon_{\bm k\mu}$ ($\varepsilon_{\bm q\mu'}$).  $E_F$ denotes the Fermi level  and $\varepsilon_f$ is the energy of the $f$ orbital. (a) The process that produces an $f^0$ virtual state; (b) The process that produces an $f^2$ virtual state. The operator $\mathcal{P}_c$ projects into the low-energy states delimited by the energy cutoff signaled by $E_F\pm\Lambda$.  }
    \label{fig:Kondo-diag}
\end{figure}

Our next step is to derive a low-energy KLM Hamiltonian through degenerate perturbation theory. We start by identifying the low-energy subspace ${ S}$ within the unperturbed Hamiltonian $\mathcal{H}_0$. For the formalism of degenerate perturbation theory to apply, 
${S}$ must constitute the ground space of $\mathcal{H}_0$. Given the gapless spectrum of $\mathcal{H}_0$ due to the presence of a Fermi surface, we introduce an energy cutoff $\Lambda$, as illustrated in Fig.~\ref{fig:Kondo-diag}. We then approximate that all band states with energies $\varepsilon_{\bm{k}, \mu }$ falling within the range 
$\left|\varepsilon_{\bm{k}, \mu }-E_F\right| \leq \Lambda$ possess approximately equal energies.

Upon making this approximation, the subspace  $S$ becomes the direct product of singly-occupied $J=J_m$ $f$ states and the subspace of states featuring any number of particle-hole excitations within the energy window  $\left|\varepsilon_{\bm{k}, \mu }-E_F\right| \leq \Lambda$. Subsequently, the projector onto  $S$ can be expressed as the direct product of two projectors:  $\mathcal{P}_S = \mathcal{P}_{J_m} \otimes \mathcal{P}_c$. Here, $\mathcal{P}_{J_m}$ denotes the projector onto the singly occupied  $f$ states with total angular momentum  $J=J_m$ that minimizes the spin-orbit coupling $\mathbb{\Lambda}$. For instance $J_m=5/2$ for Ce$^{3+}$ ions, while $J_m=7/2$ for Yb$^{3+}$ ions. 
$\mathcal{P}_c$  represents the projector onto the Fock space associated with band states satisfying the condition  $\left|\varepsilon_{\bm{k}, \mu }-E_F\right| \leq \Lambda$. 

The low-energy effective Hamiltonian to second order in the perturbation $\mathcal{V}$ reads~\cite{LindgrenI1974}
\begin{equation}
    \mathcal{H}_{\rm eff} = \mathcal{P}_{S} \mathcal{H} \mathcal{P}_{S} + \mathcal{P}_{S} \left[  \mathcal{V}  \hat G^{0}(\varepsilon_0)  \mathcal{V}   \right] \mathcal{P}_{S},
\label{eq:can}
\end{equation}
where the resolvent operator is given by
\begin{equation}\label{eq:resolvent}
    \hat G^{0}(\varepsilon_0) = \mathcal{P}_{S}^\perp \frac{1}{\varepsilon_0 - \mathcal{ H}_{\rm 0} } \mathcal{P}_{S}^\perp \ ,
\end{equation}
$\mathcal{P}_{S}^\perp = \hat{1} - \mathcal{P}_{S}$, and $\varepsilon_0$ is the ground state energy of $\mathcal{ H}_{\rm 0}$.

The projection of $\mathcal{H}_{\textrm{MO-PAM}}$ is 
\begin{align}
    \tilde{\mathcal{H}} =&  \sum_{\bm{k}} \varepsilon_{\bm{k},\mu }^{} {\bm c}^{\dagger}_{\bm{k},\mu }    {\bm c}_{\bm{k},\mu}+ \sum_{i} \tilde {\bm f}_{i}^\dag \mathbb{\tilde C} \tilde{\bm f}_{i}^{},
\end{align}
where $\tilde{\mathcal{H}} =\mathcal{P}_{S} \mathcal{H}_{\textrm{MO-PAM}} \mathcal{P}_{S} $, $\mathbb{\tilde C} = \mathcal{P}_{S} \mathbb{C} \mathcal{P}_{S} $ is  the projected CF matrix,  and $\tilde {\bm f}_{i}^\dag = (\tilde f_{i,-J_m}, \cdots, \tilde f_{i,J_m})$ is restricted to the subspace $J=J_m$. After adding the second-order contribution in the hybridization, i.e., the second term of Eq.~\eqref{eq:can}, we obtain the
KLM Hamiltonian,
\begin{align}\label{eq:KLM}
    \mathcal{H}_{\rm KL} & = \sum_{\bm{k}}' \varepsilon_{\bm{k},\mu }^{} {\bm c}^{\dagger}_{\bm{k},\mu }    {\bm c}_{\bm{k},\mu} + \sum_{i} \tilde {\bm f}_{i}^\dag \mathbb{\tilde C} \tilde{\bm f}_{i} \nonumber \\
    &- \frac{1}{N} \sum_{i,{\bm k}, {\bm q}}' e^{i(\bm q - \bm k)\cdot \bm r_{i}}  
    \mathcal{S}_{i}^{\sigma\sigma'} J_{\sigma \sigma', s s'}^{\bm k \mu , \bm q \mu'}  \ c_{\bm k \mu s}^{\dag} c_{\bm q \mu' s'}^{}.
\end{align}
The apostrophe on the summation indicates that particles and holes are created in states within the energy cut-off $E_F\pm\Lambda$, the matrix of the local U($2J_m+1$) generators is $\mathcal{S}_{i}^{\sigma\sigma'} = \tilde f_{i\sigma^{}}^{\dag} \tilde f_{i\sigma'}^{}$,
and the Kondo interaction matrix is given by 
\begin{equation}\label{eq:Kondo-coupling}
    J_{\sigma \sigma', s s'}^{\bm k \mu ,\bm q \mu'} \simeq  \frac{ [\mathbb{V}_{\bm q \mu'}]_{\sigma s'}^{}   [\mathbb{V}_{\bm k \mu}^*]_{\sigma' s}^{}  }{ \varepsilon_{f} - E_F }  
    -   \frac{ [\mathbb{V}_{\bm{q} \mu'}]_{\alpha s'}^{}  [\mathbb{V}_{\bm{k} \mu}^*]_{\alpha' s}^{} \   \gamma_{\sigma\sigma', \alpha \alpha'}^{\nu} }{ E_F -\varepsilon_{f} - \varepsilon_\nu } 
\end{equation}
The first term is due to the virtual processes that involve empty $f$ orbital ($f^0$ states) and the second term involves doubly occupied $f$-orbital ($f^2$-state) virtual processes; see Appendix~\ref{app:KLM}. 
The energies $\varepsilon_\nu$ correspond to the eigenvalues of the single-ion Hamiltonian with two $f$ particles, i.e., two $f$ electrons (4$f^2$) or two $f$ holes (4$f^{12}$), 
\begin{align}\label{eq:H_ion}
    \mathcal{H}_{{\rm SI}, 2p} = \sum_{\nu} \varepsilon_\nu |\eta_\nu \rangle \langle \eta_\nu | \,
\end{align}
\emph{in the absence of CF splitting}. 
Note that the 91 eigenstates $|\eta_\nu \rangle$ can be grouped into 13 $J$ multiplets, with values of $J$ ranging from 0 to 6 (there is more than one multiplet for some values of $J$). 
Since the $2J+1$  eigenvalues for each multiplet must be degenerate because of rotational invariance, $\varepsilon_{\nu}$ takes only 13 different values. For instance, Fig.~\ref{fig:f2-spectrum} shows the $f^2$ energy spectrum for Ce$^{2+}$.
From now on, whenever the projector $|\eta_\nu \rangle \langle \eta_\nu | $ is written for a specific $f$ ion in the site $j$, we will denote it as $|\eta_{j\nu} \rangle \langle \eta_{j \nu} |$.

The
function $\gamma_{\sigma\sigma', \alpha \alpha'}^{\nu}$ accounts for the projection of  $f^2$ states that are a direct product of two single-particle ($f^1$) states into the eigenstates $|\eta_\nu \rangle $ of the single-ion Hamiltonian, 
\begin{equation}\label{eq:gamma-def}
     \gamma_{\sigma \sigma';\alpha \alpha'}^{\nu} \equiv \langle 0 | f_{\sigma}^{}  f_{\alpha'}^{} |\eta_\nu  \rangle \langle \eta_\nu| f_{\alpha}^{\dag} f_{\sigma'}^{\dag} | 0 \rangle,
\end{equation}
where $\sigma$ and $\sigma'$ are, respectively, the final and initial $f^1$ states, while $\alpha$ ($\alpha'$) is the state where the second particle is created (annihilated).


\subsection{Kondo-Heisenberg Lattice Model} \label{sec:KHM}

The Kondo lattice model describes the effective exchange interaction between localized $f$ moments and particle-hole excitations. As pointed out by Doniach~\cite{DoniachS1977}, deep inside the perturbative regime where the excitation energy of $f^0$ and $f^2$ configurations is much higher than the hybridization, the effective Ruderman-Kittel-Kasuya-Yosida (RKKY) interaction~\cite{RudermanMA1954,KasuyaT1956,YosidaK1957} dominates over Kondo screening, and conduction band degrees of freedom can be integrated out to derive a low-energy spin Hamiltonian $\mathcal{H}_{\rm spin}$ that accounts for the effective interaction between $f$ moments mediated by conduction electrons. 

Normally, $\mathcal{H}_{\rm spin}$ is derived from $\mathcal{H}_{\mathrm{KL}}$ by implementing second-order degenerate perturbation theory in the Kondo interaction term, which corresponds to fourth-order particle-hole processes in the hybridization term of the original PAM. The resulting effective spin model is the RKKY Hamiltonian $\mathcal{H}_{\rm RKKY}$. The problem with this approach is that it neglects additional fourth-order processes in $\mathbb{V}_{\boldsymbol{k} \mu}$ that can provide comparable contributions to $\mathcal{H}_{\text {spin }}$. These processes, which  will be grouped under $\mathcal{H}_{\rm Ex}$, include any fourth-order contribution from the particle-particle channel and fourth-order contributions in the particle-hole channel involving band states outside the range $\left|\varepsilon_{\bm{k}, \mu }-E_F\right| \leq \Lambda$, and will be explicitly derived in the next section. 

In summary, the KLM must be supplemented with an exchange interaction between the localized moments, $\mathcal{H}_{\rm Ex }$, if we want to derive the same $\mathcal{H}_{\rm spin}$ that would be obtained from the PAM by direct application of  fourth order degenerate perturbation theory in $\mathbb{V}_{\boldsymbol{k} \mu}$. The resulting KHLM is
\begin{equation}
\mathcal{H}_{\mathrm{KHL}} = \mathcal{H}_{\mathrm{KL}}+     \mathcal{H}_{\rm Ex},
\end{equation}
where $\mathcal{H}_{\mathrm{KL}}$ is given in Eq.~\eqref{eq:KLM} and $\mathcal{H}_{\rm Ex}$ will be derived in the next section as part of the contributions to $\mathcal{H}_{\text {spin }}$.  We note that in general, $\mathcal{H}_{\rm Ex}$	
  may include anisotropic exchange interactions due to the interplay of spin-orbit coupling and crystalline anisotropy.

\section{Effective Spin Model} \label{sec:SPIN}

The goal of this section is to derive an effective spin model $\mathcal{H}_{\rm spin}$ by applying fourth-order degenerate perturbation theory to 
the MO-PAM,
\begin{align}
    \mathcal{H}_{\rm MO-PAM} \hspace{2mm} \xrightarrow{{\cal O}({V^4})} \hspace{2mm}  \mathcal{H}_{\rm spin}   = \mathcal{H}_{\rm RKKY}  + \mathcal{H}_{\rm Ex} 
\end{align}
 where $\mathcal{H}_{\rm RKKY}$ includes the particle-hole processes  whose middle virtual state \emph{has one particle-hole excitation within the cutoff $\left|\varepsilon_{\bm{k}, \mu }-E_F\right| \leq \Lambda$}. Once again, we note that these processes are the ones  obtained from $\mathcal{H}_{\mathrm{KL}}$ via second-order perturbation theory in the Kondo interaction.
As elucidated in the preceding section, in the strongly localized $f$-electron regime, the magnetic behavior is governed by effective spin-spin interactions mediated by the band states. These interactions can be obtained by directly applying fourth-order degenerate perturbation theory to the MO-PAM. Since the only low-energy degrees of freedom are the $f$ moments, the low-energy subspace $S_0$ simply consists of the direct product of all possible states of the $J=J_m$ $f$ moments and the Fermi sea, i.e., the ground state of the conduction electrons in the absence of hybridization with the $f$ orbitals. We denote this projector by $\mathcal{P}_{S_0}$. The effective spin Hamiltonian is then expressed as~\cite{LindgrenI1974}
\begin{equation}\label{eq:Hspin}
    \mathcal{H}_{\rm spin} = \mathcal{P}_{S_0} \left[ \mathcal{V} \hat G^{0}(\varepsilon_0) \mathcal{V} \hat G^{0}(\varepsilon_0) \mathcal{V} \hat G^{0}(\varepsilon_0) \mathcal{V}   \right] \mathcal{P}_{S_0} \ .
\end{equation}

We observe that fourth-order processes in the hybridization $\mathcal{V}$ encompass  particle-hole,  particle-particle, and hole-hole virtual excitations.

 In an attempt to ``dissect'' the contributions from each of these different channels and understand their role in the generation of effective isotropic and anisotropic exchange interactions, we introduce the following notation. Contributions that only include $f^0$ virtual states, i.e., the ones that survive when taking the limit of infinite on-site Coulomb repulsion between $f$ electrons, will be denoted as ``$f^0$ contributions''. These are the only contributions that have been included in Ref.~\cite{SimethW2023_CeIn3}. The rest of the processes containing at least one doubly occupied virtual $f^2$ state will be denoted as ``$f^2$ contributions''. We will further split these $f^2$ contributions into $f^{2:1}$ and $f^{2:2}$ processes, where the former include virtual states with \emph{only one} doubly occupied $f$ ion, while the latter include virtual states where both $f$ ions are doubly occupied in the intermediate virtual states (note that they are not necessarily simultaneously doubly occupied).

Most of the contributions from particle-hole processes can be derived from the KLM via second-order degenerate perturbation theory in the Kondo interaction. These  processes give rise to the well-known RKKY interaction. However, as shown in Ref.~\cite{SimethW2023_CeIn3}, particle-particle processes contribute comparably to the effective spin-spin interaction and are essential for accurately describing the magnon spectrum of CeIn$_3$. 
 Moreover,  the particle-particle channel also leads to effective long-range interactions that decay like $1/r^4$ (i.e. faster than the $1/r^3$ decay that is obtained from the particle-hole channel) and manifest in momentum space as a cusplike singularity at the $\Gamma$ point~\cite{SimethW2023_CeIn3}.

 We also note that the inclusion of the $f^2$ states gives rise to a hole-hole channel, which plays a similar role as the particle-particle channel. 
The particle-particle, particle-hole, and hole-hole processes are depicted in Figs.~\ref{fig:app-pp-diag}, \ref{fig:app-ph-diag} and \ref{fig:app-hh-diag} of Appendix \ref{app:spin-int}, respectively.
After adding all virtual processes, the effective spin Hamiltonian takes the form
\begin{equation}
    \mathcal{H}_{\rm spin} = \frac{1}{2} \!\! \sum_{{\bm R},{\bm R}^\prime} \!  \mathcal{S}_{{\bm R}^\prime}^{\sigma_1^{} \sigma_1'} [\mathbb{K}({\bm R})]_{\sigma_1^{} \sigma_1' , \sigma_2^{} \sigma_2'}^{} \mathcal{S}_{{\bm R} + {\bm R}^\prime}^{\sigma_2^{} \sigma_2'} + \sum_{{\bm R}} \mathcal{S}_{{\bm R}}^{\sigma \sigma'}  \mathbb{\tilde H}^{\sigma \sigma'} 
\end{equation}
where the spin-spin interaction is given by the rank-4 tensor $\mathbb{K}(\bm R)$. The effective field matrix $\mathbb{\tilde H} $
includes the CF contribution and the Zeeman coupling to an external field ${\bm B}$ and it is defined by the condition:
\begin{equation}
\mathcal{S}_{\bm R}^{\sigma \sigma'}  \mathbb{\tilde H}^{\sigma \sigma'}  = \tilde {\bm f}_{\bm R}^\dag \mathbb{\tilde C} \tilde{\bm f}_{\bm R} - g_{J_m} \mu_{B} {\bm B} \cdot {\bm J}_m
\end{equation}
where $g_{J_m}$ is the Landé factor and $\mu_{B}$ the Bohr magneton.

After the Fourier transformation of the fermionic operators as $f_{\bm R\sigma^{}}^{\dag} = \tfrac{1}{\sqrt{N}}\sum_{\bm k} e^{-i\bm k \cdot \bm R} f_{\bm k \sigma^{}}^{\dag} $, the spin Hamiltonian becomes
\begin{equation}\label{eq:Hspin-Ugenerator}
    \mathcal{H}_{\rm spin} =  \frac{1}{2} \sum_{\bm k}  \mathcal{S}_{\bm k}^{\sigma_1^{}\sigma_1'}  [\mathbb{K} (\bm k)]_{\sigma_1^{}\sigma_1',\sigma_2^{}\sigma_2'}\mathcal{S}_{-\bm k}^{\sigma_2^{}\sigma_2'} + \sum_{\bm R} \mathcal{S}_{\bm R}^{\sigma \sigma'}  \mathbb{\tilde H}^{\sigma \sigma'} 
\end{equation}
 The Fourier transform of the U($2J_m+1$) generators is given by $\mathcal{S}_{\bm k}^{\sigma\sigma'} = \frac{1}{\sqrt{N}} \sum_{\bm q}f_{\bm q + \bm k \sigma}^{\dag} f_{\bm q \sigma'}^{}$ and the rank-4 tensor of the exchange interaction in momentum space is given by
\begin{equation}
    \mathbb{K}(\bm k) \equiv \sum_{{\bm R}} e^{-i\bm k \cdot {\bm R}} \ \mathbb{K}({\bm R}).
\end{equation}
The interaction tensor can be divided in the $f^0$-state and the $f^2$-state channels,
\begin{align}
    \mathbb{K}(\bm k) = \mathbb{K}^{(f^0)}(\bm k) + \mathbb{K}^{(f^2)}(\bm k) 
\end{align}
with
\begin{align}\label{eq:f0}
  [\mathbb{K}^{(f^0)} (\bm k)]_{\sigma_1^{} \sigma_1' , \sigma_2^{} \sigma_2'} = \frac{1}{N} \sum_{ \bm q }  [\mathbb{\tilde T}_{\bm k + \bm q \mu}^{}]_{\sigma_1^{} \sigma_2'} [\mathbb{\tilde T}_{ \bm q \mu'}^{*}]_{\sigma_1' \sigma_2^{}}  g_{\varepsilon_{\bm k+\bm q \mu}, \varepsilon_{\bm q \mu'}} ^{(f^0)} 
\end{align}
where $\mathbb{\tilde T}_{\bm k \mu} = \mathbb{\tilde V}_{\bm k \mu} \mathbb{\tilde V}_{\bm k \mu}^\dag $ is the effective hopping restricted to the low-energy $J=J_m$ multiplet, with the hybridization $\mathbb{\tilde V}_{\bm k \mu}$ between the $(2J_m+1)$ $f$ states and the doublet of the band $\mu$ with momentum $\bm k$. The $g$ functions   depend only on the band energies $\varepsilon_{\bm k+\bm q \mu}$, $\varepsilon_{\bm q \mu'}$ and $\varepsilon_f$ for the $f^0$ processes, and also on the eigenenergies $\varepsilon_\nu$ of the doubly occupied single-ion state for $f^2$ processes. 

The $g$ function for $f^0$ processes can be broken down as,
\begin{equation}\label{eq:g-f0}
    g_{\varepsilon_{\bm k+\bm q \mu}, \varepsilon_{\bm q \mu'}} ^{(f^0)} = g_{\varepsilon_{\bm k+\bm q \mu}, \varepsilon_{\bm q \mu'} }^{(f^0:\text{pp})} + g_{\varepsilon_{\bm k+\bm q \mu}, \varepsilon_{\bm q \mu'} }^{(f^0:\text{ph})}
\end{equation}
where 
\begin{widetext}
\begin{align}
     g_{\varepsilon_{\bm k+\bm q \mu}, \varepsilon_{\bm q \mu'} }^{(f^0:\text{pp})} & =  \frac{ \varepsilon_{\bm k + \bm q \mu} + \varepsilon_{\bm q \mu'} - 2\varepsilon_{f}  }{ (\varepsilon_{\bm q \mu'} - \varepsilon_{f} )^2  (\varepsilon_{\bm k + \bm q \mu} - \varepsilon_{f} )^2 }  
         [1-f(\varepsilon_{\bm k + \bm q \mu})] [1- f(\varepsilon_{\bm q \mu'})]      \ ,
\end{align}
for the particle-particle channel, and

\begin{align}\label{eq:f0-ph}
     & g_{\varepsilon_{\bm k+\bm q \mu}, \varepsilon_{\bm q \mu'} }^{(f^0:\text{ph})}  =  \frac{  1 }{(\varepsilon_{\bm q \mu'} - \varepsilon_{\bm k + \bm q \mu} )}  
       \left[ \frac{ f(\varepsilon_{\bm q \mu'}) [1-f(\varepsilon_{\bm k + \bm q \mu})]  }{(\varepsilon_{\bm k + \bm q \mu} - \varepsilon_{f} )^2 } - \frac{ f(\varepsilon_{\bm k + \bm q \mu}) [1-f(\varepsilon_{\bm q \mu'})]  }{(\varepsilon_{\bm q \mu'} - \varepsilon_{f} )^2 }  \right]   
\end{align}
\end{widetext}
for the particle-hole channel. Here $f(\varepsilon_{\bm k \mu})$ is the Fermi-Dirac distribution and $\text{pp}$ ($\text{ph}$) indicates the particle-hole (particle-particle) contribution (see Appendix \ref{app:spin-int}). 

 ``Particle-particle'' $f^0$ processes ($f^0:\text{pp}$) are those  whose intermediate virtual state has two particles above the Fermi level. Note that these processes cannot be derived from a perturbative treatment of the KL model $\mathcal{H}_{\mathrm{KL}}$. ``Particle-hole'' processes are those that have one particle excitation and one hole excitation among the three virtual states. Note that  particle-hole processes that include a middle virtual state with one particle and one hole simultaneously 
(particle-hole excitation) are the ones that can be obtained from $\mathcal{H}_{\mathrm{KL}}$ by applying second-order degenerate perturbation theory in the Kondo interaction.   

The contribution to the exchange tensor from processes involving  $f^2$-states  can be further split into terms,
\begin{align}\label{eq:f2}
    \mathbb{K}^{(f^2)}(\bm k) =  \mathbb{K}^{(f^{2:1})}(\bm k) + \mathbb{K}^{(f^{2:2})}(\bm k),
\end{align}
where the first term includes contributions from processes involving the virtual state where \emph{only one} $f$ ion is doubly occupied,
\begin{widetext}
\begin{align}\label{eq:Kf2}
    [\mathbb{K}^{(f^{2:1})}(\bm k)]_{\sigma_1^{} \sigma_1',\sigma_2^{} \sigma_2'}^{} &=  \frac{1}{N} \sum_{ \bm q } 
    g_{\varepsilon_{\bm k+\bm q \mu}, \varepsilon_{\bm q \mu'} , \varepsilon_{\nu}}^{(f^{2:1})} 
     ( [\mathbb{\hat T}_{\bm k + \bm q \mu}]_{\sigma_1^{} \alpha'}^{}    [\mathbb{\hat T}_{\bm q \mu'}^*]_{\sigma_1' \alpha^{}}^{}  \gamma_{\sigma_2^{}\sigma_2',\alpha \alpha'}^{\nu}  +  
     \gamma_{\sigma_1^{}\sigma_1',\alpha \alpha'}^{\nu}  [\mathbb{\hat T}_{\bm k + \bm q \mu}^{*}]_{\sigma_2' \alpha }^{}  [\mathbb{\hat T}_{\bm q \mu'}^{}]_{\sigma_2^{} \alpha' }^{}  ),  
\end{align}
with $\mathbb{\hat T}_{\bm k \mu} = \mathbb{\tilde V}_{\bm k \mu} \mathbb{V}_{\bm k \mu}^\dag $ being the effective hopping between the low-energy $J=J_m$ multiplet in one site and the 14 possible $f$ states in the other site. 
The function of energies is broken down into three channels as
\begin{equation}\label{eq:f2:1}
    g_{\varepsilon_{\bm k+\bm q \mu}, \varepsilon_{\bm q \mu'} , \varepsilon_{\nu}}^{(f^{2:1})} = g_{\varepsilon_{\bm k+\bm q \mu}, \varepsilon_{\bm q \mu'}, \varepsilon_{\nu} }^{(f^{2:1}:\text{pp})} + g_{\varepsilon_{\bm k+\bm q \mu}, \varepsilon_{\bm q \mu'}, \varepsilon_{\nu} }^{(f^{2:1}:\text{ph})} + g_{\varepsilon_{\bm k+\bm q \mu}, \varepsilon_{\bm q \mu'}, \varepsilon_{\nu} }^{(f^{2:1}:\text{hh})},
\end{equation}
where the contribution from the particle-particle channel is
\begin{align}
    g_{\varepsilon_{\bm k+\bm q \mu}, \varepsilon_{\bm q \mu'} , \varepsilon_{\nu}}^{(f^{2:1}:\text{pp})}  = - \frac{ [1-f(\varepsilon_{\bm k+\bm q \mu})] [1-f(\varepsilon_{\bm q \mu'})] }{  \varepsilon_{\nu} ( \varepsilon_{\bm k+\bm q\mu} - \varepsilon_{f}) (\varepsilon_{\bm q \mu'} - \varepsilon_{f}) } \ ,
\end{align}
the contribution from the particle-hole channel is
\begin{align}\label{eq:f2:1-ph}
    g_{ \varepsilon_{\bm k+\bm q \mu}, \varepsilon_{\bm q \mu'} , \varepsilon_{\nu} }^{(f^{2:1}:\text{ph})} \! \! \! =  \frac{1}{\varepsilon_{\bm q \mu'} - \varepsilon_{\bm k+\bm q \mu}}    \left[ \frac{ ( \varepsilon_{\bm k+\bm q \mu} - \varepsilon_{\bm q \mu'}  + \varepsilon_\nu) f(\varepsilon_{\bm q \mu'} ) [1 - f(\varepsilon_{\bm k+\bm q \mu})] }{ (\varepsilon_{\bm k+\bm q \mu}-\varepsilon_f) (\varepsilon_{\bm q \mu'}   - \varepsilon_{f} - \varepsilon_{\nu}) \varepsilon_\nu } 
    - \frac{  ( \varepsilon_{\bm q \mu'} - \varepsilon_{\bm k+\bm q \mu} + \varepsilon_\nu)  f(\varepsilon_{\bm k+\bm q \mu} ) [1 - f(\varepsilon_{\bm q \mu'})] }{ (\varepsilon_{\bm q \mu'}-\varepsilon_{f}) (\varepsilon_{\bm k+\bm q \mu} - \varepsilon_{f} - \varepsilon_{\nu}) \varepsilon_\nu }    \right],
\end{align}
and the contribution from the hole-hole channels is 
\begin{align}\label{eq:f2:1-hh}
    g_{\varepsilon_{\bm k+\bm q \mu}, \varepsilon_{\bm q \mu'} , \varepsilon_{\nu}}^{(f^{2:1}:\text{hh})} = - \frac{  f(\varepsilon_{\bm q \mu'}) \ f(\varepsilon_{\bm k+\bm q \mu}) }{ (\varepsilon_{\bm k+\bm q \mu}-\varepsilon_f-\varepsilon_{\nu}) (\varepsilon_{\bm q \mu'}-\varepsilon_f-\varepsilon_{\nu}) \varepsilon_{\nu} } \ .
\end{align}
\end{widetext}


In this context, ``particle-particle'' ($f^{2:1}:\text{pp}$) processes involve two intermediate states with one excited particle, as shown in Fig.~\ref{fig:app-pp-diag} for the sequences $abcd$ and $cdab$. Similarly, ``hole-hole'' ($f^{2:1}:\text{hh}$) processes entail two intermediate states with one excited hole, as in Fig.~\ref{fig:app-hh-diag} for the sequences $abcd$ and $cdab$.


The second term in Eq.~\eqref{eq:f2} represents the virtual processes that involve doubly occupied states in \emph{both} $f$ ions. These contributions can be expressed as
\begin{widetext}
\begin{align}\label{eq:Kf2f2}
    [\mathbb{K}^{(f^{2:2})}(\bm k)]_{\sigma_1^{} \sigma_1',\sigma_2^{} \sigma_2'}^{}  = \frac{1}{N}  \sum_{\bm q} & \ g_{\varepsilon_{\bm k+\bm q \mu}, \varepsilon_{\bm q \mu'} , \varepsilon_{\nu_1},\varepsilon_{\nu_2} }^{(f^{2:2})}  
     \gamma_{\sigma_1^{}\sigma_1',\alpha \alpha'}^{\nu_1} [\mathbb{T}_{\bm k + \bm q \mu}]_{\alpha \beta'}^{}  [\mathbb{T}_{\bm q \mu'}^*]_{\alpha'\beta}^{}  \gamma_{\sigma_2^{}\sigma_2',\beta \beta'}^{\nu_2},   
%
\end{align}
where $\mathbb{T}_{\bm k \mu}^{} = \mathbb{V}_{\bm k \mu}^{}   \mathbb{V}_{\bm k \mu}^{\dag}$ is the $14\times 14$ matrix of the effective hopping that connects all the $f^2$ states in both sites.
The function of energies is separated into two channels,
\begin{eqnarray}\label{eq:g-f22}
    g_{\varepsilon_{\bm k+\bm q \mu}, \varepsilon_{\bm q \mu'} , \varepsilon_{\nu_1},\varepsilon_{\nu_2} }^{(f^{2:2})} = g_{\varepsilon_{\bm k+\bm q \mu}, \varepsilon_{\bm q \mu'} , \varepsilon_{\nu_1},\varepsilon_{\nu_2} }^{(f^{2:2}:\text{ph})} + g_{\varepsilon_{\bm k+\bm q \mu}, \varepsilon_{\bm q \mu'} , \varepsilon_{\nu_1},\varepsilon_{\nu_2} }^{(f^{2:2}:\text{hh})},
\end{eqnarray}
where the contribution from the particle-hole channel is
\begin{align}\label{eq:f2:2-ph}
    & g_{\varepsilon_{\bm k+\bm q \mu}, \varepsilon_{\bm q \mu'} , \varepsilon_{\nu_1},\varepsilon_{\nu_2} }^{(f^{2:2}:\text{ph})} =  \frac{1}{\varepsilon_{\bm q\mu'} - \varepsilon_{\bm k+\bm q \mu}}    
    \left[ \frac{ f(\varepsilon_{\bm q \mu'}) [1 - f(\varepsilon_{\bm k+\bm q \mu})] }{ (\varepsilon_{\bm q \mu'}-\varepsilon_f-\varepsilon_{\nu_1}) (\varepsilon_{\bm q \mu'}-\varepsilon_f-\varepsilon_{\nu_2}) } 
    - \frac{ f(\varepsilon_{\bm k+\bm q\mu}) [1 - f(\varepsilon_{\bm q \mu'})] }{ (\varepsilon_{\bm k+\bm q \mu}-\varepsilon_f-\varepsilon_{\nu_1}) (\varepsilon_{\bm k+\bm q \mu}-\varepsilon_f-\varepsilon_{\nu_2}) }   \right], 
\end{align}
and the contribution from the  hole-hole channels is
\begin{align}\label{eq:f2:2-hh}
     & g_{\varepsilon_{\bm k+\bm q \mu}, \varepsilon_{\bm q \mu'} , \varepsilon_{\nu_1},\varepsilon_{\nu_2} }^{(f^{2:2}:\text{hh})} 
     =  -  \frac{ (\varepsilon_{\bm q \mu'} + \varepsilon_{\bm k+\bm q \mu} - 2\varepsilon_f - \varepsilon_{\nu_1} - \varepsilon_{\nu_2}) }{ (\varepsilon_{\bm k+\bm q \mu}-\varepsilon_f-\varepsilon_{\nu_1}) (\varepsilon_{\bm q \mu'}-\varepsilon_f-\varepsilon_{\nu_1})  } 
     \frac{ f(\varepsilon_{\bm q \mu'}) \ f(\varepsilon_{\bm k+\bm q \mu}) }{ (\varepsilon_{\bm k+\bm q \mu}-\varepsilon_f-\varepsilon_{\nu_2}) (\varepsilon_{\bm q \mu'}-\varepsilon_f-\varepsilon_{\nu_2}) } \ .
\end{align}
\end{widetext}

\subsection{Symmetry of the spin Hamiltonian}

\begin{figure}[h!]
    \centering
    \includegraphics[width=0.48\textwidth]{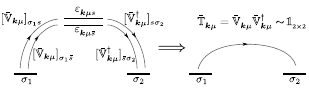}
    \caption{Schematic representation of the emergence of effective hopping $\mathbb{\bar T}_{\bm k \mu}$. When the Hamiltonian commutes with the combination of time-reversal and spatial-inversion operator, the effective hopping between the low-energy doublet conserves the spin. The indices $\sigma_1$ and $\sigma_2$ belong to the same Kramers doublet.}
    \label{fig:effective-hopp}
\end{figure}

In this section, we show that when the hybridization is restricted to the lowest-energy Kramers doublet, denoted by $\mathbb{\bar V}_{\bm k \mu}$, the effective hopping 
to second order in the hybridization, $\mathbb{\bar T}_{\bm k \mu}=\mathbb{\bar V}_{\bm k \mu}\mathbb{\bar V}_{\bm k \mu}^\dag$,  conserves the spin (see  Fig.~\ref{fig:effective-hopp}).
This property is derived from the invariance of the MO-PAM under time-reversal symmetry and spatial inversion relative to the center of the bond under consideration.

Let us  consider the $2\times2$ hybridization matrix 
$\bar{\mathbb{V}}^{\alpha}_{j_{\alpha} n_{\alpha}}({\bm R} - {\bm R}^{\alpha}_{j_{\alpha}})$ %
between the $f$ doublet and the $n$-th Wannier orbital of the basis that was extracted from the first-principles calculation. Due to the time-reversal symmetry, the hybridization matrix fulfills 
\begin{equation}\label{eq:hybr-unitary}
    \bar{\mathbb{V}}^{\alpha}_{j_{\alpha} n_{\alpha}}({\bm R} - {\bm R}^{\alpha}_{j_{\alpha}})  = v^{\alpha}_{j_{\alpha} n_{\alpha} {\bm R}}  \mathbb{U}^{\alpha}_{j_{\alpha} n_{\alpha}}({\bm R} - {\bm R}^{\alpha}_{j_{\alpha}}).
\end{equation}
 where the scalar $v^{\alpha}_{j_{\alpha} n_{\alpha} {\bm R}} $ can be chosen to be real via a gauge transformation, and $\mathbb{U}^{\alpha}_{j_{\alpha} n_{\alpha}}({\bm R} - {\bm R}^{\alpha}_{j_{\alpha}})$ is a $2 \times 2$ unitary matrix with a determinant equal to one. 
In the presence of spatial-inversion symmetry $\bar{\mathbb{V}}^{\alpha}_{j_{\alpha} n_{\alpha}}({\bm R} - {\bm R}^{\alpha}_{j_{\alpha}}) = \pm \bar{\mathbb{V}}^{\alpha}_{j_{\alpha} n_{\alpha}}({\bm R}^{\alpha}_{j_{\alpha}} -{\bm R}  )$, where the $+$ ($-$) sign holds for the odd (even) $n$-th Wannier orbital. The Fourier transform of the projected hybridization is 
\begin{eqnarray}
\mathbb{\bar V}^{\alpha, {\bm k}}_{ j_{\alpha}, n_{\alpha}} &=& \sum_{\bm R } e^{i {\bm k} \cdot ({\bm R}-{\bm R}^{\alpha}_{j_{\alpha}})} \;\; \bar{\mathbb{V}}^{\alpha}_{j_{\alpha} n_{\alpha}}({\bm R} - {\bm R}^{\alpha}_{j_{\alpha}}) 
\nonumber \\
&=&  \sum_{\bm R } \cos{[\bm k \cdot ({\bm R}-{\bm R}^{\alpha}_{j_{\alpha}})]} \;\; \bar{\mathbb{V}}^{\alpha}_{j_{\alpha} n_{\alpha}}({\bm R} - {\bm R}^{\alpha}_{j_{\alpha}})
\nonumber \\
\label{eq:inv1}
\end{eqnarray}
for $\bar{\mathbb{V}}^{\alpha}_{j_{\alpha}, n_{\alpha}}({\bm R} - {\bm R}^{\alpha}_{j_{\alpha}}) =  \bar{\mathbb{V}}^{\alpha}_{j_{\alpha} n_{\alpha}}({\bm R}^{\alpha}_{j_{\alpha}} -{\bm R}  )$ and
\begin{eqnarray}
\mathbb{\bar V}^{\alpha, {\bm k}}_{ j_{\alpha}, n_{\alpha}} &=& \sum_{\bm R } e^{i {\bm k} \cdot ({\bm R}-{\bm R}^{\alpha}_{j_{\alpha}})}  \;\; \bar{\mathbb{V}}^{\alpha}_{j_{\alpha} n_{\alpha}}({\bm R} - {\bm R}^{\alpha}_{j_{\alpha}}) 
\nonumber \\
&=& i \sum_{\bm R } \sin{[\bm k \cdot ({\bm R}-{\bm R}^{\alpha}_{j_{\alpha}})]} \;\;  \bar{\mathbb{V}}^{\alpha}_{j_{\alpha} n_{\alpha}}({\bm R} - {\bm R}^{\alpha}_{j_{\alpha}})
\nonumber \\
\label{eq:inv2}
\end{eqnarray}
for $\bar{\mathbb{V}}^{\alpha}_{j_{\alpha} n_{\alpha}}({\bm R} - {\bm R}^{\alpha}_{j_{\alpha}}) = - \bar{\mathbb{V}}^{\alpha}_{j_{\alpha} n_{\alpha}}({\bm R}^{\alpha}_{j_{\alpha}} -{\bm R}  )$. Equations~\eqref{eq:hybr-unitary}-\eqref{eq:inv2} imply that 
the determinant of $\mathbb{\bar V}^{\alpha,{\bm k}}_{ j_{\alpha}, n_{\alpha}}$ is a real number,
\begin{equation}
{\rm Det} \mathbb{\bar V}^{\alpha, {\bm k}}_{j_{\alpha}, n_{\alpha}} = ({\rm Det} \mathbb{\bar V}^{\alpha, {\bm k}}_{ j_{\alpha}, n_{\alpha}})^*.
\label{eq:redet}
\end{equation}

Now we are ready to introduce the effective hopping matrix,
\begin{widetext}
\begin{eqnarray}
&&    \mathbb{\bar V}^{\alpha,{\bm k}}_{ j_{\alpha}, n_{\alpha}} [\mathbb{\bar V}^{\alpha,{\bm k}}_{j_{\alpha}, n_{\alpha}}]^\dag =  \sum_{\bm R {\bm R}^{\prime}} e^{-i\bm k \cdot (\bm R - {\bm R}^{\prime})} \bar{\mathbb{V}}^{\alpha}_{j_{\alpha} n_{\alpha}}({\bm R} - {\bm R}^{\alpha}_{j_{\alpha}})
[\bar{\mathbb{V}}^{\alpha}_{j_{\alpha} n_{\alpha}}({\bm R}^{\prime} - {\bm R}^{\alpha}_{j_{\alpha}})]^{\dagger}
\nonumber \\
 &&= \frac{1}{2 } \sum_{\bm R {\bm R}^{\prime}} e^{-i\bm k \cdot (\bm R - {\bm R}^{\prime})} ( \bar{\mathbb{V}}^{\alpha}_{j_{\alpha} n_{\alpha}}({\bm R} - {\bm R}^{\alpha}_{j_{\alpha}})
[\bar{\mathbb{V}}^{\alpha}_{j_{\alpha} n_{\alpha}}({\bm R}^{\prime} - {\bm R}^{\alpha}_{j_{\alpha}})]^{\dagger} + \bar{\mathbb{V}}^{\alpha}_{j_{\alpha} n_{\alpha}}( {\bm R}^{\alpha}_{j_{\alpha}} -{\bm R}^{\prime}  ) 
[\bar{\mathbb{V}}^{\alpha}_{j_{\alpha} n_{\alpha}}({\bm R}^{\alpha}_{j_{\alpha}} -{\bm R}  )]^{\dagger})
\nonumber \\
    &&= \frac{1}{2 } \sum_{\bm R {\bm R}^{\prime}} e^{-i\bm k \cdot (\bm R - {\bm R}^{\prime})} ( \bar{\mathbb{V}}^{\alpha}_{j_{\alpha} n_{\alpha}}({\bm R} - {\bm R}^{\alpha}_{j_{\alpha}})
[\bar{\mathbb{V}}^{\alpha}_{j_{\alpha} n_{\alpha}}({\bm R}^{\prime} - {\bm R}^{\alpha}_{j_{\alpha}})]^{\dagger} + \bar{\mathbb{V}}^{\alpha}_{j_{\alpha} n_{\alpha}}( {\bm R}^{\prime} - {\bm R}^{\alpha}_{j_{\alpha}}   ) 
[\bar{\mathbb{V}}^{\alpha}_{j_{\alpha} n_{\alpha}}({\bm R} -{\bm R}^{\alpha}_{j_{\alpha}}   )]^{\dagger}) \nonumber \\
   & &= \frac{1}{2 } \sum_{\bm R {\bm R}^{\prime}} e^{-i\bm k \cdot (\bm R - {\bm R}^{\prime})} 
   v^{\alpha}_{j_{\alpha} n_{\alpha} {\bm R}}  v^{\alpha}_{j_{\alpha} n_{\alpha} {\bm R}^{\prime}}
   ( \bar{\mathbb{U}}^{\alpha}_{j_{\alpha} n_{\alpha}}({\bm R} - {\bm R}^{\alpha}_{j_{\alpha}})
[\bar{\mathbb{U}}^{\alpha}_{j_{\alpha} n_{\alpha}}({\bm R}^{\prime} - {\bm R}^{\alpha}_{j_{\alpha}})]^{\dagger} + \bar{\mathbb{U}}^{\alpha}_{j_{\alpha} n_{\alpha}}( {\bm R}^{\prime} - {\bm R}^{\alpha}_{j_{\alpha}}   ) 
[\bar{\mathbb{U}}^{\alpha}_{j_{\alpha} n_{\alpha}}({\bm R} -{\bm R}^{\alpha}_{j_{\alpha}}   )]^{\dagger})
\nonumber \\
    & &=  \sum_{\bm R {\bm R}^{\prime}} e^{-i\bm k \cdot (\bm R - {\bm R}^{\prime})}  v^{\alpha}_{j_{\alpha} n_{\alpha} {\bm R}}  v^{\alpha}_{j_{\alpha} n_{\alpha} {\bm R}^{\prime}} \cos{\phi^{\alpha}_{n_{\alpha} j_{\alpha}}({\bm R}, {\bm R}^{\prime})} \ \mathbb{1}_{2 \times 2} 
    = |{\rm Det} \mathbb{\bar V}_{{\bm k} n}| \ \mathbb{1}_{2 \times 2}
    \label{eq:iden1}
\end{eqnarray}
\end{widetext}
where $\mathbb{1}_{2 \times 2}$ is the $2 \times 2$ identity matrix and we have used 
\begin{eqnarray}
&&\bar{\mathbb{U}}^{\alpha}_{j_{\alpha} n_{\alpha}}({\bm R} - {\bm R}^{\alpha}_{j_{\alpha}})
[\bar{\mathbb{U}}^{\alpha}_{j_{\alpha} n_{\alpha}}({\bm R}^{\prime} - {\bm R}^{\alpha}_{j_{\alpha}})]^{\dagger} 
\nonumber \\
&&=
\mathbb{1} \cos \phi^{\alpha}_{j_{\alpha} n_{\alpha}}({\bm R}, {\bm R}^{\prime})  
+ i (\bm n_{j_\alpha,n_\alpha}^{\alpha} \!\! \cdot \bm \sigma)  \sin \phi^{\alpha}_{j_{\alpha} n_{\alpha}}({\bm R}, {\bm R}^{\prime}),
\nonumber \\
\end{eqnarray} 
where $\bm n_{j_\alpha,n_\alpha}^{\alpha}$ is a unit vector and $\bm \sigma$ is the vector of Pauli matrices. 

Equation~\eqref{eq:iden1} implies that
\begin{equation}
\mathbb{\bar V}^{\alpha, {\bm k}}_{ j_{\alpha}, n_{\alpha}} = \sqrt{{\rm Det} \mathbb{\bar V}^{\alpha, {\bm k}}_{ j_{\alpha}, n_{\alpha}}}  \mathbb{\bar U}^{\alpha, {\bm k}}_{j_{\alpha}, n_{\alpha}}, 
\end{equation}
where $\mathbb{\bar U}^{\alpha, {\bm k}}_{ j_{\alpha}, n_{\alpha}} \in $ SU(2).
After projecting the unitary transformation given in Eq.~\eqref{eq:unitary} into the 
lowest-energy Kramers doublet, we obtain
\begin{equation}
\bar{\mathbb{V}}_{{\bm k}  \mu } =  \bar{b}^{{\bm k}, \mu}_{\alpha, j_{\alpha}, n_{\alpha}}  \bar{\mathbb{V}}^{\alpha, {\bm k}}_{ j_{\alpha} n_{\alpha}} [\mathbb{U}^{{\bm k}, \mu}_{\alpha, j_{\alpha}, n_{\alpha}}]^{\dagger}. 
\label{eq:unitary2}
\end{equation}
The coefficients $\bar{b}^{{\bm k}, \mu}_{\alpha, j_{\alpha}, n_{\alpha}}$ are real, directly following from Eq.~\eqref{eq:redet}.
From Eq.~\eqref{eq:unitary2}, we obtain
\begin{widetext}
    \begin{eqnarray}
\mathbb{\bar T}_{\bm k \mu}^{} \!  &=& \!   \mathbb{\bar V}_{{\bm k} \mu} \mathbb{\bar V}^{\dagger}_{{\bm k} \mu} 
= \frac{1}{2}  \bar{b}^{{\bm k}, \mu}_{\alpha, j_{\alpha}, n_{\alpha}} \bar{b}^{{\bm k}, \mu}_{\beta, j_{\beta}, n_{\beta}} ( \bar{\mathbb{V}}^{\alpha, {\bm k}}_{ j_{\alpha} n_{\alpha}} [\mathbb{U}^{{\bm k}, \mu}_{\alpha, j_{\alpha}, n_{\alpha}}]^{\dagger}
\mathbb{U}^{{\bm k}, \mu}_{\beta, j_{\beta}, n_{\beta}} [\bar{\mathbb{V}}^{\beta, {\bm k}}_{ j_{\beta} n_{\beta}}]^{\dagger}
+ \bar{\mathbb{V}}^{\beta, {\bm k}}_{ j_{\beta} n_{\beta}} [\mathbb{U}^{{\bm k}, \mu}_{\beta, j_{\beta}, n_{\beta}}]^{\dagger}
\mathbb{U}^{{\bm k}, \mu}_{\alpha, j_{\alpha}, n_{\alpha}} [\bar{\mathbb{V}}^{\alpha, {\bm k}}_{ j_{\beta} n_{\alpha}}]^{\dagger})
\nonumber \\ 
&=& \!  \frac{1}{2}  \bar{b}^{{\bm k}, \mu}_{\alpha, j_{\alpha}, n_{\alpha}} \bar{b}^{{\bm k}, \mu}_{\beta, j_{\beta}, n_{\beta}} \sqrt{{\rm Det} \mathbb{\bar V}^{\alpha, {\bm k}}_{ j_{\alpha}, n_{\alpha}}} \sqrt{{\rm Det} \mathbb{\bar V}^{\beta, {\bm k}}_{ j_{\beta}, n_{\beta}}}
( \bar{\mathbb{U}}^{\alpha, {\bm k}}_{ j_{\alpha} n_{\alpha}} [\mathbb{U}^{{\bm k}, \mu}_{\alpha, j_{\alpha}, n_{\alpha}}]^{\dagger}
\mathbb{U}^{{\bm k}, \mu}_{\beta, j_{\beta}, n_{\beta}} [\bar{\mathbb{U}}^{\beta, {\bm k}}_{ j_{\beta} n_{\beta}}]^{\dagger}
\!\!   + \!  \bar{\mathbb{U}}^{\beta, {\bm k}}_{ j_{\beta} n_{\beta}} [\mathbb{U}^{{\bm k}, \mu}_{\beta, j_{\beta}, n_{\beta}}]^{\dagger}
\mathbb{U}^{{\bm k}, \mu}_{\alpha, j_{\alpha}, n_{\alpha}} [\bar{\mathbb{U}}^{\alpha, {\bm k}}_{ j_{\alpha} n_{\alpha}}]^{\dagger}) 
\nonumber \\
&=& \!  |{\rm Det}(\mathbb{\bar V}_{\bm k \mu})| \mathbb{1}_{2 \times 2}.
\label{eq:diag-hopp}
\end{eqnarray}
\end{widetext}
where we have used the fact that $(\mathbb{U}+ \mathbb{U}^{\dagger}) \propto  \mathbb{1}_{2 \times 2}$ for 
$\mathbb{U} \in$ SU(2).

As we show below, Eq.~\eqref{eq:diag-hopp} implies that the magnetic interaction that emerges from  processes \emph{within} the lowest-energy Kramers doublet is isotropic. In other words, the anisotropic exchange interactions originate from virtual processes involving single-particle $f$ states that are orthogonal to the Kramers doublet.

\subsection{Isotropic contribution}
From now on, we reduce the low-energy Hamiltonian to the lowest-energy Kramers doublet  (e.g. $\{\Gamma_7^+, \Gamma_7^- \}$ for the case of CeIn$_3$). By using Eq.~\eqref{eq:diag-hopp}, we can simplify the expression of the exchange coupling in Eqs.~\eqref{eq:f0}, \eqref{eq:Kf2}, and \eqref{eq:Kf2f2} as follows:
\begin{align}\label{eq:f0-simp}
 [\mathbb{\bar K}^{(f^0)}(\bm k)]_{\sigma_1^{} \sigma_1',\sigma_2^{} \sigma_2'}^{} = & \frac{1}{N} \sum_{ \bm q } v_{\bm k + \bm q \mu}^2 v_{\bm q \mu'}^2 g_{\varepsilon_{\bm k+\bm q \mu}, \varepsilon_{\bm q \mu'} }^{(f^0)}
  \delta_{\sigma_1^{} \sigma_2'}  \delta_{\sigma_1' \sigma_2^{}} ,   
\end{align}
where $v_{\bm k \mu}^2 \equiv |{\rm Det}(\bar{\mathbb{V}}_{\bm k \mu}) |  $.

For the $f^2$ channel, we can distinguish two contributions,
\begin{equation}
    \mathbb{\bar K}^{(f^{2:\lambda)}}(\bm k) = \mathbb{\bar K}_{\rm in}^{(f^{2:\lambda})}(\bm k)  + \mathbb{\bar K}_{\rm out}^{(f^{2:\lambda})}(\bm k),
\end{equation}
where $\lambda=1,2$ and the subscript ``${\rm in}$'' indicates that the $f^2$ virtual states are built with two electrons occupying states within the low-energy doublet. Specifically, it signifies that the second electron introduced into the preexisting low-energy $f^1$ configuration occupies a state \emph{within the doublet} that is orthogonal to the one occupied by the first electron.
In contrast, the subscript ``${\rm out}$'' indicates that the second electron occupies a state ``outside of the doublet'', i.e., orthogonal to the subspace spanned by the two states of the lowest-energy doublet.

The \emph{intra-doublet} $f^2$ contributions are given by
\begin{align}\label{eq:Kf2-simp}
    & [\mathbb{\bar K}_{\rm in}^{(f^{2:1})}(\bm k)]_{\sigma_1^{} \sigma_1',\sigma_2^{} \sigma_2'}^{} = \nonumber \\
    & \ \ \ \ \ = \frac{1}{N} \sum_{\bm q} 2 v_{\bm k + \bm q \mu}^2 v_{\bm q \mu'}^2   \gamma_{\sigma_1^{}\sigma_1',\sigma_2' \sigma_2^{}}^{\nu}  \  g_{\varepsilon_{\bm k+\bm q \mu}, \varepsilon_{\bm q \mu'} ,\varepsilon_{\nu} }^{(f^{2:1})} \ ,
\end{align}
and
\begin{align}\label{eq:Kf2f2-simp}
    &[\mathbb{\bar K}_{\rm in}^{(f^{2:2})}(\bm k)]_{\sigma_1^{} \sigma_1',\sigma_2^{} \sigma_2'}^{} =  \nonumber \\
    & \ \  =   \frac{1}{N} \sum_{\bm q} v_{\bm k + \bm q \mu}^2 v_{\bm q \mu'}^2   \Gamma_{\sigma_1^{} \sigma_1'; \sigma_2' \sigma_2^{} }^{\nu_1 \nu_2}  \  g_{\varepsilon_{\bm k+\bm q \mu}, \varepsilon_{\bm q \mu'} ,\varepsilon_{\nu_1},\varepsilon_{\nu_2}}^{(f^2:2)}, 
\end{align}
where $\Gamma_{\sigma_1^{} \sigma_1'; \sigma_2' \sigma_2^{} }^{\nu_1 \nu_2} \equiv \sum_{s,s'} \gamma_{\sigma_1^{}\sigma_1', s s'}^{\nu_1}   \gamma_{\sigma_2^{}\sigma_2',s' s}^{\nu_2}$, with $s$ and $s'$ running over the low-energy doublet.

The expressions  for $\mathbb{\bar K}_{\rm out}^{(f^{2:\lambda})}(\bm k)$ are the same as in Eqs.~\eqref{eq:Kf2} and \eqref{eq:Kf2f2} with the indices $\sigma_1^{}$, $\sigma_1'$, $\sigma_2^{}$, and $\sigma_2'$ running over the low-energy doublets, and with at least one of the indices $\alpha$, $\alpha'$, $\beta$, and $\beta'$ running over the out-of-doublet states; this is $\mathbb{\bar K}_{\rm out}^{(f^{2:\lambda})}(\bm k) = \mathbb{\bar K}_{}^{(f^{2:\lambda})}(\bm k) - \mathbb{\bar K}_{\rm in}^{(f^{2:\lambda})}(\bm k) $.

The resulting spin Hamiltonian for zero external magnetic field can be written in the spin operator basis as
\begin{equation}\label{eq:Hspin-doublet}
    \mathcal{H}_{\rm spin} = \frac{1}{2} \sum_{\bm k} I_{\bm k}^{\alpha \beta} \ S_{\bm k}^\alpha S_{-\bm k}^\beta 
\end{equation}
where ${S}_{\bm k}^{\alpha} = \frac{1}{2} \tfrac{1}{\sqrt{N}} \sum_{\bm q} \bm f_{\bm q+\bm k}^{\dag} \sigma^\alpha  \bm f_{\bm q}^{}$
is the spin-1/2 operator in momentum space,  $\bm f_{\bm k}^{\dag}=(f_{\bm k \uparrow}^{\dag},f_{\bm k \downarrow}^{\dag})$ is the $f$-spinor and $\sigma^\alpha$ is the Pauli matrix with $\alpha=x,y,z$. The magnetic interaction components are given by
\begin{subequations}
\begin{align}
I_{\bm{k}}^{xx} &= 2 {\rm Re} \left( [\mathbb{\bar K}(\bm{k})]_{\uparrow\downarrow\downarrow\uparrow}+[\mathbb{\bar K}(\bm{k})]_{\uparrow\downarrow\uparrow\downarrow}\right) ,  \\
I_{\bm{k}}^{yy} &=  2 {\rm Re} \left([\mathbb{\bar K}(\bm{k})]_{\uparrow\downarrow\downarrow\uparrow}-[\mathbb{\bar K}(\bm{k})]_{\uparrow\downarrow\uparrow\downarrow}\right) , \\
I_{\bm{k}}^{zz} &= 2 \left([\mathbb{\bar K}(\bm{k})]_{\uparrow\uparrow\uparrow\uparrow}-[\mathbb{\bar K}(\bm{k})]_{\uparrow\uparrow\downarrow\downarrow}\right) , \\
I_{\bm{k}}^{xy} &= I_{\bm{k}}^{yx} = 2 {\rm Im} \left([\mathbb{\bar K}(\bm{k})]_{\uparrow\downarrow\downarrow\uparrow}-[\mathbb{\bar K}(\bm{k})]_{\uparrow\downarrow\uparrow\downarrow}\right) , \\
I_{\bm{k}}^{yz} & =I_{\bm{k}}^{zy} = -4 {\rm Im} [\mathbb{\bar K}(\bm{k})]_{\uparrow\downarrow\uparrow\uparrow}, \\
I_{\bm{k}}^{zx} &= I_{\bm{k}}^{xz} = 4 {\rm Re} [\mathbb{\bar K}(\bm{k})]_{\uparrow\uparrow\uparrow\downarrow} . 
\end{align}\label{eq:mag-int-total}
\end{subequations}
where we have used the Hermiticity condition, 
\begin{align}
     & [\mathbb{\bar K}(\bm{k})]_{\sigma_1^{}\sigma_1'\sigma_2^{}\sigma_2'} = [\mathbb{\bar K}(-\bm{k})]_{\sigma_1' \sigma_1^{} \sigma_2' \sigma_2^{}}^*
\end{align}
and spatial inversion combined with time-reversal symmetry, 
\begin{align}
     & [\mathbb{\bar K}(\bm{k})]_{\sigma_1^{}\sigma_1'\sigma_2^{}\sigma_2'} \nonumber \\
     & \ = {\rm sgn}(\sigma_1^{}) {\rm sgn}(\sigma_1') {\rm sgn}(\sigma_2^{}) {\rm sgn}(\sigma_2') [\mathbb{\bar K}(\bm{k})]_{\bar \sigma_1^{}\bar \sigma_1'\bar \sigma_2^{}\bar \sigma_2'}^*,
\end{align}
where ${\rm sgn}(\sigma)=+1(-1)$ if $\sigma=\uparrow (\downarrow)$, and $\bar \sigma$ is the time-reversed state of $\sigma$. 

The symmetry of the exchange interaction is dictated by $\delta_{\sigma_1^{} \sigma_2'} \delta_{\sigma_1' \sigma_2^{}}$, $\gamma_{\sigma_1^{} \sigma_1'; \sigma_2'\sigma_2^{}}^{\nu}$ and $\Gamma_{\sigma_1^{} \sigma_1'; \sigma_2' \sigma_2^{} }^{\nu_1 \nu_2}$ in Eq. \eqref{eq:f0-simp}, \eqref{eq:Kf2-simp} and \eqref{eq:Kf2f2-simp}, respectively.

For the $f^0$ channel, $\delta_{\sigma_1^{} \sigma_2'} \delta_{\sigma_1' \sigma_2^{}}$ yields the non-null elements,
\begin{equation}
    [\mathbb{\bar K}_{}^{(f^0)}(\bm k)]_{\uparrow \uparrow \uparrow \uparrow}^{} = [\mathbb{\bar K}_{}^{(f^0)}(\bm k)]_{\uparrow \downarrow \downarrow \uparrow}^{} \ ,
\end{equation}
and the other two non-null elements ($\downarrow\downarrow\downarrow\downarrow$ and $\downarrow\uparrow\uparrow\downarrow$) are related by time-reversal symmetry.

Due to Pauli's exclusion principle, Eq.~\eqref{eq:gamma-def} includes  only four non-zero elements of $\gamma_{\sigma_1^{} \sigma_1'; \sigma_2'\sigma_2^{}}^{\nu}$, which are real and related by
\begin{align}\label{eq:non-null_gamma}
     \gamma_{\uparrow \uparrow ; \downarrow \downarrow}^{\nu} = \gamma_{\downarrow \downarrow ; \uparrow \uparrow}^{\nu} = - \gamma_{\uparrow \downarrow ; \uparrow \downarrow}^{\nu} = - \gamma_{\downarrow \uparrow ; \downarrow \uparrow}^{\nu} \ .
\end{align}
This relation is derived by the anticommutation relation of the fermionic operators in Eq.~\eqref{eq:gamma-def}. Then, the nonzero contributions from the $f^{2:1}$ intradoublet channel are
\begin{equation}
    [\mathbb{\bar K}_{\rm in}^{(f^{2:1})}(\bm k)]_{\uparrow \downarrow \downarrow \uparrow}^{} = - [\mathbb{\bar K}_{\rm in}^{(f^{2:1})}(\bm k)]_{\uparrow \uparrow \downarrow \downarrow}^{} \ .
\end{equation}

From Eq.~\eqref{eq:non-null_gamma}, the function $\Gamma_{\sigma_1^{} \sigma_1'; \sigma_2' \sigma_2^{} }^{\nu_1 \nu_2}$ yields four nonzero components related by 
\begin{align}
     \Gamma_{\uparrow \uparrow ; \uparrow \uparrow}^{\nu_1 \nu_2}  = \Gamma_{\downarrow \downarrow ; \downarrow \downarrow}^{\nu_1 \nu_2}  =
     \Gamma_{\uparrow \downarrow ; \uparrow \downarrow}^{\nu_1 \nu_2}  = \Gamma_{\downarrow \uparrow; \downarrow  \uparrow}^{\nu_1 \nu_2} \ ,
\end{align}
which implies the following condition for the $f^{2:2}$ contribution from the intradoublet channel:
\begin{equation}
    [\mathbb{\bar K}_{\rm in}^{(f^{2:2})}(\bm k)]_{\uparrow \uparrow \uparrow \uparrow}^{} = [\mathbb{\bar K}_{\rm in}^{(f^{2:2})}(\bm k)]_{\uparrow \downarrow \downarrow \uparrow}^{}
\end{equation}
These properties imply that the three channels give rise to an isotropic magnetic interaction,
\begin{equation}\label{eq:int-iso}
    I_{\bm k}^{\alpha \beta} = I^{(f^0)\alpha\beta}_{\bm k} + I^{(f^2)\alpha\beta}_{{\rm in}, \ \bm k} ,
\end{equation}
where
\begin{equation}
    I^{(f^0)\alpha\beta}_{\bm k} = 2 [\mathbb{\bar K}_{}^{(f^0)}(\bm k)]_{\uparrow \downarrow \downarrow \uparrow}^{} \delta_{\alpha\beta}
\end{equation}
and
\begin{equation}
    I^{(f^2)\alpha\beta}_{{\rm in}, \ \bm k} = 2 ([\mathbb{\bar K}_{\rm in}^{(f^{2:1})}(\bm k)]_{\uparrow \downarrow \downarrow \uparrow}^{} \! + \! [\mathbb{\bar K}_{\rm in}^{(f^{2:2})}(\bm k)]_{\uparrow \downarrow \downarrow \uparrow}^{} ) \delta_{\alpha\beta} .
\end{equation}

To illustrate the relative contributions from each channel in Eq.~\eqref{eq:int-iso} with a concrete example, we will consider the specific case of the heavy-fermion metal CeIn$_3$. In this case, the spin-orbit coupling yields a $J_m=5/2$ low-energy subspace, which is further split into a low-energy Kramers doublet spanned by $\{ |\Gamma_7^+ \rangle , |\Gamma_7^- \rangle \}$, and high-energy quartet spanned by $\{ |\Gamma_{8a}^{+} \rangle , |\Gamma_{8a}^{-} \rangle , |\Gamma_{8b}^{+} \rangle , |\Gamma_{8b}^{-} \rangle \}$. The doublet is given by
\begin{subequations}
\begin{align}
    &|\Gamma_7^{+} \rangle = \sqrt{\tfrac{1}{6}} |+\tfrac{5}{2} \rangle - \sqrt{\tfrac{5}{6}} |-\tfrac{3}{2} \rangle = |\downarrow \rangle, \\
    &|\Gamma_7^{-} \rangle = \sqrt{\tfrac{1}{6}} |-\tfrac{5}{2} \rangle - \sqrt{\tfrac{5}{6}} |+\tfrac{3}{2} \rangle = |\uparrow \rangle,
\end{align}
\end{subequations}
and the quartet by
\begin{subequations}
\begin{align}
    &|\Gamma_{8a}^{+} \rangle = \sqrt{\tfrac{5}{6}} |+\tfrac{5}{2} \rangle + \sqrt{\tfrac{1}{6}} |-\tfrac{3}{2} \rangle,  \\
    &|\Gamma_{8a}^{-} \rangle = \sqrt{\tfrac{5}{6}} |-\tfrac{5}{2} \rangle + \sqrt{\tfrac{1}{6}} |+\tfrac{3}{2} \rangle,  \\
    &|\Gamma_{8b}^{+} \rangle = |+\tfrac{1}{2} \rangle,   \\
    &|\Gamma_{8b}^{-} \rangle = |-\tfrac{1}{2} \rangle,
\end{align}
\end{subequations}
where $|m_J\rangle \equiv |J=\frac{5}{2},m_J \rangle $.

\begin{figure}[t!]
    \centering
    \includegraphics[width=0.48\textwidth, trim=0 0 0 0,clip]{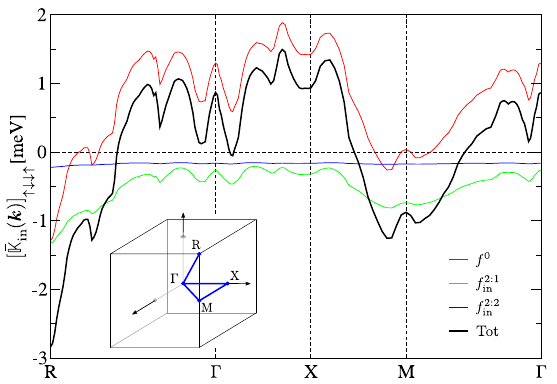}
    \caption{Contributions from different channels to the interaction in the low-energy doublet of CeIn$_3$. The main contribution originates from the $f^0$ and the $f_{\rm in}^{2:1}$ intradoublet channels. The superscripts $0$, $2\!:\!1$ and $2\!:\!2$ indicate the configuration of the intermediate virtual state, while the subscripts ``$in$" stands for intradoublet. The label ``Tot" refers to the total contribution. The energy of the $f$ orbital is given by $E_F-\varepsilon_f=\SI{0.65}{eV}$, and $F^0=\SI{6.4}{eV}$. The inset shows the path in the Brillouin zone.}
    \label{fig:int-channels}
\end{figure}

Figure~\ref{fig:int-channels} illustrates the contribution from the $f^0$, $f^{2:1}$, and $f^{2:2}$ intradoublet channels to the interaction $[\mathbb{\bar K}_{}^{}(\bm k)]_{\uparrow \downarrow \downarrow \uparrow}^{} $. From Eq.~\eqref{eq:int-iso} it is clear that the magnetic interaction $I_{\bm k}$ can be extracted from Fig.~\ref{fig:int-channels} by applying a scale factor of two. For an energy difference between the Fermi level and the $f$-orbital level $E_F-\varepsilon_f=\SI{0.65}{eV}$ and for the Slater integral $F^0=\SI{6.4}{eV}$ (see discussion in Sec.~\ref{sec:VAL}), the relative contributions are roughly 0.73 and 0.26 from the $f^0$ and $f^{2:1}$ intradoublet channels. The $f^{2:2}$ intradoublet contribution is negligible in this case. The out-of-doublet contribution arising from a nondegenerate $f^2$ configuration will be discussed below.

\section{Origin of Exchange Anisotropy \label{sec:exan}}

 In the previous section, we established that when the system is time-reversal invariant and the center of the bonds has spatial inversion symmetry, the effective spin-spin interaction between Kramers doublets is isotropic if the virtual processes are reduced to $f^0$ and \emph{intra-doublet} $f^2$ states. It is thus natural to inquire about the source of anisotropic exchange between Kramers doublets, particularly in scenarios where the centers of the bonds connecting pairs of magnetic ions exhibit inversion symmetry.

The objective of this section is to derive the anisotropic exchange stemming from the out-of-doublet $f^2$ virtual processes. Additionally, we aim to delineate conditions under which this contribution remains weak in comparison to the isotropic contribution previously derived.

\subsection{Out-of-doublet contributions: degenerate case}\label{subsec:degenerate-case}
 First, we will examine the scenario in which all eigenvalues of the single-ion Hamiltonian are degenerate for the $f^2$ configuration, meaning that the Slater integrals are
 \begin{equation}
\varepsilon_\nu = F^0, \quad {\rm and} \quad F^2 =  F^4 = F^6 =0.  
 \end{equation}
In this limit, we will demonstrate that the contribution from the out-of-doublet $f^2$ virtual processes vanishes exactly.  In other words, the out-of-doublet processes do not produce an effective spin-spin interaction, but only an irrelevant density-density interaction. 

For $\varepsilon_\nu = F^0$, the summation over all the eigenstates labeled by $\nu$ in Eq.~\eqref{eq:gamma-def} leads to
\begin{align}\label{eq:gamma-degen}
    \sum_{\nu}  \gamma_{\sigma \sigma';\alpha \alpha'}^{\nu} &= \sum_{\nu} \langle 0 | f_{\sigma}^{}  f_{\alpha'}^{} |\eta_\nu  \rangle \langle \eta_\nu|  f_{\alpha}^{\dag} f_{\sigma'}^{\dag} | 0 \rangle \nonumber \\
    &= (\delta_{\sigma^{}\sigma'} \delta_{\alpha^{} \alpha'} - \delta_{\sigma \alpha} \delta_{\sigma' \alpha'} ),
\end{align}
where we have used $\sum_\nu |\eta_\nu  \rangle \langle \eta_\nu| = \mathbb{1}_{91\times 91}  $.
Physically, this relationship indicates that the spin of the incoming 
$f$ electron does not interact with the spin of the preexisting one because the energy of the double-occupied state is independent of the relative spin orientation.


The out-of-doublet contribution to the $zz$ component of the magnetic interaction for wave vector ${\bm k}$ is 
\begin{equation}
    I_{{\rm out},\bm k}^{zz} = 2( [\mathbb{\bar K}_{\rm out}^{(f^{2:1})}(\bm k)]_{\uparrow \uparrow \uparrow \uparrow} - [\mathbb{\bar K}_{\rm out}^{(f^{2:1})}(\bm k)]_{\uparrow \uparrow \downarrow \downarrow} ).
\end{equation}
By using the doublet projected hopping $\mathbb{\check T}_{\bm k \mu} = \mathbb{\bar V}_{\bm k  \mu} \mathbb{ V}_{\bm k \mu}^\dag $ of dimension $2\times14$ in Eq.~\eqref{eq:Kf2}, we obtain
\begin{widetext}
\begin{align}\label{eq:a-b}
    [\mathbb{\bar K}_{\rm out}^{(f^{2:1})}(\bm k)]_{\uparrow \uparrow \uparrow \uparrow} - [\mathbb{\bar K}_{\rm out}^{(f^{2:1})}(\bm k)]_{\uparrow \uparrow \downarrow \downarrow} =& \frac{1}{N} \sum_{\bm q}  g_{\varepsilon_{\bm k+\bm q \mu}, \varepsilon_{\bm q \mu',F^0} }^{(f^{2:1}) }  
     [( [\mathbb{\check T}_{\bm k + \bm q \mu}]_{\uparrow \alpha'}^{}    [\mathbb{\check T}_{\bm q \mu'}^*]_{\uparrow \alpha^{}}^{}  \gamma^\nu_{\uparrow\uparrow,\alpha \alpha'}   +  [\mathbb{\check  T}_{\bm k + \bm q \mu}^{*}]_{\uparrow \alpha }^{}  [\mathbb{\check  T}_{\bm q \mu'}^{}]_{\uparrow \alpha' }^{} \gamma^\nu_{\uparrow\uparrow,\alpha \alpha'} ) \nonumber \\
    &  - ( [\mathbb{\check  T}_{\bm k + \bm q \mu}]_{\uparrow \alpha'}^{}    [\mathbb{\check  T}_{\bm q \mu'}^*]_{\uparrow \alpha^{}}^{}  \gamma_{\downarrow\downarrow,\alpha \alpha'}^\nu  +   [\mathbb{\check  T}_{\bm k + \bm q \mu}^{*}]_{\downarrow \alpha }^{}  [\mathbb{\check  T}_{\bm q \mu'}^{}]_{\downarrow \alpha' }^{}  \gamma_{\uparrow\uparrow,\alpha \alpha'}^\nu  ) ] 
    \nonumber \\
     =& \frac{1}{N} \sum_{\bm q}  g_{\varepsilon_{\bm k+\bm q \mu}, \varepsilon_{\bm q \mu',F^0} }^{(f^{2:1}) }   
  ( [\mathbb{\check T}_{\bm k + \bm q \mu}^{*}]_{\uparrow \alpha }^{}  [\mathbb{\check  T}_{\bm q \mu'}^{}]_{\uparrow \alpha' }^{}  -  [\mathbb{\check  T}_{\bm k + \bm q \mu}^{*}]_{\downarrow \alpha }^{}  [\mathbb{\check T}_{\bm q \mu'}^{}]_{\downarrow \alpha' }^{} ) \delta_{\alpha \alpha'} = 0 
\end{align}
 where $\alpha$ and $\alpha'$ run over the 12 out-of-doublet $f$ states.
The cancellation of Eq.~\eqref{eq:a-b} arises from the relation 
\begin{align}\label{eq:hopp-TR}
     \sum_{\bm q} [\mathbb{\check T}_{\bm k + \bm q \mu}^{*}]_{\sigma \alpha }^{}  [\mathbb{\check T}_{\bm q \mu'}^{}]_{\sigma' \alpha }^{} \  g_{\varepsilon_{\bm k+\bm q \mu}, \varepsilon_{\bm q \mu',F^0} }^{(f^{2:1}) }  
     = {\rm sgn}(\sigma) {\rm sgn}(\sigma')  \sum_{\bm q} [\mathbb{\check T}_{\bm k + \bm q \mu}^{*}]_{\bar \sigma' \bar\alpha }^{}  [\mathbb{\check T}_{\bm q \mu'}^{}]_{\bar \sigma \bar\alpha }^{} \ g_{\varepsilon_{\bm k+\bm q \mu}, \varepsilon_{\bm q \mu',F^0} }^{(f^{2:1}) },
\end{align}
which is derived in Appendix \ref{app:aux-calculation} and is  a direct consequence of  time-reversal and spatial-inversion symmetries.  Note, also, that the cancellation holds for each independent process involving a particular pair of bands $(\mu, \mu')$ and a particular \emph{out-of-doublet} Kramers conjugate $(\alpha,\bar\alpha )$.

The remaining components of the magnetic interaction $I_{{\rm out},\bm k}^{\alpha\beta}$ also cancel out because the other elements of $\mathbb{\bar K}_{\rm out}^{(f^{2:1})}(\bm k)$ are null, as demonstrated below.

The interaction tensor component,
\begin{align}
    &[\mathbb{\bar K}_{\rm out}^{(f^{2:1})}(\bm k)]_{\uparrow \downarrow \downarrow \uparrow} = \frac{1}{N} \sum_{\bm q}  g_{\varepsilon_{\bm k+\bm q \mu}, \varepsilon_{\bm q \mu'} , F^0}^{(f^{2:1})} \ (-\delta_{\uparrow \alpha} \delta_{\downarrow \alpha'} )  
     ( [\mathbb{\check T}_{\bm k + \bm q \mu}^{}]_{\uparrow \alpha }^{}  [\mathbb{\check T}_{\bm q \mu'}^{*}]_{\downarrow \alpha' }^{}  +  [\mathbb{\check T}_{\bm k + \bm q \mu}^{*}]_{\uparrow \alpha }^{}  [\mathbb{\check T}_{\bm q \mu'}^{}]_{\downarrow \alpha' }^{} ) = 0,
\end{align}
\end{widetext}
cancels out because $\alpha$ or $\alpha'$ are out-of-doublet states,  meaning that they are different from $\uparrow, \downarrow$. Similarly, $[\mathbb{\bar K}_{\rm out}^{(f^{2:1})}(\bm k)]_{\uparrow \downarrow \uparrow \downarrow} = 0 $.
For $[\mathbb{\bar K}_{\rm out}^{(f^{2:1})}  (\bm k)]_{\uparrow \uparrow \uparrow \downarrow } $, we have
\begin{align}
    & [\mathbb{\bar K}_{\rm out}^{(f^{2:1})}  (\bm k)]_{\uparrow \uparrow \uparrow \downarrow }^{} = \frac{1}{N} \sum_{\bm q}  ( - [\mathbb{\check T}_{\bm k + \bm q \mu}^{}]_{\uparrow \alpha }^{}  [\mathbb{\check T}_{\bm q \mu'}^{*}]_{\uparrow \alpha' }^{} \ \delta_{\uparrow \alpha} \delta_{\downarrow \alpha'}   \nonumber \\
    & \ +  [\mathbb{\check T}_{\bm k + \bm q \mu}^{*}]_{\downarrow \alpha }^{}  [\mathbb{\check T}_{\bm q \mu'}^{}]_{\uparrow \alpha' }^{} \ \delta_{\alpha \alpha'} ) \  g_{\varepsilon_{\bm k+\bm q \mu}, \varepsilon_{\bm q \mu'} , F^0}^{(f^{2:1})}  = 0,
\end{align}
where the first term vanishes because $\delta_{\uparrow \alpha} \delta_{\downarrow \alpha'} = 0$ when $\alpha$ and $\alpha'$ run over out-of-doublet states. The second term is equivalent to Eq.~\eqref{eq:hopp-TR}, which vanishes for $\sigma=\bar \sigma'$. A similar demonstration leads to $[\mathbb{\bar K}_{\rm out}^{(f^{2:1})}(\bm k)]_{\uparrow \downarrow \uparrow \uparrow } = 0$.

For processes that involve $f^2$ virtual states on both sites, the cancellation of the components,
\begin{eqnarray}
&& [\mathbb{\bar K}_{\rm out}^{(f^{2:2})}(\bm k)]_{\uparrow \downarrow \downarrow \uparrow} = 0,
\quad \mathbb{\bar K}_{\rm out}^{(f^{2:2})}(\bm k)]_{\uparrow \uparrow \uparrow \downarrow} = 0,
\nonumber \\
&& [\mathbb{\bar K}_{\rm out}^{(f^{2:2})}(\bm k)]_{\uparrow \downarrow \uparrow \uparrow} = 0,
\quad [\mathbb{\bar K}_{\rm out}^{(f^{2:2})}(\bm k)]_{\uparrow \downarrow \uparrow \downarrow} = 0,
\end{eqnarray}
 arises from $\gamma_{\uparrow\downarrow,\alpha\alpha'}^\nu = - \delta_{\uparrow\alpha} \delta_{\downarrow\alpha'} $, which vanishes when $\alpha$ or $\alpha'$ runs over out-of-doublet states. Finally, the elements $[\mathbb{\bar K}_{\rm out}^{(f^{2:2})}(\bm k)]_{\uparrow \uparrow \uparrow \uparrow}$ and $[\mathbb{\bar K}_{\rm out}^{(f^{2:2})}(\bm k)]_{\uparrow \uparrow \downarrow \downarrow}$ do not vanish, but the difference does. From Eq.~\eqref{eq:Kf2f2}, we have
\begin{align}\label{eq:f2:2_a-b}
    &[\mathbb{\bar K}_{\rm out}^{(f^{2:2})}(\bm k)]_{\uparrow \uparrow \uparrow \uparrow} - [\mathbb{\bar K}_{\rm out}^{(f^{2:2})}(\bm k)]_{\uparrow \uparrow \downarrow \downarrow} = \frac{1}{N} \sum_{\bm q}  g_{\bm k+\bm q \mu, \bm q \mu',F^0,F^0}^{(f^{2:2})} \ \nonumber \\
    & \ \ \times  [\mathbb{T}_{\bm k + \bm q \mu}^{}]_{\alpha \beta' }^{}  [\mathbb{T}_{\bm q \mu'}^{*}]_{\alpha' \beta}^{} ( \gamma_{\uparrow \uparrow; \alpha \alpha' }^{\nu_1 }  \gamma_{\uparrow \uparrow; \beta \beta' }^{\nu_2 }  - \gamma_{\uparrow \uparrow; \alpha \alpha' }^{\nu_1 }  \gamma_{\downarrow \downarrow; \beta \beta' }^{\nu_2 }) \nonumber \\
    &= \frac{1}{N} \sum_{\bm q}  g_{\bm k+\bm q \mu, \bm q \mu',F^0,F^0}^{(f^{2:2})}  \  [\mathbb{T}_{\bm k + \bm q \mu}^{}]_{\alpha \beta' }^{}  [\mathbb{T}_{\bm q \mu'}^{*}]_{\alpha' \beta}^{}  \nonumber \\
    & \ \ \ \times 
     (\delta_{\alpha\alpha'} \delta_{\beta \beta'} - \delta_{\alpha\alpha'} \delta_{\beta \beta'} ) = 0 .
\end{align}
This implies that the contributions to $I_{{\rm out},\bm k}^{zz}$ from the $f^{2:2}$ out-of-doublet vanish when the $f^2$ multiplets are degenerate. Note that despite Eqs.~\eqref{eq:a-b} and \eqref{eq:f2:2_a-b} indicating that there is no contribution to $I_{{\rm out},\bm k}^{zz}$, there is a finite contribution from $[\mathbb{\bar K}_{\rm out}(\bm k)]_{\uparrow \uparrow \uparrow \uparrow} + [\mathbb{\bar K}_{\rm out}(\bm k)]_{\uparrow \uparrow \downarrow \downarrow}$, which is related to an effective density-density interaction. As anticipated, this contribution is irrelevant for the present work because  the ground space $S_0$ of the unperturbed Hamiltonian contains only states with one $f$ electron in each $f$ ion.

\subsection{Non-degenerate case }

In this section, we identify two mechanisms 
that reduce the out-of-doublet $f^2$ contributions relative to the sum of the $f^0$ and intradoublet $f^2$ contributions.
Since the anisotropic exchange arises from out-of-doublet $f^2$ contributions,
these two mechanisms are responsible for reducing the strength of the anisotropic exchange relative to the isotropic contribution. The mechanisms consist of a compensation  between positive and negative contributions arising  from different $f^2$ multiplets and from different bands.

\begin{figure}[h!]
    \centering
    \includegraphics[width=0.48\textwidth, trim=0 0 0 0,clip]{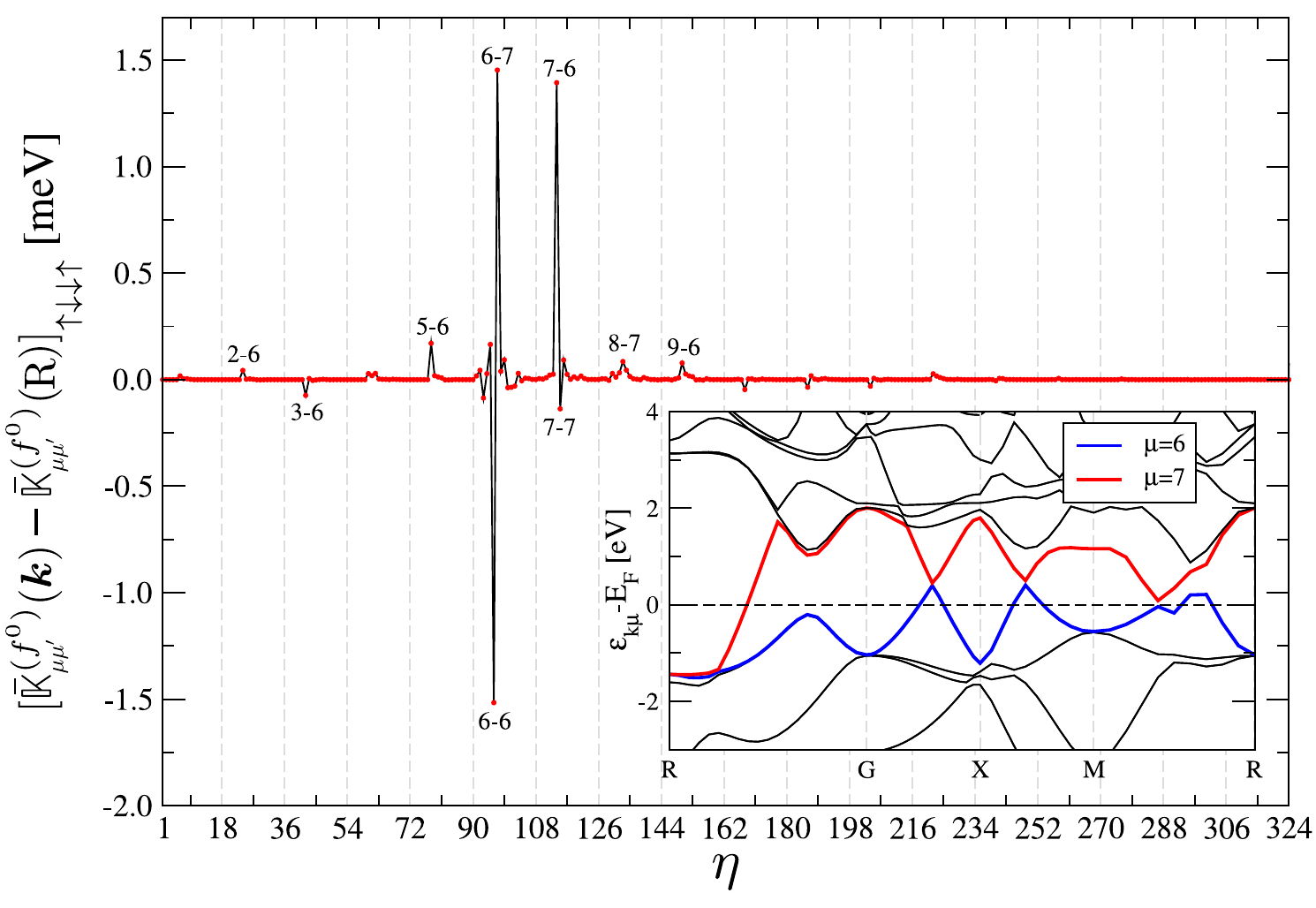}
    \caption{Contributions to $[\mathbb{\bar K}_{\mu\mu'}^{(f^0)}(\bm k)]_{\uparrow \downarrow \downarrow \uparrow } - [\mathbb{\bar K}_{\mu\mu'}^{(f^0)}({\rm R})]_{\uparrow \downarrow \downarrow \uparrow }$ involving different  pairs of conduction bands  for $\bm k=(0.083,0.083,0.083)$. The virtual process involves hybridizations with two bands labeled by $\mu$ and $\mu'$, and each pair of bands is denoted by the following combination of the two indices: $\eta =18(\mu-1)+\mu'$. The ``$ph$"  contributions are always negative. In contrast, contributions from ``$pp$" processes are positive and only exist for $\eta \geq96$. The labels next to the peaks indicate the band pair index $(\mu,\mu')$. The inset shows the conduction bands near the Fermi level for the path R$\Gamma$XM$\Gamma$. The bands indexed by $\mu=6$ and $\mu=7$ are the sole ones intersecting the Fermi level, thus constituting the primary contributors.}
    \label{fig:f0-bands}
\end{figure}

\begin{figure}[h!]
    \centering
    \includegraphics[width=0.48\textwidth, trim=0 0 0 0,clip]{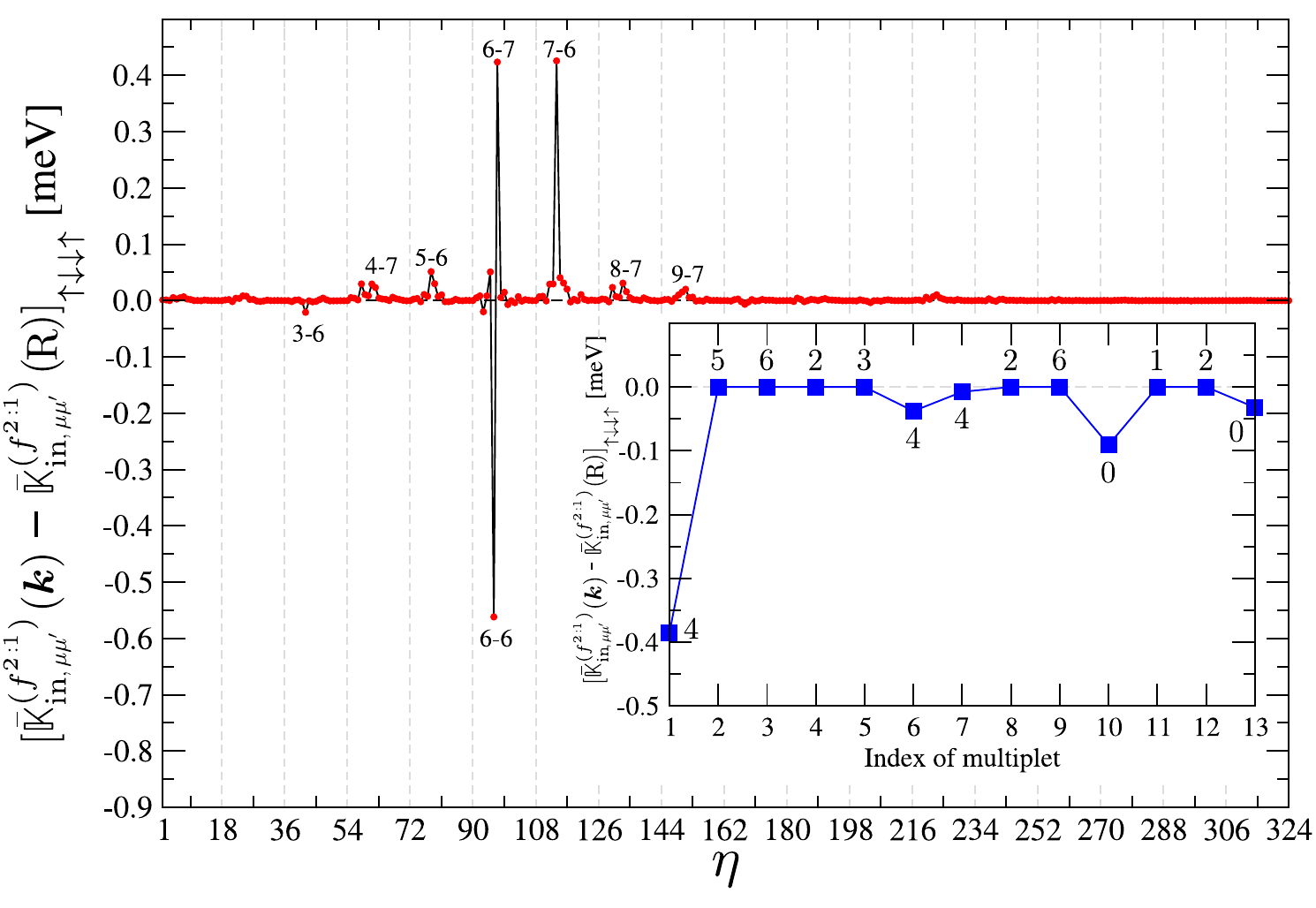}
    \caption{Contributions to  $[\mathbb{\bar K}_{\rm in}^{(f^{2:1})}(\bm k)-\mathbb{\bar K}_{\rm in}^{(f^{2:1})}({\rm R})]_{\uparrow \downarrow \downarrow \uparrow }$ involving different  pairs of conduction bands  for $\bm k=(0.083,0.083,0.083)$. The virtual process involves hybridizations with two bands labeled by $\mu$ and $\mu'$ and each pair of bands is denoted by the following combination of the two indices: $\eta =18(\mu-1)+\mu'$. The ``$hh$"  processes exist only for $\eta < 116$ and give
    positive contributions, while ``$pp$" processes exist only for $\eta \geq96$ and give a positive contribution. The labels next to the peaks indicate the band pair index $(\mu,\mu')$. The inset shows the separate contributions from each $f^2$ multiplet for  the band pair $(\mu,\mu')=(6,6)$, which gives the main contribution. The angular momentum of each multiplet is indicated next to each point. The selection rule derived in the text states that only the $J=4$ and $J=0$ multiplets are excited for  \emph{intra-doublet} $f^2$ processes. Note that all contributions from different multiplets have the same sign.}
    \label{fig:f2in-bands}
\end{figure}


\begin{figure}[h!]
    \centering
    \includegraphics[width=0.48\textwidth, trim=0 0 0 0,clip]{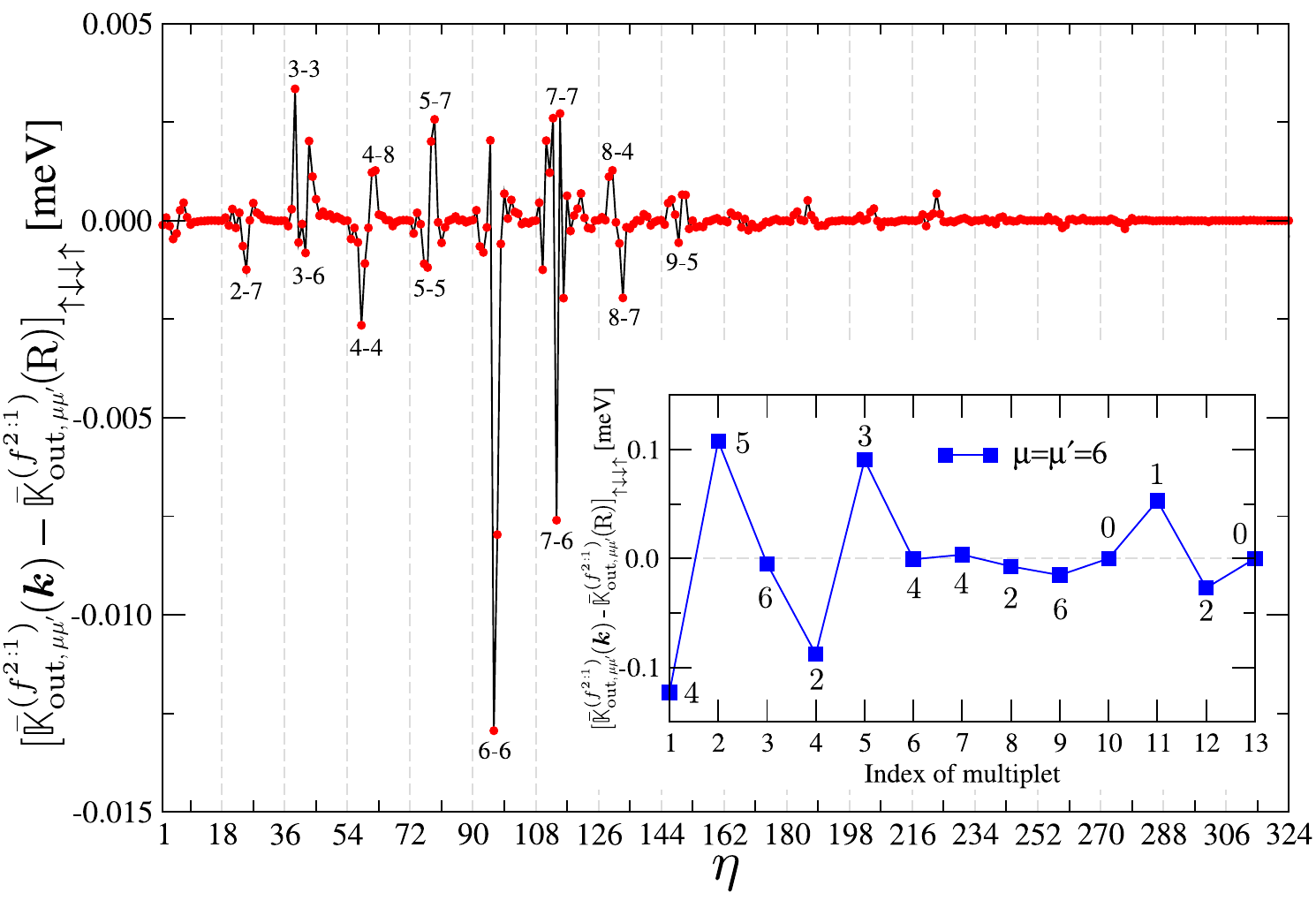}
    \caption{Contributions to $[\mathbb{\bar K}_{\rm out}^{(f^{2:1})}(\bm k) - \mathbb{\bar K}_{\rm out}^{(f^{2:1})}({\rm R})]_{\uparrow \downarrow \downarrow \uparrow }$ involving different  pairs of conduction bands  for $\bm k=(0.083,0.083,0.083)$. The virtual process involves hybridizations with two bands labeled by $\mu$ and $\mu'$ and each pair of bands is denoted by the following combination of the two indices: $\eta =18(\mu-1)+\mu'$. The ``$hh$" (``$pp$") processes contribute only for $\eta < 116$ ($\eta \geq96$ ), while the ``$ph$" processes contribute for any $\eta$ but mostly near $E_F$. The labels next to the peaks indicate the band pair index $(\mu,\mu')$. The inset showcases the multiplet contributions for the band pair $(\mu,\mu')=(6,6)$, with the total angular momentum $J$ of each multiplet indicated adjacent to its respective point.   }
    \label{fig:f2out-bands}
\end{figure}


 
We begin by observing a reduction of the interaction due to similar-magnitude, alternating-sign contributions from various band pairs within the $f^0$, $f^2$ intradoublet, and $f^2$ out-of-doublet channels. Again, we use CeIn$_3$ as a concrete example to illustrate this effect. This phenomenon is illustrated in Figs.~\ref{fig:f0-bands}-\ref{fig:f2out-bands}, where $\mathbb{\bar K}_{\mu\mu'}^{(f^0)}(\bm k) - \mathbb{\bar K}_{\mu\mu'}^{(f^0)}(R)$, $\mathbb{\bar K}_{{\rm in},\mu\mu'}^{(f^{2:1})}(\bm k) - \mathbb{\bar K}_{{\rm in},\mu\mu'}^{(f^{2:1})}(R)$ and $\mathbb{\bar K}_{{\rm out},\mu\mu'}^{(f^{2:1})}(\bm k) - \mathbb{\bar K}_{{\rm out},\mu\mu'}^{(f^{2:1})}(R)$ refer to the band pair $(\mu,\mu')$ contribution to the interaction bandwidth $W\sim \mathbb{\bar K}_{}^{}(\Gamma) - \mathbb{\bar K}_{}^{}(R)\sim$ \SI{4}{meV} (see Fig.~\ref{fig:int-channels}). Notably, across these channels, a significant contribution originates from a band intersecting the Fermi level, yet it remains uncompensated.

As shown in Fig.~\ref{fig:f0-bands}, in the $f^0$ channel, band compensation arises between particle-hole ($ph$) processes, which contribute negatively, and the particle-particle ($pp$) processes, which contributes positively.

Regarding the $f^{2:1}$ intradoublet and $f^{2:1}$ out-of-doublet channels, band contributions from $ph$ processes are offset by contributions from both $pp$ and hole-hole ($hh$) processes. Similarly, in the $f^{2:2}$ intradoublet and $f^{2:2}$ out-of-doublet channels, the $hh$ and $ph$ processes give contributions of opposite sign.

By comparing the summation over the band contributions with the summation of the absolute value of each band contribution, we find out that the reduction factor is 7.7 for the $f^{2:1}$ out-of-doublet channel, 2.7 for the $f^0$ channel, and 2.3 for the $f^{2:1}$ intradoublet channel. Since the anisotropic contributions are contained in the $f^{2:1}$ out-of-doublet channel, this result implies that the sign alternation in the sum over contributions from different band pairs reduces the relative strength of the anisotropic contributions by a factor of $\sim 3$.

\begin{figure}[h!]
    \centering
    \includegraphics[width=0.46\textwidth, trim=0 0 0 0,clip]{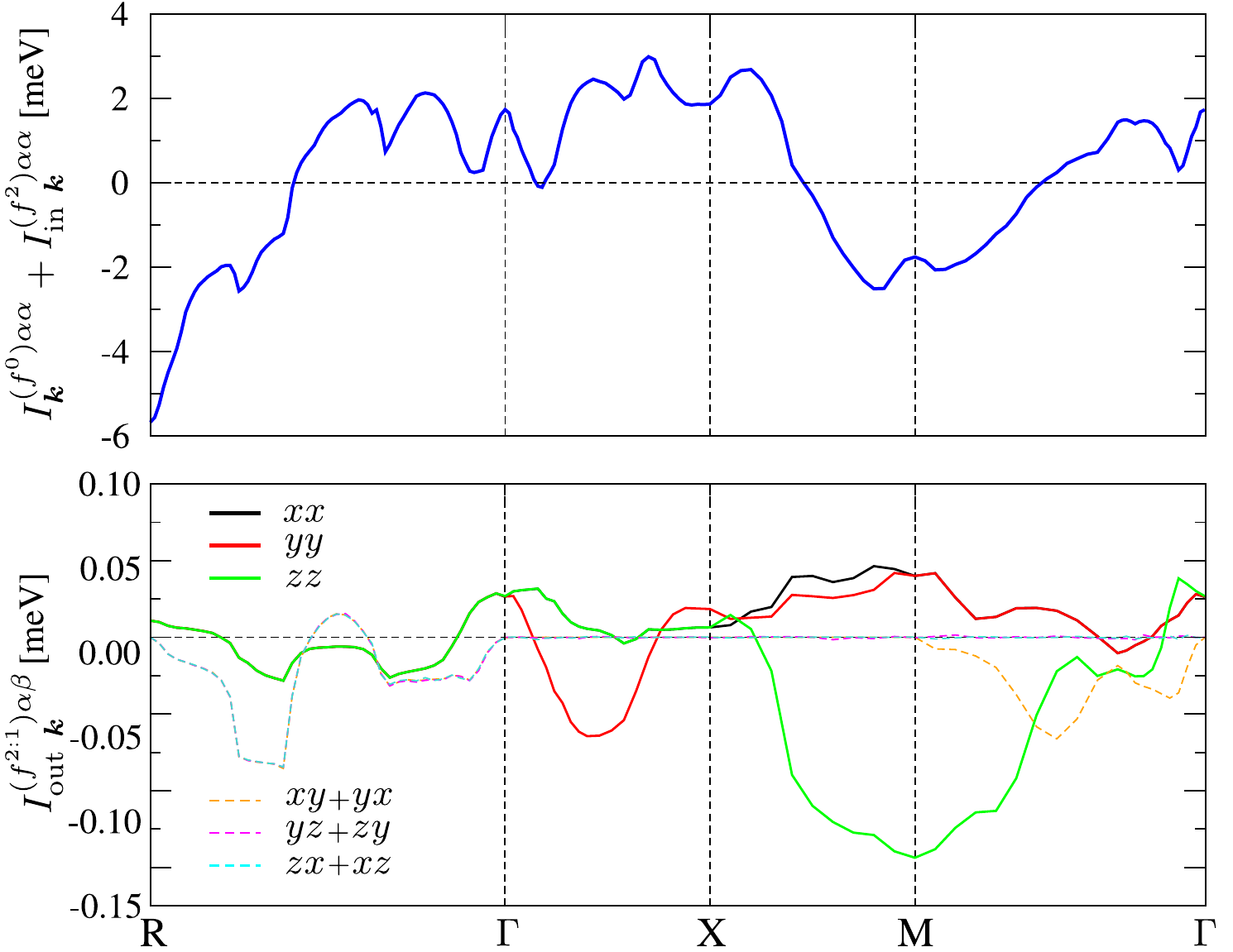}
    \caption{The upper panel shows the contribution from $f^0$ and \emph{intradoublet} $f^2$ virtual states to the magnetic interaction $I_{\bm k}^{(f^0)\alpha \alpha} + I_{{\rm in} \ \bm k}^{(f^2)\alpha \alpha}$ along the path R$\Gamma$XM$\Gamma$. The lower panel shows the magnetic interaction $I_{{\rm out} \ \bm k}^{(f^{2:1}) \alpha \beta}$ derived from the \emph{out-of-doublet} $f^{2:1}$ virtual states. Note that the anisotropic exchange interactions (lower panel) are roughly 50 times smaller than the isotropic exchange (upper panel). Contributions from \emph{out-of-doublet} $f^{2:2}$ virtual states are even smaller than $I_{{\rm out} \ \bm k}^{(f^{2:1}) \alpha \beta}$. } 
    \label{fig:net-interaction}
\end{figure}

However, looking at the net result  (see Fig.~\ref{fig:net-interaction}) that accounts for both contributions (band $+$ multiplet), the overall magnitudes of the $f^0$ and $f^{2:1}$ \emph{intradoublet} contributions are comparable, whereas the net contribution to   the $f^{2:1}$ \emph{out-of-doublet} channel is approximately 50 times smaller. The extra reduction factor of $\sim 50/3$ stems from the cancellation effects between contributions with alternating signs from different multiplets which only occur in the $f^{2:1}$ \emph{out-of-doublet} channel.


The inset of Fig.~\ref{fig:f2in-bands} reveals that the multiplet contributions within the $f^{2:1}$ \emph{intradoublet} channel exhibit the same sign. 
The consistent sign across all multiplets for $f^{2:1}$ intradoublet virtual processes stems from
\begin{equation}
\gamma_{\uparrow \uparrow \downarrow \downarrow}^\nu =   \langle 0 | f_{\uparrow}^\dag f_{\downarrow}^\dag |\eta_\nu \rangle \langle \eta_\nu |  f_{\downarrow}^{} f_{\uparrow}^{} |0\rangle = |  \langle 0 | f_{\uparrow}^\dag f_{\downarrow}^\dag |\eta_\nu \rangle  |^2 ,
\end{equation}
which is positive for any multiplet labeled by $\nu$.  The same holds true for $f^{2:2}$ intra-doublet virtual processes: $\Gamma_{\uparrow \uparrow \uparrow \uparrow}^{\nu_1 \nu_2}>0$ for any $\nu_1 \nu_2$.
Back to the $f^{2:1}$ \emph{intradoublet} processes, contributions with the same sign arise solely from the three $J=4$ multiplets, as well as from the two $J=0$ singlets, as shown in the inset of Fig.~\ref{fig:f2in-bands}. This selection rule can be derived as follows.
The representation of the intradoublet $f^2$ configuration can be decomposed as $\Gamma_7 \times \Gamma_7 = \Gamma_1 + \Gamma_4$, implying that the intradoublet $f^2$ state belongs to the $\Gamma_1$ irrep (singlet of the cubic group) (the $\Gamma_4$ states are symmetric under exchange of the two particles). The only $J$ multiplets that contain the $\Gamma_1$ irrep upon reducing the symmetry from SO(3) to cubic are  $J=0,4$ and $6$. However, since the $f^2$ state is built from two $J=5/2$ particles, the projection on the $J=6$ multiplet must be equal to zero.


For the $f^{2:1}$ \emph{out-of-doublet} channel, contributions from virtual states involving different multiplets are of a similar order of magnitude as in the $f^{2:1}$ intradoublet case. However, as illustrated in the inset of Fig.~\ref{fig:f2out-bands}, there is a significant reduction due to the alternating sign, which is a necessary consequence of the cancellation of the \emph{out-of-doublet} contributions in the degenerate limit that was derived in Sec.~\ref{subsec:degenerate-case}.
Comparing the net contribution against the sum of the absolute values of the contribution from each multiplet, we obtain a  reduction factor of $\sim 50$. 
The same mechanism works in the reduction of the $f^{2:2}$ out-of-doublet channel.

Since the \emph{out-of-doublet} $f^2$ configurations consist of one electron in the $\Gamma_7$ ($J=5/2$) doublet and a second electron in any orthogonal state, the possible configurations are $\Gamma_7$ ($J=5/2$) $\times \Gamma_6$ ($J=7/2$), $\Gamma_7$ ($J=5/2$) $\times \Gamma_7$ ($J=7/2$) or $\Gamma_7$ ($J=5/2$) $\times \Gamma_8$ ($J=5/2,7/2$). Since a singlet $J=0$ can only be obtained when both electrons occupy $J=5/2$ states, and $\Gamma_7 \times \Gamma_8 = \Gamma_3 + \Gamma_4 + \Gamma_5$, the virtual out-of-doublet $f^2$ configuration has zero projection on $J=0$ states. 






Finally, we can compare the relative contribution between the $f^0$ ($I_{\bm k}^{(f^0)\alpha\alpha}$) and $f^2$ \emph{intradoublet} ($I_{{\rm in \ }\bm k}^{(f^{2})\alpha\alpha}$), which are isotropic, with the contribution from the $f^2$ \emph{out-of-doublet} ($I_{{\rm out \ }\bm k}^{\alpha\beta}$). In Fig.~\ref{fig:net-interaction}, the upper panel shows the isotropic component, which is 50 times bigger than the anisotropic component shown in the lower panel.

 We also considered the effective magnetic interaction between nearest-neighbor ions arising from direct hopping between the  $f$ ion. The calculation in Appendix \ref{app:Hff} shows that this interaction, of the order of $\SI{1}{\mu eV}$, is very anisotropic, but extremely weak in comparison to the contribution from  fourth-order processes in the hybridization between $f$ orbitals and band states. This smallness can be understood after realizing that the effective $f-f$ hopping between $J=5/2$ multiplets vanishes exactly for a single molecule of two $f$ ions under the assumption of uniaxial symmetry
(i.e., invariance under continuous rotations along the axis of the molecule)~\cite{BatistaCD2009}. Since this continuous symmetry is weakly broken by the crystal environment, the $f-f$ hopping matrix is finite but strongly suppressed when projected onto the lowest-energy $J=5/2$ multiplet.

 Finally, it is important to emphasize that nearly isotropic exchange interactions are rare in $f$-electron magnets. This rarity is one reason we chose CeIn$_3$ to illustrate conditions where exchange anisotropy can be over an order of magnitude smaller than the isotropic (Heisenberg) contribution. In addition to requiring Kramers doublets and centrosymmetric bonds, there must be dominant intradoublet contributions compared to out-of-doublet ones. This requirement is met in CeIn$_3$ due to the alternating signs of contributions from different $f^2$ multiplets with similar energies and from different bands near the Fermi level. However, this property does not necessarily hold for other $f$-electron materials. For instance, preliminary calculations for CeRhIn$_5$ indicate that the exchange anisotropy for bonds connecting different CeIn$_2$ layers is comparable to the isotropic contribution when the $f$ energy is \SI{2.5}{eV} below the Fermi level. In general, we expect the relative exchange anisotropy to become stronger when the weight of the magnetic (out-of-doublet) $f^2$ fluctuations increases relative to the $f^0$ and intradoublet $f^2$ fluctuations.

\section{Experimental Validation}\label{sec:VAL}

In this section, we validate the procedure to obtain a low-energy spin Hamiltonian derived in previous sections through comparison with INS experiments conducted on CeIn$_3$~\cite{SimethW2023_CeIn3}. Unlike the comparison presented in Ref.~\cite{SimethW2023_CeIn3}, our effective spin model incorporates $f^2$ configurations to the virtual processes.
Although these processes do not significantly alter the overall magnon dispersion, they do augment the single-magnon bandwidth. Consequently, to maintain the experimentally observed magnon bandwidth, the energy $\varepsilon_f$ of the $f$ levels must be downwardly adjusted relative to the Fermi level $E_F$. This adjustment yields a new value of $\varepsilon_f$  that better aligns  with findings from photoemission experiments~\cite{KimHD1997,ZhangY2016}, but is still not sufficient to reproduce the photoemission data. 

To derive the MO-PAM for CeIn$_3$, 
DFT calculations were performed using the \textsc{Quantum ESPRESSO} package, assuming itinerant $f$ electrons and including spin-orbit coupling~\cite{GiannozziP2017}. Fully relativistic projector augmented-wave (PAW) pseudopotentials and the Perdew-Burke-Ernzerhof (PBE) exchange-correlation functional from PSlibrary~\cite{DalCorsoA2014} were implemented. A 25-orbital tight-binding model was then constructed using the Wannier90 package to match the DFT-derived electronic structure~\cite{PizziG2020}. As emphasized by Simeth \textit{et al}.~\cite{SimethW2023_CeIn3}, accurately describing the electronic structure of the conduction bands near the Fermi level requires incorporating 18 conduction electron orbitals per spin: 9 In-$p$, 3 In-$s$, 5 Ce-$d$, and 1 Ce-$s$ orbitals. This comprehensive approach ensures proper localization of the $f$ orbitals. The calculated Fermi level is 
$E_F=$ \SI{12.588}{eV}.

\begin{figure}[h!]
    \centering
    \includegraphics[width=0.48\textwidth, trim=0 0 0 0,clip]{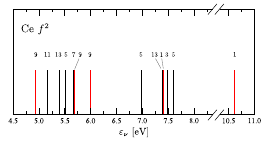}
    \caption{Energy spectrum of Ce${}^{2+}$ ion (4$f^2$ configuration). These results are taken from Table \ref{tab:table1} for $F^0=\SI{6.4}{eV}$~\cite{GamzaM2008}. The numbers above the vertical lines indicate the number of states in each multiplet. The red color marks the multiplets involved in the $f^2$ \emph{intradoublet} processes.}\label{fig:f2-spectrum}
\end{figure}

The Gunnarsson-Schönhammer method is commonly used to determine the value of Slater integral $F^0$ from x-ray photoelectron spectroscopy (XPS) data. For Ce compounds, $F^0$ typically falls within the range of 5.5 to \SI{8}{eV}, although it tends to be closer to \SI{6}{eV} for intermetallic compounds~\cite{SchneiderWD1985}. An experimental investigation of CeIn$_3$ has reported $F^0=$ \SI{6.4}{eV}~\cite{GamzaM2008}. 
In this work, we adopt the following values for the Slater integrals~\cite{Jang2023}: $F^0=$\SI{6.4}{eV}, $F^2=$\SI{8.3436}{eV}, $F^4=$\SI{5.57482}{eV}, and $F^6=$\SI{4.12446}{eV},
which align with the findings of the aforementioned studies. Additionally, we utilize $\zeta_{\rm SO}=$\SI{280}{meV} for the spin-orbit coupling.
We opted to disregard the CF in the single-ion Hamiltonian, given that the energy scale of the splitting (of the order of $\Delta=$\SI{10} {meV}) pales in comparison to the gaps generated by the spin-orbit coupling~\cite{KnafoW2003}. The energy spectrum is displayed in Fig.~\ref{fig:f2-spectrum}.

\begin{table}[b]
\caption{\label{tab:table1}%
Spectrum of Ce$^{2+}$ computed by using the free-ion model of Eq.~\eqref{eq:Hion} with the parameters in Eq.~\eqref{eq:Hion-param}.
}
\begin{ruledtabular}
\begin{tabular}{llcc}
$\nu$ &
\textrm{$\varepsilon_\nu - F^0$ (eV)
}&
\textrm{Degeneracy }&
\textrm{Composition}\\
\colrule 
\\
1 &  $-1.460198$ & 9    & ${}^{3}H_4$ \\
2 &  $-1.235430$ & 11  & ${}^{3}H_5$\\
3 &  $-1.000293$ & 13  & ${}^{3}H_6$ \\
4 &  $-0.886112$ & 5   & ${}^{3}F_2$ \\
5 &  $-0.739996$ & 7   & ${}^{3}F_3$\\
6 &  $-0.715614$ & 9   & ${}^{3}F_4$,  ${}^{1}G_4$  \\
7 &  $-0.411830$ & 9   & ${}^{3}F_4$,  ${}^{1}G_4$ \\
8 &  $+0.575542$ & 5   & ${}^{1}D_2$ \\
9 &  $+0.978563$ & 13  & ${}^{1}I_6$  \\
10 & $+1.012500$ & 1   & ${}^{3}P_0$ \\
11 & $+1.076594$ & 3   & ${}^{3}P_1$\\
12 & $+1.205526$ & 5   & ${}^{3}P_2$ \\
13 & $+4.224071$ & 1   & ${}^{1}S_0$\\
\end{tabular}
\end{ruledtabular}
\end{table}



As discussed in more detail in the next section, we choose to leave the $f$-energy level $\varepsilon_f$ as a free parameter to fit the experimental magnon bandwidth $W_{\rm exp} =$ \SI{2.75(03)}{meV}. The optimized value is found to be 
$E_F-\varepsilon_f=$ \SI{0.65}{eV}, which is smaller than the experimental value deduced to be between 1.5 and \SI{2}{eV} from photoemission measurements~\cite{KimHD1997,ZhangY2016}.
We note that charge fluctuations appear to second order in the $f-c$ hybridization and that these Hamiltonian parameters lead to charge fluctuation gaps of $E_F-\varepsilon_f =$ \SI{0.65}{eV} for $f^0$ and $\varepsilon_{\nu=1} - (E_F-\varepsilon_f) =$ 4.94 - \SI{0.65}{eV} = \SI{4.29}{eV}  for $f^2$. These gaps are consistent with the observed predominance  of $f^0$ fluctuations~\cite{LawrenceJ1979, FuggleJC1983, SundermannM2016}.





To test the theoretically calculated exchange interaction $I_{\bm k}$ quantitatively against experiment, we compute the magnetic excitations for the spin Hamiltonian in Eq.~\eqref{eq:Hspin-doublet}. Since contributions from the $f^2$ \emph{out-of-doublet} channel are negligible, we use the magnetic interaction obtained in Eq.~\eqref{eq:int-iso}, which is shown in the upper panel of Fig.~\ref{fig:net-interaction}. The single-magnon dispersion is given by
\begin{equation}\label{eq:magnon-disp}
    E_{\bm k} = S \sqrt{(I_{\bm Q} - I_{\bm k}) (I_{\bm Q} - I_{\bm Q-\bm k})}
\end{equation}
with $S=1/2$ and the magnetic ordering wave vector $\bm Q = (\frac{1}{2},\frac{1}{2},\frac{1}{2})$.

\begin{figure}[t!]
    \centering
    \includegraphics[width=0.48\textwidth, trim=0 0 0 0,clip]{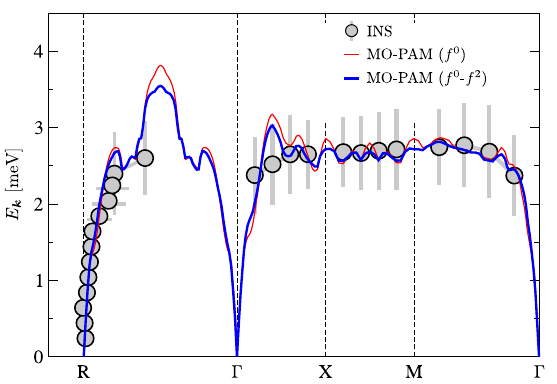}
    \caption{Comparison between the single-magnon dispersion $E_{\bm k}$ obtained from $\mathcal{H}_{\rm spin}$ and the dispersion extracted from INS data for CeIn$_3$ along the path R$\Gamma$XM$\Gamma$. The blue solid line shows the magnon dispersion derived from the MO-PAM with $f^0$ and \emph{intradoublet} $f^2$ configurations for $E_F-\varepsilon_f=$ \SI{0.65}{eV}. The red solid line is the magnon-derived dispersion from the MO-PAM with only the $f^0$ channel for $E_F-\varepsilon_f=$ \SI{0.512}{eV} as in Ref.~\cite{SimethW2023_CeIn3}.}
    \label{fig:magnon-disp}
\end{figure}

Figure~\ref{fig:magnon-disp} shows a comparison between the magnon dispersion $E_{\bm k}$ calculated from Eq.~\eqref{eq:magnon-disp} and the one extracted from INS experiments. Remarkably, the calculated dispersion quantitatively reproduces the measured dispersion, including the magnon velocity.
From the INS data, the averaged value of the magnon velocity at the $R$ point along the directions $R\Gamma$, $RX$ and $RM$ is $v_{\rm exp}=$ \SI{38.6(8)}{meV}/r.l.u, with 1 r.l.u. $\simeq$ \SI{1.340}{\r{A}}, while the average value obtained for $\mathcal{H}_{\rm spin}$ is $v_{}=$\SI{42(1)}{meV}/r.l.u. For comparison, in Fig.~\ref{fig:magnon-disp}, we have also included the magnon dispersion computed in Ref.~\cite{SimethW2023_CeIn3} with only the $f^0$ channel contributions. 

\begin{figure}[h!]
    \centering
    \includegraphics[width=0.48\textwidth, trim=0 0 0 0,clip]{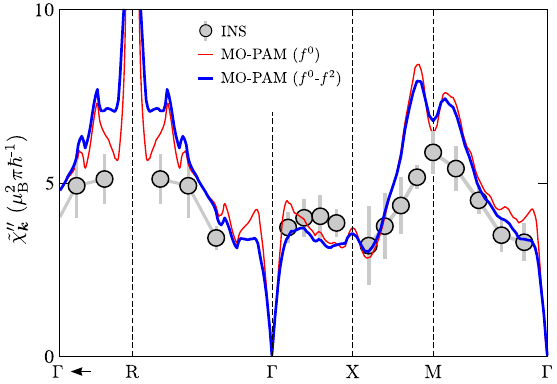}
    \caption{Comparison between the SSF of the MO-PAM and the integral over energy of the imaginary part of the dynamical magnetic susceptibility, $\tilde \chi_{\bm k}''$, extracted from the INS data for CeIn$_3$ along the path R$\Gamma$XM$\Gamma$. See Ref.~\cite{SimethW2023_CeIn3} for more details. The blue line shows the SSF derived from the MO-PAM with $f^0$ and \emph{intradoublet} $f^2$ configurations for $E_F-\varepsilon_f=$ \SI{0.65}{eV}. The red line is the SSF derived from MO-PAM with only the $f^0$ channel for $E_F-\varepsilon_f=$ \SI{0.512}{eV} as in Ref.~\cite{SimethW2023_CeIn3}.}
    \label{fig:SSF}
\end{figure}

To go further in the quantitative comparison with the INS experiment, we test the spectral weight distribution of the magnetic excitations. The agreement in the spectral weight distribution can be quantified by comparing the calculated static structure factor (SSF) with the integral over energy of the imaginary part of the dynamic magnetic susceptibility extracted from the INS experiment (see Ref.~\cite{SimethW2023_CeIn3} for details). The SSF is given by the expression
\begin{equation}\label{eq:SSF}
    \mathcal{S}^{} (\bm k) = S \sqrt{\frac{I_{\bm Q} - I_{\bm k + \bm Q}}{I_{\bm Q} - I_{\bm k}}} \ .
\end{equation}
In Fig.~\ref{fig:SSF}, we show the comparison between the spectral weight integrated over energy from INS data, and the SSF obtained from Eq.~\eqref{eq:SSF}. The blue line shows the SSF for the $\mathcal{H}_{\rm spin}$ derived with $f^0$ and $f^2$ channels, and the red line show the SSF derived with only the $f^0$ channel as in Ref.~\cite{SimethW2023_CeIn3}.


\subsection{Fitting of the Ce-4f energy level.}

In this section, we study the sensitivity of the magnon dispersion to different values of the $f$-orbital energy $\varepsilon_f$ and the Coulomb integral $F^0$, which controls the relative strength of the $f^0$ and $f^2$ processes. We note that these parameters primarily adjust the magnon bandwidth without significantly altering the momentum dependence. Therefore, we will concentrate on fitting the magnon bandwidth to the INS experiment.

The range of possible values of $\varepsilon_f$ is constrained by the validity of the perturbation theory, i.e., the energy differences that appear in the denominator must remain much larger than the matrix elements of the $f-c$ hybridization that appear in the numerator of the functions given 
in Eqs.~\eqref{eq:g-f0}, (\ref{eq:f2:1}), and (\ref{eq:g-f22}). 
Note that $g_{\varepsilon_{\bm k +\bm q \mu},\varepsilon_{\bm q \mu'}}^{(f^0)}$ diverges when $E_F-\varepsilon_f \rightarrow 0$, while $g_{\varepsilon_{\bm k +\bm q \mu},\varepsilon_{\bm q \mu'},\varepsilon_{\nu}}^{(f^{2:1})}$ diverges when either $E_F-\varepsilon_f \rightarrow 0$ or $E_F-\varepsilon_f  \rightarrow \varepsilon_{\nu=1}$, and the function  $g_{\varepsilon_{\bm k +\bm q \mu},\varepsilon_{\bm q \mu'},\varepsilon_{\nu_1},\varepsilon_{\nu_2}}^{(f^{2:2})}$ diverges when $E_F-\varepsilon_f  \rightarrow \varepsilon_{\nu=1}$. Thus,  $\varepsilon_f$ must satisfy $0 < E_F-\varepsilon_f < \varepsilon_{\nu=1}$.

\begin{figure}[h!]
    \centering
    \includegraphics[width=0.48\textwidth, trim=0 0 0 0,clip]{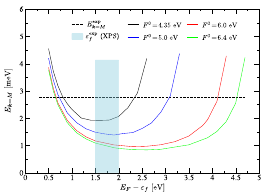}
    \caption{Magnon energy  at the $M$ point vs $E_F-\varepsilon_f$ for different values of $F^0$. The magnon bandwidth equals the experimental value when $E_F-\varepsilon_f=0.65$ or \SI{4.53}{eV}, for the case with $F^0=$ \SI{6.4}{eV} (green line). The shaded area shows the range values of 
    $E_F-\varepsilon_f$ according to spectroscopic experiments.}
    \label{fig:bandwidth-vs-ef}
\end{figure}

The magnon bandwidth can be parameterized by the value of the dispersion at the $M$ point. Then, Fig.~\ref{fig:bandwidth-vs-ef} shows the magnon energy at the $M$ point $E_{\bm k =M}$ for different values of $F^0$ and $E_F-\varepsilon_f$ in the range $[0:\varepsilon_{\nu=1}]$. 
For $F^0 =$ \SI{6.4}{eV}, two values fit the experimental bandwidth: $E_F - \varepsilon_f =$ \SI{0.65}{eV} and $E_F - \varepsilon_f =$ \SI{4.53}{eV}.  Although both solutions yield nearly identical magnon dispersion, the solution $E_F - \varepsilon_f =$ \SI{4.53}{eV} must be discarded as it implies that $f^2$ charge fluctuations dominate over $f^0$ ones.
{For smaller values of $F^0$, the two values $E_F-\varepsilon_f$ that fit the magnon bandwidth get closer to the experimental $\varepsilon_f^{\rm exp}$ range. Nevertheless, note that for $F^0=$\SI{4.35}{eV}, which is significantly below the experimental range, it is not possible to fit the magnon bandwidth for $E_F-\varepsilon_f$ in the range 1.5 - \SI{2}{eV}.}
 Moreover, values $F^0$ that better align with the magnon bandwidth for $E_F-\varepsilon_f=1.5$ - \SI{2.0}{eV}, will result in  $f^2$ charge fluctuations predominating over the $f^0$.  

 In summary, the inclusion of the $f^2$ processes is insufficient to account for the measured energy difference of $E_F-\varepsilon_f=1.5$ - \SI{2.0}{eV}.  If this energy difference were forced to match the experimental value, the resulting magnon bandwidth would be approximately half of the observed value  (see Fig.~\ref{fig:bandwidth-vs-ef}). This result suggests that the  $f-c$  hybridization is underestimated by the band structure calculations.  The results shown in Fig.~\ref{fig:bandwidth-vs-ef} indicate that an upward renormalization of the hybridization by a factor $\approx 1.22$ will reproduce the measured magnon bandwidth for $E_F-\varepsilon_f=1.5$ - \SI{2}{eV} (see shaded area in Fig.~\ref{fig:bandwidth-vs-ef}). This is because the magnon bandwidth increases as the fourth power of the hybridization, and the factor $(1.22)^4 = 2.2$, when multiplied by the bandwidth of \SI{1.25}{meV} obtained for $E_F - \varepsilon_f =$ \SI{1.5}{eV}, results in the experimental value of approximately \SI{2.75}{meV}. This observation underscores the importance of accurately determining the $f-c$ hybridization amplitudes ${\mathbb{V}}_{m j_{\alpha} n_{\alpha}}^{\alpha}$  in first-principles calculations for deriving low-energy effective models. 

The discrepancy between the theoretical and experimental values of $\varepsilon_f$ could, in principle, be attributed to magnon damping and renormalization caused by the Stoner continuum. An upward renormalization of the magnon would bring the theoretical prediction in line with the observed magnon bandwidth. However, damping would also result in a broadening of the magnon peak and a redistribution of spectral weight. Since no peak broadening is observed in the INS experiment and the spectral weight distribution is well captured by spin wave theory, we conclude that the damping effect from particle-hole excitations is too weak to significantly alter the magnon bandwidth.

\section{Conclusions \label{sec:CONC}}

Developing minimal microscopic models that can quantitatively account for the properties of specific $f$-electron materials remains a longstanding challenge in correlated electron physics. Pure first-principles calculations struggle to account for strongly correlated effects, such as the localization of $4f$ electrons. This localization arises from the combination of strong intra-atomic Coulomb interactions and weak hybridization of $4f$ orbitals with the orbitals of neighboring ions. On the other hand, typical many-body approaches to this problem often start from phenomenological theories or oversimplified microscopic models, aiming to capture qualitative aspects such as heavy-fermion behavior, non-Fermi-liquid fixed points, and unconventional superconductivity. While these approaches are undoubtedly valuable for exploring and explaining new states of matter, their tenuous connection to specific materials hinders experimental validation.

This work attempts to combine both approaches: first-principles calculations for accurately describing the non-$f$-orbital band dispersion, and a many-body treatment for addressing the strongly correlated effects induced by the strong intra-atomic Coulomb interaction of electrons in the $f$ orbitals. The resulting MO-PAM serves as the starting point for deriving low-energy Hamiltonians via degenerate perturbation theory, including a Kondo-Heisenberg lattice model and an effective spin Hamiltonian $\mathcal{H}_{\rm spin}$. This approach, valid in the localized regime, has been validated against INS data in CeIn$_3$.

An important difference relative to more traditional perturbative approaches is that we incorporate all the fourth-order processes in the hybridization between the $f$ orbitals and the band states in $\mathcal{H}_{\rm spin}$. As emphasized in our previous work on CeIn$_3$~\cite{SimethW2023_CeIn3}, the inclusion of particle-particle processes (and hole-hole processes when $f^2$ virtual states are also included) in addition to the more standard particle-hole processes that 
lead to the effective RKKY interaction is crucial to explain the material's magnetic ordering and single-magnon dispersion.
However, in Ref.~\cite{SimethW2023_CeIn3} we included only $f^0$ virtual processes, assuming suppressed $f^2$ charge fluctuations in CeIn$_3$. Formally, this simplification was achieved by setting the Coulomb integral $F^0 \to \infty$, effectively
eliminating the $f^2$ virtual states.

While considering that only the $f^0$ virtual states may be a reasonable approximation for CeIn$_3$, it is not necessarily sufficient for other compounds containing Ce$^{3+}$ (one $f$-electron) or Yb$^{3+}$ (one $f$-hole) ions that can be described using the formalism presented in this work. Moreover, including $f^2$ virtual states is essential for obtaining anisotropic exchange interactions when the bond center under consideration is a center of spatial inversion and the lowest-energy CF multiplet is a Kramers doublet. As demonstrated in this work, the combination of spatial inversion and time-reversal symmetries can lead to predominantly isotropic exchange interactions. More specifically, interactions mediated by $f^0$ processes ($f$ electrons leave their $f$ orbitals to occupy band states) and intra doublet $f^2$ processes ($f$ orbitals are virtually double occupied with both electrons in the lowest-energy doublet) are purely isotropic under these conditions.

Exchange anisotropy arises from \emph{out-of-doublet} $f^2$ virtual processes, where the second $f$ electron occupies an $f$ state that is orthogonal to the lowest-energy doublet in which the first electron resides. More specifically, we have observed that the net contribution from these out-of-doublet $f^2$ processes can be significantly suppressed due to a combination of two effects that are present in CeIn$_3$. Firstly, the projection of these out-of-doublet $f^2$ states into the different multiplets (eigenstates of the single-ion Hamiltonian) leads to contributions with alternating signs that exactly cancel out in the limit when the multiplets are degenerate.
Second, the sign also alternates for contributions involving different pairs of bands, where one $f$ electron virtually hops via one band to double occupy another $f$ ion, and then returns to the original ion through a second band state. This sign alternation not only reduces the exchange anisotropy, but also the isotropic exchange contribution arising from out-of-doublet $f^2$ processes. 

The strong reduction in exchange anisotropy is consistent with the weak anisotropy observed experimentally in CeIn$_3$. The strong suppression of out-of-doublet $f^2$ processes,
contributing approximately 50 times less than the other processes combined,
allows for approximated treatments that include only $f^0$ and $f^2$ intradoublet processes. 
This approximation significantly reduces the computational cost of calculating $\mathcal{H}_{\rm spin}$, as the number of out-of-doublet $f^2$ processes (the second $f$ electron can occupy any of the 12 available $f$ orbitals) far exceeds that of intradoublet  $f^2$ processes (the second electron can only occupy the empty state of the lowest-energy doublet). 

 While the effective spin Hamiltonian derived in this work reproduces the magnetic properties of CeIn$_3$, 
the value of $E_F-\varepsilon_f=0.65$ that fits the measured magnon bandwidth does not agree with the photoemission data suggesting that $E_F-\varepsilon_f=1.5$ - \SI{2.0}{eV}~\cite{KimHD1997,ZhangY2016}. As explained in the previous section, this disagreement can be eliminated if the $f-c$ hybridization amplitudes ${\mathbb{V}}_{m j_{\alpha} n_\alpha}^\alpha$ extracted from first-principle calculations are re-scaled by a factor of $\approx 1.22$. Since ${\mathbb{V}}_{m j_{\alpha} n_\alpha}^\alpha$  enters to fourth order in the effective spin-spin interactions, more work should be devoted in the future to refine the estimate of ${\mathbb{V}}_{m j_{\alpha} n_\alpha}^\alpha$ from first-principles calculations

The results derived in this work also offer insight into the possible origin of quasi-isotropic super-exchange interactions in insulators containing Yb$^{3+}$ ions with a low-energy CF doublet~\cite{SalaG2021,DingL2019,BordelonMM2019,BordelonMM2020,XingJ2019_CsYbSe2,XieT2023,VillanovaJW2023,ScheieAO2024}. According to our findings, in the absence of intra-atomic interaction on the ligand field (e.g., oxygen $p$ orbitals in the case of oxides), the anisotropic contribution to the super-exchange interaction on centrosymmetric bonds arises from  out-of-doublet $f^2$ processes, where the extra $f$ electron occupies an $f$ orbital that is orthogonal to the lowest-energy doublet. Notably, the net contribution from these out-of-doublet $f^2$ processes vanishes in the limit where the $f^2$ $J$ multiplets are degenerate.  This observation suggests that the contribution from out-of-doublet $f^2$ processes can be much smaller than the  isotropic one from intradoublet $f^2$ processes.  This occurs due to partial cancellation between virtual processes involving different $f^2$ $J$ multiplets, as the existing splitting between different $f^2$ $J$ multiplets turns the exact cancellation into a partial one.

 In summary, the results presented in this work demonstrate the feasibility of deriving microscopic effective models for $4f$-electron intermetallic compounds. This achievement is particularly remarkable given the three orders of magnitude separating the energy scales of the bare interactions (of the order of eV) from the effective spin-spin interactions (of the order of meV). Moreover, the approach successfully accounts for the magnitude of exchange anisotropy, which plays a crucial role in stabilizing various magnetic textures such as skyrmion crystals~\cite{WangZ2020_RKKY}, and quantum spin-liquid phases such as the Kitaev model~\cite{KitaevA2006}. These results suggest that $f$-electron materials are entering a new era, where speculative statements based on oversimplified phenomenological models can now be tested against quantitatively accurate microscopic approaches.


\begin{acknowledgments}
We thank Jeffrey G. Rau for useful discussions. This work was carried out under the auspices of the U.S. Department of Energy, Office of Science, Basic Energy Sciences, Materials Sciences and Engineering Division project ``Quantum Fluctuations in Narrow Band Systems'' (F.R.) and LANL LDRD Program (J.X.Z. and E.A.G.). C.D.B. acknowledges support from the Center for Nonlinear Studies (CNLS) through the Stanislaw M. Ulam Distinguished Scholar position funded by LANL's LDRD program. 
Z.\nobreak\,W. acknowledges support of the National Natural Science Foundation of China Grant No.~12374124.
L. M. Ch. was supported by the U.S. Department of Energy, Office of Science, Basic Energy Sciences, Materials Science and Engineering Division.
W.S. was supported through funding from the European Union's Horizon 2020 research and innovation programme under the Marie Sklodowska-Curie Grant Agreement No. 884104 (PSI-FELLOW-III-3i). M.J. acknowledges funding by the Swiss National Science Foundation through the Project No.~200650. Y.N. acknowledges the support from JSPS KAKENHI (Grant No. JP23H04869).
\end{acknowledgments}

\appendix

\section{Kondo Lattice Model}\label{app:KLM}
\setcounter{figure}{0}                       
\renewcommand\thefigure{A.\arabic{figure}}   

In this appendix, we show the steps for the derivation of the KLM Hamiltonian given in Eq.~\eqref{eq:KLM}. 
The KLM is obtained through a second-order perturbation in the MO-PAM presented in Eq.~\eqref{eq:AndersonModel}. The low-energy effective Hamiltonian is given by~\cite{LindgrenI1974}
\begin{equation}
    \mathcal{H}_{\rm eff} = \mathcal{P}_{\rm S} \mathcal{H} \mathcal{P}_{\rm S} + \mathcal{P}_{\rm S} \left[  \mathcal{V}  \hat G^{0}(\varepsilon_0)  \mathcal{V}   \right] \mathcal{P}_{\rm S} 
\end{equation}
where the resolvent operator is given by
\begin{equation}
    \hat G^{0}(\varepsilon_0) = \mathcal{P}_{\rm S}^\perp \frac{1}{\varepsilon_0 - \mathcal{ H}_{\rm 0} } \mathcal{P}_{\rm S}^\perp \ ,
\end{equation}
$\mathcal{P}_{\rm S}^\perp = \hat{1} - \mathcal{P}_{\rm S}$, and $\varepsilon_0$ is the ground-state energy of $\mathcal{ H}_{\rm 0}$. The projector $\mathcal{P}_{\rm S}$ is already defined in the main text.

\begin{figure}[h!]
    \centering
    \includegraphics[width=0.48\textwidth]{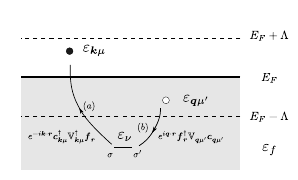}
    \caption{Diagram of virtual processes involved in the derivation of the KLM. Particle-hole pairs are created near the Fermi level $E_F$. The energy of the $f$ orbital is $\varepsilon_f$. The indices $\sigma$ and $\sigma'$ indicate, respectively, the final and initial states of the $f$ momentum. The particles created in the $c$ bands are labeled by the energy $\varepsilon_{\bm k \mu}$ and the hole by $\varepsilon_{\bm q \mu'}$. Each arrow, labeled by a letter, represents a virtual process that is given by the perturbation written next to it. There are two different processes given by the orders $ab$ or $ba$. The former produces a $f^0$ intermediate state of the $f$ orbital, while the latter produces a $f^2$ intermediate state with energy $2\varepsilon_f + \varepsilon_\nu$. The dashed lines indicate the energy cutoff $E_F\pm \Lambda$ used to project into the low-energy states. See the definition of $\mathcal{P}_c$ in the main text.}
    \label{fig:app:KML-diag}
\end{figure}

At second order in perturbation theory, there are two different sequences that the virtual processes can take, as depicted in Fig.~\ref{fig:app:KML-diag}. The first sequence of virtual processes happens in $(a)$ before $(b)$ (see Fig.~\ref{fig:app:KML-diag}), which involve the $f^0$ configuration of the $f$ orbital. This contribution to the Kondo interaction is
\begin{align}
    &\quad \mathcal{P}_{\rm S} \left[  \mathcal{V}  \hat G^{0}(\varepsilon_0)  \mathcal{V}   \right] \mathcal{P}_{\rm S} \nonumber \\
    &= \sum_{\bm r,\bm k, \bm q}' \frac{e^{i(\bm q - \bm k)\cdot \bm r}}{N} \frac{  \bm f^{\dag}_{\bm r} \mathbb{V}_{\bm{q}\mu'}^{} \bm c_{\bm{q}\mu'}^{}    \bm c_{\bm{k}\mu}^{\dag} \mathbb{V}_{\bm{k} \mu}^{\dag} \bm f_{\bm r}^{}   }{\varepsilon_f - \varepsilon_{\bm k \mu}} \nonumber \\
    &=-\frac{1}{N} \sum_{\bm r,\bm k, \bm q}' e^{i(\bm q - \bm k)\cdot \bm r} \frac{[\mathbb{V}_{\bm{q}\mu'}^{}]_{\sigma s'}^{}  [\mathbb{V}_{\bm{k} \mu}^{*}]_{\sigma' s}^{} }{\varepsilon_f - \varepsilon_{\bm k \mu}^{}}  \nonumber \\
    &\quad \quad \times f_{\bm r\sigma}^\dag f_{\bm r\sigma'}^{} c_{\bm{k}\mu s}^{\dag} c_{\bm{q}\mu' s'}^{},
\end{align}
where the apostrophe on the summation means it is restricted to the states within the energy cutoff, $E_F\pm \Lambda$.

The contribution to the Kondo coupling from the $f^2$ channel is given by the virtual processes in the order $(b)$ and then $(a)$,
\begin{align}
    &\quad \mathcal{P}_{\rm S}  \left[  \mathcal{V}  \hat G^{0}(\varepsilon_0)  \mathcal{V}   \right] \mathcal{P}_{\rm S} \nonumber \\
    &=  \sum_{\bm r,\bm k, \bm q}' \frac{e^{i(\bm q - \bm k)\cdot \bm r}}{N} \frac{    \bm c_{\bm{k}\mu}^{\dag} \mathbb{V}_{\bm{k} \mu}^{\dag} \bm f_{\bm r}^{}  |\eta_\nu \rangle \langle \eta_\nu |  \bm f^{\dag}_{\bm r} \mathbb{V}_{\bm{q}\mu'}^{} \bm c_{\bm{q}\mu'}^{}    }{\varepsilon_{\bm q \mu'} - \varepsilon_f - \varepsilon_\nu } \nonumber \\
    &=  \frac{1}{N} \sum_{\bm r,\bm k, \bm q}' e^{i(\bm q - \bm k)\cdot \bm r} \frac{[\mathbb{V}_{\bm{q}\mu'}^{}]_{\alpha s'}^{}  [\mathbb{V}_{\bm{k} \mu}^{*}]_{\alpha' s}^{} }{ \varepsilon_{\bm q \mu'} - \varepsilon_f - \varepsilon_\nu } \nonumber \\
    & \quad \quad \quad  \times f_{\bm r\alpha'}^{}  |\eta_\nu \rangle \langle \eta_\nu |  f_{\bm r\alpha}^{\dag} \ c_{\bm{k}\mu s}^{\dag} c_{\bm{q}\mu' s'}^{} \nonumber \\
    &=  \frac{1}{N} \sum_{\bm r,\bm k, \bm q}' e^{i(\bm q - \bm k)\cdot \bm r} \frac{[\mathbb{V}_{\bm{q}\mu'}^{}]_{\alpha s'}^{}  [\mathbb{V}_{\bm{k} \mu}^{*}]_{\alpha' s}^{} }{ \varepsilon_{\bm q \mu'} - \varepsilon_f - \varepsilon_\nu }   \nonumber \\
    & \quad \quad \quad \times \gamma_{\sigma\sigma',\alpha\alpha'}^{\nu} f_{\bm r\sigma}^\dag  f_{\bm r\sigma'}^{} \ c_{\bm{k}\mu s}^{\dag} c_{\bm{q}\mu' s'}^{},
\end{align}
where we have used
\begin{align}
    &\quad f_{\bm r\alpha'}^{}  |\eta_\nu \rangle  \langle \eta_\nu |  f_{\bm r\alpha}^{\dag} \nonumber \\
    &= \sum_{\sigma \sigma'} |\sigma\rangle\langle\sigma| f_{\bm r\alpha'}^{}  |\eta_\nu \rangle \langle \eta_\nu |  f_{\bm r\alpha}^{\dag} |\sigma'\rangle\langle\sigma'| \nonumber \\
    &=  \sum_{\sigma \sigma'} \gamma_{\sigma\sigma',\alpha\alpha'}^{\nu}   |\sigma\rangle \langle\sigma'|  = \sum_{\sigma \sigma'} \gamma_{\sigma\sigma',\alpha\alpha'}^{\nu} f_{\bm r\sigma}^\dag  f_{\bm r\sigma'}^{}  ,
\end{align}
with $\gamma_{\sigma\sigma',\alpha\alpha'}^{\nu}$ given by Eq.~\eqref{eq:gamma-def}.

By adding up the contributions from the $f^0$ and $f^2$ channels, the exchange interaction is given by
\begin{align}
    J_{\sigma \sigma', s s'}^{\bm k \mu , \bm q \mu'} &= \frac{[\mathbb{V}_{\bm{q}\mu'}^{}]_{\sigma s'}^{}  [\mathbb{V}_{\bm{k} \mu}^{*}]_{\sigma' s}^{} }{\varepsilon_f - \varepsilon_{\bm k \mu}^{}} \nonumber \\
&\quad     - \frac{[\mathbb{V}_{\bm{q}\mu'}^{}]_{\alpha s'}^{}  [\mathbb{V}_{\bm{k} \mu}^{*}]_{\alpha' s}^{} }{ \varepsilon_{\bm q \mu'} - \varepsilon_f - \varepsilon_\nu }  \gamma_{\sigma\sigma',\alpha\alpha'}^{\nu} .
\end{align}
Finally, by approximating the energy of the particles and the holes by the energy of the Fermi level, we get Eq.~\eqref{eq:Kondo-coupling}.

\section{Derivation of the spin Hamiltonian}\label{app:spin-int}
\setcounter{figure}{0}                       
\renewcommand\thefigure{B.\arabic{figure}}   
In this appendix, we give a detailed derivation of the spin Hamiltonian by implementing the degenerate perturbation theory introduced in the main text.

Figures~\ref{fig:app-pp-diag}-\ref{fig:app-hh-diag} show all the virtual processes that contribute to the spin-spin interaction. We classify these into three types based on the intermediate states of the conduction bands, that is, particle-particle, particle-hole, and hole-hole, as shown in Figs.~\ref{fig:app-pp-diag}, \ref{fig:app-ph-diag}, and \ref{fig:app-hh-diag}, respectively. We note that within each type, only a few are independent - we can obtain many others by interchanging sites $i$ and $j$ or by interchanging the particles and holes in the conduction bands. 

The low-energy spin Hamiltonian is given by the degenerate perturbation theory~\cite{LindgrenI1974},
\begin{equation}
    \mathcal{H}_{\rm spin} = \mathcal{P}_{S_0} \left[ \mathcal{V} \hat G^{0}(\varepsilon_0) \mathcal{V} \hat G^{0}(\varepsilon_0) \mathcal{V} \hat G^{0}(\varepsilon_0) \mathcal{V}   \right] \mathcal{P}_{S_0},
\end{equation}
where $\mathcal{P}_{S_0}$ projects the $f$ moments into the low-energy states of the $J=J_m$ $f$ moments. The resolvent $\hat G^{0}$ and the hybridization $\mathcal{V}$ are given, respectively, in Eq.~\eqref{eq:resolvent} and in Eq.~\eqref{eq:hyb}.

\subsection{Particle-particle channel}
\begin{figure}[h!]
    \centering
    \includegraphics[width=0.4\textwidth]{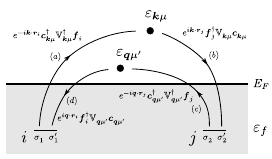}
    \caption{Diagram of virtual processes where the hopping between the $f$ orbital in sites $i$ and $j$ is mediated by the creation of a particle above the Fermi level $E_F$. The energy of the $f$ orbital is $\varepsilon_f$. The indices $\sigma_1^{}$ ($\sigma_2^{}$) and $\sigma_1'$ ($\sigma_2'$) indicate, respectively, the final and initial states of the site $i$ ($j$). The particles created in the $c$ bands are labeled by their energy, $\varepsilon_{\bm k \mu}$ and  $\varepsilon_{\bm q \mu'}$. Each arrow, labeled by a letter, represents a virtual process that is given by the perturbation written next to it. There are six different processes given by the orders $acbd$, $acdb$, $cabd$, $cadb$, $abcd$, $cdab$.}
    \label{fig:app-pp-diag}
\end{figure}

In the particle-particle channel shown in Fig.~\ref{fig:app-pp-diag}, there are six different virtual processes.

The virtual process $\mathcal{P}_{\rm pp}^{f^0(1)}$ given in the order $acbd$ in  Fig.~\ref{fig:app-pp-diag} involves $f^0$ intermediate states,
\begin{align}
    \mathcal{P}_{\rm pp}^{f^0(1)} \ (acbd) : & f_i^1 c_{k}^0 c_{q}^0 f_j^1 \rightarrow f_i^0 c_{k}^1 c_{q}^0 f_j^1 \rightarrow \nonumber \\
    & f_i^0 c_{k}^1 c_{q}^1 f_j^0 \rightarrow f_i^0 c_{k}^0 c_{q}^1 f_j^1 \rightarrow f_i^1 c_{k}^0 c_{q}^0 f_j^1
\end{align}
The contribution of this process reads as
\begin{align}
    &\mathcal{P}_{S_0} \left[ \mathcal{V} \hat G^{0}(\varepsilon_0) \mathcal{V} \hat G^{0}(\varepsilon_0) \mathcal{V} \hat G^{0}(\varepsilon_0) \mathcal{V}   \right] \mathcal{P}_{S_0} = \nonumber \\
    &= \frac{1}{N^2}  \sum_{\bm r_i, \bm r_j,\bm k, \bm q}  
     e^{i(\bm q - \bm k)\cdot (\bm r_i - \bm r_j)} \ g_{\rm pp}^{f^0(1)} \nonumber \\
    & \times \mathcal{P}_{S_0} \bm f^{\dag}_{\bm r_i} \mathbb{V}_{\bm{q}\mu'}^{} \bm c_{\bm{q}\mu'}^{} 
    \bm f^{\dag}_{\bm r_j} \mathbb{V}_{\bm{k}\mu}^{} \bm c_{\bm{k}\mu}^{}
    \bm c_{\bm{q}\mu'}^{\dag} \mathbb{V}_{\bm{q} \mu'}^{\dag} \bm f_{\bm r_j}^{}
    \bm c_{\bm{k}\mu}^{\dag} \mathbb{V}_{\bm{k} \mu}^{\dag} \bm f_{\bm r_i}^{} \mathcal{P}_{S_0} \nonumber \\
    & = - \frac{1}{N^2}  \sum_{\bm r_i, \bm r_j,\bm k, \bm q}  e^{i(\bm q - \bm k)\cdot (\bm r_i - \bm r_j)} \ g_{\rm pp}^{f^0(1)} \nonumber \\
    & \times  [\mathbb{V}_{\bm{q}\mu'}^{}]_{\sigma_1^{}s'} 
    [\mathbb{V}_{\bm{k}\mu}^{} ]_{\sigma_2^{} s} 
    [\mathbb{V}_{\bm{q} \mu'}^{*} ]_{\sigma_2' s'} 
    [\mathbb{V}_{\bm{k} \mu}^{*}]_{\sigma_1' s} \nonumber \\
    & \times  [1-f(\varepsilon_{\bm k \mu})] [1-f(\varepsilon_{\bm q \mu'})]  (  f_{\bm r_i \sigma_1^{}}^{\dag} f_{\bm r_i \sigma_1'}^{} \ f_{\bm r_j \sigma_2^{}}^{\dag} f_{\bm r_j \sigma_2'}^{} )
\end{align}
where $[1-f(\varepsilon_{\bm k \mu})]$ comes from the projection of the conduction bands, $\mathcal{P}_{S_0} c_{\bm k \mu}^{} c_{\bm k \mu}^{\dag} \mathcal{P}_{S_0}$. The function of energy is given by
\begin{equation}
    g_{\rm pp}^{f^0(1)} = \frac{ 1  }{ (\varepsilon_f - \varepsilon_{\bm q \mu'}) (2\varepsilon_f - \varepsilon_{\bm k \mu} - \varepsilon_{\bm q \mu'} ) (\varepsilon_f - \varepsilon_{\bm k \mu})} .
\end{equation}

The virtual process $\mathcal{P}_{\rm pp}^{f^0(2)}$ is given by the order $acdb$ in  Fig.~\ref{fig:app-pp-diag},
\begin{align}
    \mathcal{P}_{\rm pp}^{f^0(2)}\ (acdb): & f_i^1 c_{k}^0 c_{q}^0 f_j^1 \rightarrow f_i^0 c_{k}^1 c_{q}^0 f_j^1 \rightarrow \nonumber \\
    & f_i^0 c_{k}^1 c_{q}^1 f_j^0 \rightarrow f_i^1 c_{k}^1 c_{q}^0 f_j^0 \rightarrow f_i^1 c_{k}^0 c_{q}^0 f_j^1 
\end{align}
The contribution of this process involves $f^0$ intermediate states and only differs from $\mathcal{P}_{\rm pp}^{f^0(1)}$ in the energy,  
\begin{align}
     g_{\rm pp}^{f^0(2)} = \frac{ 1 }{ (\varepsilon_f - \varepsilon_{\bm k \mu}) (2\varepsilon_f - \varepsilon_{\bm k \mu} - \varepsilon_{\bm q \mu'} ) (\varepsilon_f - \varepsilon_{\bm k \mu})}
\end{align}
The virtual process $\mathcal{P}_{\rm pp}^{f^0(3)}$ is given by the order  $cabd$ in Fig.~\ref{fig:app-pp-diag},
\begin{align}
    \mathcal{P}_{\rm pp}^{f^0(3)} \ (cabd): & f_i^1 c_{k}^0 c_{q}^0 f_j^1 \rightarrow f_i^1 c_{k}^0 c_{q}^1 f_j^0 \rightarrow \nonumber \\
    & f_i^0 c_{k}^1 c_{q}^1 f_j^0 \rightarrow f_i^0 c_{k}^0 c_{q}^1 f_j^1 \rightarrow f_i^1 c_{k}^0 c_{q}^0 f_j^1
\end{align}
The energy part of this process is
\begin{align}
     g_{\rm pp}^{f^0(3)} = \frac{ 1 }{ (\varepsilon_f - \varepsilon_{\bm q \mu'}) (2\varepsilon_f - \varepsilon_{\bm k \mu} - \varepsilon_{\bm q \mu'} ) (\varepsilon_f - \varepsilon_{\bm q \mu'})}
\end{align}
The last process that involves $f^0$ states is $\mathcal{P}_{\rm pp}^{f^0(4)}$, which is given by the order $cadb$. That is,
\begin{align}
    \mathcal{P}_{\rm pp}^{f^0(4)} (cadb): & f_i^1 c_{k}^0 c_{q}^0 f_j^1 \rightarrow f_i^1 c_{k}^0 c_{q}^1 f_j^0 \rightarrow  \nonumber \\
    &f_i^0 c_{k}^1 c_{q}^1 f_j^0 \rightarrow f_i^1 c_{k}^1 c_{q}^0 f_j^0 \rightarrow f_i^1 c_{k}^0 c_{q}^0 f_j^1
\end{align}
And the energy part reads as
\begin{align}
     g_{\rm pp}^{f^0(4)} = \frac{ 1 }{ (\varepsilon_f - \varepsilon_{\bm k \mu}) (2\varepsilon_f - \varepsilon_{\bm k \mu} - \varepsilon_{\bm q \mu'} ) (\varepsilon_f - \varepsilon_{\bm q \mu'})} .
\end{align}

The contribution of all the processes with $f^0$ intermediate states is obtained by summing up the four $f^0$ processes, 
\begin{align}
    &\mathcal{P}_{S_0} \left[ \mathcal{V} \hat G^{0}(\varepsilon_0) \mathcal{V} \hat G^{0}(\varepsilon_0) \mathcal{V} \hat G^{0}(\varepsilon_0) \mathcal{V}   \right] \mathcal{P}_{S_0} = \nonumber \\
    & = \frac{1}{N^2}  \sum_{\bm r_i, \bm r_j,\bm k, \bm q} e^{i(\bm q - \bm k)\cdot (\bm r_i - \bm r_j)}   \frac{ \varepsilon_{\bm k \mu} + \varepsilon_{\bm q \mu'} -2\varepsilon_f  }{ (\varepsilon_{\bm k \mu} - \varepsilon_f )^2  ( \varepsilon_{\bm q \mu'} - \varepsilon_f)^2} \nonumber \\
    & \times  [\mathbb{V}_{\bm{q}\mu'}^{}]_{\sigma_1^{}s'} 
    [\mathbb{V}_{\bm{k}\mu}^{} ]_{\sigma_2^{} s} 
    [\mathbb{V}_{\bm{q} \mu'}^{*} ]_{\sigma_2' s'} 
    [\mathbb{V}_{\bm{k} \mu}^{*}]_{\sigma_1' s} \nonumber \\
    & \times  [1-f(\varepsilon_{\bm k \mu})] [1-f(\varepsilon_{\bm q \mu'})] \ f_{\bm r_i \sigma_1^{}}^{\dag}  f_{\bm r_i \sigma_1^{}}^{\dag} f_{\bm r_i \sigma_1'}^{} \ f_{\bm r_j \sigma_2^{}}^{\dag} f_{\bm r_j \sigma_2'}^{}
\end{align}
Then, the exchange interaction is
\begin{align}
    &[\mathbb K^{(f^0:\text{pp})}(i,j)]_{\sigma_1^{}\sigma_1' \sigma_2^{}\sigma_2'} = \frac{1}{N^2}  \sum_{\bm r_i, \bm r_j,\bm k, \bm q} e^{i(\bm q - \bm k)\cdot (\bm r_i - \bm r_j)}  \nonumber \\
    &\times \frac{ \varepsilon_{\bm k \mu} + \varepsilon_{\bm q \mu'} -2\varepsilon_f  }{ (\varepsilon_{\bm k \mu} - \varepsilon_f )^2  ( \varepsilon_{\bm q \mu'} - \varepsilon_f)^2}  [1-f(\varepsilon_{\bm k \mu})] [1-f(\varepsilon_{\bm q \mu'})] \nonumber \\
    & \times  [\mathbb{V}_{\bm{q}\mu'}^{}]_{\sigma_1^{}s'} 
    [\mathbb{V}_{\bm{k}\mu}^{} ]_{\sigma_2^{} s} 
    [\mathbb{V}_{\bm{q} \mu'}^{*} ]_{\sigma_2' s'} 
    [\mathbb{V}_{\bm{k} \mu}^{*}]_{\sigma_1' s}
\end{align}

The exchange interaction in momentum space is obtained through
\begin{equation}
    \mathbb{K}(\bm k) \equiv \sum_{\bm r_i - \bm r_j} e^{-i\bm k \cdot (\bm r_i - \bm r_j)} \ \mathbb{K}(i,j)
\end{equation}
and reads as,
\begin{align}\label{eq:app:f0pp}
    &[\mathbb K^{(f^0:\text{pp})}(\bm k)]_{\sigma_1^{}\sigma_1' \sigma_2^{}\sigma_2'} = \frac{1}{N}  \sum_{\bm q}   \nonumber \\
    &\times \frac{ \varepsilon_{\bm q \mu'} + \varepsilon_{\bm q +\bm k \mu} -2\varepsilon_f  }{ (\varepsilon_{\bm q \mu'} - \varepsilon_f )^2  ( \varepsilon_{\bm q+\bm k \mu} - \varepsilon_f)^2}  [1-f(\varepsilon_{\bm q \mu'})] [1-f(\varepsilon_{\bm q+\bm k \mu})] \nonumber \\
    & \times  [\mathbb{V}_{\bm{q}+\bm k\mu}^{}]_{\sigma_1^{}s} 
    [\mathbb{V}_{\bm{q} \mu'}^{*}]_{\sigma_1' s'}
    [\mathbb{V}_{\bm{q} \mu'}^{} ]_{\sigma_2^{} s'} 
    [\mathbb{V}_{\bm{q}+\bm k \mu}^{*} ]_{\sigma_2' s} 
\end{align}
(we have used the transformation $\mu \longrightarrow \mu'$ and $\mu' \longrightarrow \mu$).

There are two virtual processes that involve $f^2$ intermediate states. The first process $\mathcal{P}_{\rm pp}^{f^2(5)}$ is given by the order $abcd$ in  Fig.~\ref{fig:app-pp-diag},
\begin{align}
    \mathcal{P}_{\rm pp}^{f^2(5)}\ (abcd): & f_i^1 c_{k}^0 c_{q}^0 f_j^1 \rightarrow  f_i^0 c_{k}^1 c_{q}^0 f_j^1 \rightarrow \nonumber \\
    & f_i^0 c_{k}^0 c_{q}^0 f_j^2 \rightarrow f_i^0 c_{k}^0 c_{q}^1 f_j^1 \rightarrow f_i^1 c_{k}^0 c_{q}^0 f_j^1
\end{align}
and the contribution to the exchange interaction is
\begin{align}
    &\mathcal{P}_{S_0} \left[ \mathcal{V} \hat G^{0}(\varepsilon_0) \mathcal{V} \hat G^{0}(\varepsilon_0) \mathcal{V} \hat G^{0}(\varepsilon_0) \mathcal{V}   \right] \mathcal{P}_{S_0} = \nonumber \\
    &= \frac{1}{N^2}  \sum_{\bm r_i, \bm r_j,\bm k, \bm q} \frac{ e^{i(\bm q - \bm k)\cdot (\bm r_i - \bm r_j)} }{ (\varepsilon_f - \varepsilon_{\bm q \mu'}) (-\varepsilon_\nu ) (\varepsilon_f - \varepsilon_{\bm k \mu})}  \nonumber  \\
    & \ \ \ \times  \mathcal{P}_{S_0} \bm f^{\dag}_{\bm r_i} \mathbb{V}_{\bm{q}\mu'}^{} \bm c_{\bm{q}\mu'}^{}  
    \bm c_{\bm{q}\mu'}^{\dag} \mathbb{V}_{\bm{q} \mu'}^{\dag} \bm f_{\bm r_j}^{}
    | \eta_{j\nu} \rangle  \langle \eta_{j\nu} | \nonumber  \\
    & \ \ \  \times 
    \bm f^{\dag}_{\bm r_j} \mathbb{V}_{\bm{k}\mu}^{} \bm c_{\bm{k}\mu}^{}
    \bm c_{\bm{k}\mu}^{\dag} \mathbb{V}_{\bm{k} \mu}^{\dag} \bm f_{\bm r_i}^{} \mathcal{P}_{S_0}  \nonumber \\
    & = \frac{1}{N^2}  \sum_{\bm r_i, \bm r_j,\bm k, \bm q} e^{i(\bm q - \bm k)\cdot (\bm r_i - \bm r_j)}  \frac{ [1-f(\varepsilon_{\bm k \mu})] [1-f(\varepsilon_{\bm q \mu'})]  }{ (\varepsilon_f - \varepsilon_{\bm q \mu'}) (-\varepsilon_\nu ) (\varepsilon_f - \varepsilon_{\bm k \mu})}    \nonumber \\
    & \times   [\mathbb{V}_{\bm{q}\mu'}^{}]_{\sigma_1^{}s'} 
    [\mathbb{V}_{\bm{q} \mu'}^{*} ]_{\alpha' s'} 
    [\mathbb{V}_{\bm{k}\mu}^{} ]_{\alpha s} 
    [\mathbb{V}_{\bm{k} \mu}^{*}]_{\sigma_1' s} \  \gamma_{\sigma_2^{}\sigma_2',\alpha\alpha'}^{\nu}  \nonumber \\
    & \times  f_{\bm r_i \sigma_1^{}}^{\dag} f_{\bm r_i \sigma_1'}^{}  \ f_{\bm r_j \sigma_2^{}}^{\dag} f_{\bm r_j \sigma_2'}^{}  
\end{align}
The virtual process $\mathcal{P}_{\rm pp}^{(6)}$ is given by the order $cdab$, which is equivalent to swap sites $i$ and $j$,
\begin{align}
    \mathcal{P}_{\rm pp}^{f^2(6)} \ (cdab): & f_i^1 c_{k}^0 c_{q}^0 f_j^1 \rightarrow f_i^1 c_{k}^0 c_{q}^1 f_j^0 \rightarrow \nonumber \\
    & f_i^2 c_{k}^0 c_{q}^0 f_j^0 \rightarrow f_i^1 c_{k}^1 c_{q}^1 f_j^0 \rightarrow f_i^1 c_{k}^0 c_{q}^0 f_j^1
\end{align}
Then, the contribution to the exchange interaction is
\begin{align}
    &\mathcal{P}_{S_0} \left[ \mathcal{V} \hat G^{0}(\varepsilon_0) \mathcal{V} \hat G^{0}(\varepsilon_0) \mathcal{V} \hat G^{0}(\varepsilon_0) \mathcal{V}   \right] \mathcal{P}_{S_0} = \nonumber \\
    & = \frac{1}{N^2}  \sum_{\bm r_i, \bm r_j,\bm k, \bm q}  e^{i(\bm q - \bm k)\cdot (\bm r_i - \bm r_j)} \frac{ [1-f(\varepsilon_{\bm k \mu})] [1-f(\varepsilon_{\bm q \mu'})]  }{ (\varepsilon_f - \varepsilon_{\bm q \mu'}) (-\varepsilon_\nu ) (\varepsilon_f - \varepsilon_{\bm k \mu})} \nonumber \\
    & \times  [\mathbb{V}_{\bm{q}\mu'}^{}]_{\alpha s'} 
    [\mathbb{V}_{\bm{q} \mu'}^{*} ]_{\sigma_2' s'} 
    [\mathbb{V}_{\bm{k}\mu}^{} ]_{\sigma_2^{} s} 
    [\mathbb{V}_{\bm{k} \mu}^{*}]_{\alpha' s} \  \gamma_{\sigma_1^{}\sigma_1',\alpha\alpha'}^{\nu}  \nonumber \\
    & \times  f_{\bm r_i \sigma_1^{}}^{\dag} f_{\bm r_i \sigma_1'}^{}  \ f_{\bm r_j \sigma_2^{}}^{\dag} f_{\bm r_j \sigma_2'}^{}  
\end{align}

The total contribution of the processes with $f^2$ intermediate states is
\begin{align}
    & [\mathbb K^{(f^2:\text{pp})}(i,j)]_{\sigma_1^{}\sigma_1' \sigma_2^{}\sigma_2'}  = \nonumber \\
    & = \frac{-1}{N^2}  \sum_{\bm r_i, \bm r_j,\bm k, \bm q}  e^{i(\bm q - \bm k)\cdot (\bm r_i - \bm r_j)} \frac{ [1-f(\varepsilon_{\bm k \mu})] [1-f(\varepsilon_{\bm q \mu'})]  }{ (\varepsilon_f - \varepsilon_{\bm q \mu'}) (\varepsilon_f - \varepsilon_{\bm k \mu})  \varepsilon_\nu } \nonumber \\
    & \times \big( [\mathbb{V}_{\bm{q}\mu'}^{}]_{\alpha s'} 
    [\mathbb{V}_{\bm{q} \mu'}^{*} ]_{\sigma_2' s'} 
    [\mathbb{V}_{\bm{k}\mu}^{} ]_{\sigma_2^{} s} 
    [\mathbb{V}_{\bm{k} \mu}^{*}]_{\alpha' s} \  \gamma_{\sigma_1^{}\sigma_1',\alpha\alpha'}^{\nu} \nonumber \\
    & \ \,  + [\mathbb{V}_{\bm{q}\mu'}^{}]_{\sigma_1^{}s'} 
    [\mathbb{V}_{\bm{q} \mu'}^{*} ]_{\alpha' s'} 
    [\mathbb{V}_{\bm{k}\mu}^{} ]_{\alpha s} 
    [\mathbb{V}_{\bm{k} \mu}^{*}]_{\sigma_1' s} \  \gamma_{\sigma_2^{}\sigma_2',\alpha\alpha'}^{\nu} \big)
\end{align}

In momentum space,
\begin{align}\label{eq:app:f2pp}
    & [\mathbb K^{(f^2:\text{pp})}(\bm k)]_{\sigma_1^{}\sigma_1' \sigma_2^{}\sigma_2'} \!  =  \! \frac{-1}{N} \!  \sum_{\bm q}   \frac{ [1-f(\varepsilon_{\bm q \mu'})] [1-f(\varepsilon_{\bm k+\bm q \mu})]  }{ (\varepsilon_f - \varepsilon_{\bm k+\bm q \mu}) (\varepsilon_f - \varepsilon_{\bm q \mu'})  \varepsilon_\nu } \nonumber \\
    & \times \big( [\mathbb{V}_{\bm k+\bm{q}\mu}^{}]_{\alpha s'} 
    [\mathbb{V}_{\bm{q} \mu'}^{*}]_{\alpha' s}
    [\mathbb{V}_{\bm{q}\mu'}^{} ]_{\sigma_2^{} s} 
    [\mathbb{V}_{\bm k+\bm{q} \mu}^{*} ]_{\sigma_2' s'}  \  \gamma_{\sigma_1^{}\sigma_1',\alpha\alpha'}^{\nu} \nonumber \\
    & \ \,  + [\mathbb{V}_{\bm k+\bm{q}\mu}^{}]_{\sigma_1^{}s'} 
    [\mathbb{V}_{\bm{q} \mu'}^{*}]_{\sigma_1' s}
    [\mathbb{V}_{\bm{q}\mu'}^{} ]_{\alpha s} 
    [\mathbb{V}_{\bm k+\bm{q} \mu}^{*} ]_{\alpha' s'}  \  \gamma_{\sigma_2^{}\sigma_2',\alpha\alpha'}^{\nu} \big)
\end{align}

\subsection{Particle-hole channel}
\begin{figure}[h!]
    \centering
    \includegraphics[width=0.4\textwidth]{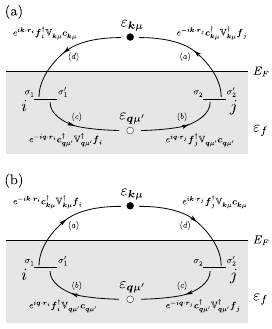}
    \caption{Diagram of virtual processes where the hopping between the $f$ orbital in sites $i$ and $j$ is mediated by the creation of a particle in one direction and a hole in the other direction. The Fermi level is indicated by $E_F$ and the energy of the $f$ orbital is $\varepsilon_f$. The indices $\sigma_1^{}$ ($\sigma_2^{}$) and $\sigma_1'$ ($\sigma_2'$) indicate, respectively, the final and initial states of the site $i$ ($j$). The particle created in the $c$ bands is labeled by its energy $\varepsilon_{\bm k \mu}$, and the hole by $\varepsilon_{\bm q \mu'}$. Each arrow, labeled by a letter, represents a virtual process that is given by the perturbation written next to it. In panel (a), there are 6 different processes given by the orders $abcd$, $abdc$, $adbc$, $badc$, $bacd$, $bcad$. The processes in panel (b) are obtained by swapping the sites $i$ with $j$. Then, there are 12 processes in total.}
    \label{fig:app-ph-diag}
\end{figure}

Here we consider those virtual processes that involve virtual states  with one excited particle and one excited  hole. 
  Note that the application of second order degenerate perturbation theory the KL model only includes particle-hole processes whose virtual state has one particle-hole excitation. The ``particle-hole channel", as defined in this subsection includes 
more fourth-order processes in the hybridization, where one of the three intermediate states has a particle excitation, while another one has a hole-excitation, but the particle and the hole do not appear simultaneously.
All the processes of this kind are given by different orders of the processes depicted in Fig.~\ref{fig:app-ph-diag}.

The virtual process $\mathcal{P}_{\rm ph}^{(1)}$ comes from the order $abcd$ in Fig.~\ref{fig:app-ph-diag}(a),
\begin{align}
    \mathcal{P}_{\rm ph}^{f^0(1)} \ (abcd) : & f_i^1 c_{k}^0 c_{q}^1 f_j^1 \rightarrow f_i^1 c_{k}^1 c_{q}^1 f_j^0 \rightarrow \nonumber \\
    & f_i^1 c_{k}^1 c_{q}^0 f_j^1 \rightarrow f_i^0 c_{k}^1 c_{q}^1 f_j^1 \rightarrow f_i^1 c_{k}^0 c_{q}^1 f_j^1 .
\end{align}

The contribution to the interaction is
\begin{align}
    &\mathcal{P}_{S_0} \left[ \mathcal{V} \hat G^{0}(\varepsilon_0) \mathcal{V} \hat G^{0}(\varepsilon_0) \mathcal{V} \hat G^{0}(\varepsilon_0) \mathcal{V}   \right] \mathcal{P}_{S_0} = \nonumber \\
    &= \frac{1}{N^2}  \sum_{\bm r_i, \bm r_j,\bm k, \bm q}  
     e^{i(\bm k - \bm q)\cdot (\bm r_i - \bm r_j)} \frac{1}{(\varepsilon_{\bm q \mu'}- \varepsilon_{\bm k \mu})(\varepsilon_f - \varepsilon_{\bm k \mu})^2 }      \nonumber \\
    & \times  \mathcal{P}_{S_0} \bm f^{\dag}_{\bm r_i} \mathbb{V}_{\bm{k}\mu}^{} \bm c_{\bm{k}\mu}^{}
    \bm c_{\bm{q}\mu'}^{\dag} \mathbb{V}_{\bm{q} \mu'}^{\dag} \bm f_{\bm r_i}^{}
    \bm f^{\dag}_{\bm r_j} \mathbb{V}_{\bm{q}\mu'}^{} \bm c_{\bm{q}\mu'}^{}
    \bm c_{\bm{k}\mu}^{\dag} \mathbb{V}_{\bm{k} \mu}^{\dag} \bm f_{\bm r_j}^{} \mathcal{P}_{S_0} \nonumber \\
    & = - \frac{1}{N^2}  \sum_{\bm r_i, \bm r_j,\bm k, \bm q}  e^{i(\bm k - \bm q)\cdot (\bm r_i - \bm r_j)}   \frac{f(\varepsilon_{\bm q \mu'}) [1-f(\varepsilon_{\bm k \mu})]}{(\varepsilon_{\bm q \mu'}- \varepsilon_{\bm k \mu})(\varepsilon_f - \varepsilon_{\bm k \mu})^2 }  \nonumber \\
    & \times [\mathbb{V}_{\bm{k}\mu}^{} ]_{\sigma_1^{} s}  
    [\mathbb{V}_{\bm{q} \mu'}^{*} ]_{\sigma_1' s'} 
    [\mathbb{V}_{\bm{q}\mu'}^{}]_{\sigma_2^{}s'} 
    [\mathbb{V}_{\bm{k} \mu}^{*}]_{\sigma_2' s}  \nonumber \\
    & \times  (  f_{\bm r_i \sigma_1^{}}^{\dag} f_{\bm r_i \sigma_1'}^{} \ f_{\bm r_j \sigma_2^{}}^{\dag} f_{\bm r_j \sigma_2'}^{} )
\end{align}

The virtual process $\mathcal{P}_{\rm ph}^{(2)}$ is given by the order $abdc$ in Fig.~\ref{fig:app-ph-diag}(a),
\begin{align}
    \mathcal{P}_{\rm ph}^{f^2(2)} (abdc):& f_i^1 c_{k}^0 c_{q}^1 f_j^1 \rightarrow f_i^1 c_{k}^1 c_{q}^1 f_j^0 \rightarrow \nonumber \\ 
    & f_i^1 c_{k}^1 c_{q}^0 f_j^1 \rightarrow f_i^2 c_{k}^0 c_{q}^0 f_j^1 \rightarrow f_i^1 c_{k}^0 c_{q}^1 f_j^1
\end{align}
The contribution to the interaction is
\begin{align}
    &\mathcal{P}_{S_0} \left[ \mathcal{V} \hat G^{0}(\varepsilon_0) \mathcal{V} \hat G^{0}(\varepsilon_0) \mathcal{V} \hat G^{0}(\varepsilon_0) \mathcal{V}   \right] \mathcal{P}_{S_0} = \nonumber \\
    &= \frac{1}{N^2}  \sum_{\bm r_i, \bm r_j,\bm k, \bm q}  
     e^{i(\bm k - \bm q)\cdot (\bm r_i - \bm r_j)} \ g_{\rm ph}^{f^2(2)} \  \mathcal{P}_{S_0} 
    \bm c_{\bm{q}\mu'}^{\dag} \mathbb{V}_{\bm{q} \mu'}^{\dag} \bm f_{\bm r_i}^{}
    |\eta_{i\nu} \rangle   \nonumber \\
    & \times   \langle \eta_{i\nu} |
    \bm f^{\dag}_{\bm r_i} \mathbb{V}_{\bm{k}\mu}^{} \bm c_{\bm{k}\mu}^{}
    \bm f^{\dag}_{\bm r_j} \mathbb{V}_{\bm{q}\mu'}^{} \bm c_{\bm{q}\mu'}^{}
    \bm c_{\bm{k}\mu}^{\dag} \mathbb{V}_{\bm{k} \mu}^{\dag} \bm f_{\bm r_j}^{} \mathcal{P}_{S_0} \nonumber \\
    & = -\frac{1}{N^2}  \sum_{\bm r_i, \bm r_j,\bm k, \bm q}  e^{i(\bm k - \bm q)\cdot (\bm r_i - \bm r_j)} \ g_{\rm ph}^{f^2(2)} f(\varepsilon_{\bm q \mu'}) [1-f(\varepsilon_{\bm k \mu})]   \nonumber \\
    & \times [\mathbb{V}_{\bm{q} \mu'}^{*} ]_{\alpha' s'} 
    [\mathbb{V}_{\bm{k}\mu}^{} ]_{\alpha^{} s}    
    [\mathbb{V}_{\bm{q}\mu'}^{}]_{\sigma_2^{}s'} 
    [\mathbb{V}_{\bm{k} \mu}^{*}]_{\sigma_2' s} \ \gamma_{\sigma_1^{}\sigma_1',\alpha\alpha'}^\nu \nonumber \\
    & \times  (  f_{\bm r_i \sigma_1^{}}^{\dag} f_{\bm r_i \sigma_1'}^{} \ f_{\bm r_j \sigma_2^{}}^{\dag} f_{\bm r_j \sigma_2'}^{} )
\end{align}
with the function of energy
\begin{equation}
    g_{\rm ph}^{f^2(2)} = \frac{1}{(\varepsilon_f - \varepsilon_{\bm k \mu}) (\varepsilon_{\bm q \mu'}- \varepsilon_{\bm k \mu}) (\varepsilon_{\bm q \mu'} - \varepsilon_f - \varepsilon_{\nu}) }
\end{equation}

The virtual process $\mathcal{P}_{\rm ph}^{(3)}$ corresponds to the order $adbc$ in Fig.~\ref{fig:app-ph-diag}(a),
\begin{align}
    \mathcal{P}_{\rm ph}^{f^2(3)} (adbc): & f_i^1 c_{k}^0 c_{q}^1 f_j^1 \rightarrow f_i^1 c_{k}^1 c_{q}^1 f_j^0 \rightarrow \nonumber \\
    & f_i^2 c_{k}^0 c_{q}^1 f_j^0 \rightarrow f_i^2 c_{k}^0 c_{q}^0 f_j^1 \rightarrow f_i^1 c_{k}^0 c_{q}^1 f_j^1
\end{align}
The contribution to the exchange interaction differs from $\mathcal{P}_{\rm ph}^{(2)}$ in the function of energy, which is given by
\begin{equation}
    g_{\rm ph}^{f^2(3)} = \frac{1}{(\varepsilon_{\bm q \mu'}-\varepsilon_f - \varepsilon_\nu ) (- \varepsilon_{\nu}) (\varepsilon_f - \varepsilon_{\bm k \mu}) } .
\end{equation}
This process does not create a particle-hole pair, and as a result it cannot be obtained by the RKKY interaction. Note that the denominator in $g_{\rm ph}^{f^2(3)}$ does not have the energy difference $(\varepsilon_{\bm q \mu'} - \varepsilon_{\bm k \mu})$.

The virtual process $\mathcal{P}_{\rm ph}^{(4)}$ is given by the order $badc$ in Fig.~\ref{fig:app-ph-diag}(a),
\begin{align}
    \mathcal{P}_{\rm ph}^{f^2(4)} (badc) :& f_i^1 c_{k}^0 c_{q}^1 f_j^1 \rightarrow f_i^1 c_{k}^0 c_{q}^0 f_j^2 \rightarrow \nonumber \\
    & f_i^1 c_{k}^1 c_{q}^0 f_j^1 \rightarrow f_i^2 c_{k}^0 c_{q}^0 f_j^1 \rightarrow f_i^1 c_{k}^0 c_{q}^1 f_j^1 
\end{align}
Observe that $\mathcal{P}_{\rm ph}^{(1)}$ and $\mathcal{P}_{\rm ph}^{(4)}$ are related by $ph$ symmetry.

The contribution to the interaction is
\begin{align}
    &\mathcal{P}_{S_0} \left[ \mathcal{V} \hat G^{0}(\varepsilon_0) \mathcal{V} \hat G^{0}(\varepsilon_0) \mathcal{V} \hat G^{0}(\varepsilon_0) \mathcal{V}   \right] \mathcal{P}_{S_0} = \nonumber \\
    &= \frac{1}{N^2}  \sum_{\bm r_i, \bm r_j,\bm k, \bm q}  
     e^{i(\bm k - \bm q)\cdot (\bm r_i - \bm r_j)} \ g_{\rm ph}^{f^2(4)} \  \mathcal{P}_{S_0} 
    \bm c_{\bm{q}\mu'}^{\dag} \mathbb{V}_{\bm{q} \mu'}^{\dag} \bm f_{\bm r_i}^{}
    |\eta_{i \nu_1} \rangle   \nonumber \\
    & \times    \langle \eta_{i \nu_1} |
    \bm f^{\dag}_{\bm r_i} \mathbb{V}_{\bm{k}\mu}^{} \bm c_{\bm{k}\mu}^{}
    \bm c_{\bm{k}\mu}^{\dag} \mathbb{V}_{\bm{k} \mu}^{\dag} \bm f_{\bm r_j}^{}
    |\eta_{j\nu_2} \rangle \langle \eta_{j\nu_2} |
    \bm f^{\dag}_{\bm r_j} \mathbb{V}_{\bm{q}\mu'}^{} \bm c_{\bm{q}\mu'}^{} \mathcal{P}_{S_0} \nonumber \\
    & = \frac{1}{N^2}  \sum_{\bm r_i, \bm r_j,\bm k, \bm q}  e^{i(\bm k - \bm q)\cdot (\bm r_i - \bm r_j)} \ g_{\rm ph}^{f^2(4)} f(\varepsilon_{\bm q \mu'}) [1-f(\varepsilon_{\bm k \mu})]   \nonumber \\
    & \times [\mathbb{V}_{\bm{q} \mu'}^{*} ]_{\alpha' s'} 
    [\mathbb{V}_{\bm{k}\mu}^{} ]_{\alpha^{} s} 
    [\mathbb{V}_{\bm{k} \mu}^{*}]_{\beta' s}
    [\mathbb{V}_{\bm{q}\mu'}^{}]_{\beta s'}  \ 
    \gamma_{\sigma_1^{}\sigma_1',\alpha\alpha'}^{\nu_1} \ \gamma_{\sigma_2^{}\sigma_2',\beta\beta'}^{\nu_2} \nonumber \\
    & \times  (  f_{\bm r_i \sigma_1^{}}^{\dag} f_{\bm r_i \sigma_1'}^{} \ f_{\bm r_j \sigma_2^{}}^{\dag} f_{\bm r_j \sigma_2'}^{} )
\end{align}
where the function of energy is
\begin{equation}
    g_{\rm ph}^{f^2(4)} = \frac{1}{(\varepsilon_{\bm q \mu'}-\varepsilon_f - \varepsilon_{\nu_1} ) (\varepsilon_{\bm q \mu'}-\varepsilon_f - \varepsilon_{\nu_2} ) (\varepsilon_{\bm q \mu'} - \varepsilon_{\bm k \mu}) }
\end{equation}

The virtual process $\mathcal{P}_{\rm ph}^{(5)}$ is obtained by the order $bacd$ in Fig.~\ref{fig:app-ph-diag}(a),
\begin{align}
    \mathcal{P}_{\rm ph}^{f^2(5)} (bacd):& f_i^1 c_{k}^0 c_{q}^1 f_j^1 \rightarrow f_i^1 c_{k}^0 c_{q}^0 f_j^2 \rightarrow \nonumber \\
    & f_i^1 c_{k}^1 c_{q}^0 f_j^1 \rightarrow f_i^0 c_{k}^1 c_{q}^1 f_j^1 \rightarrow f_i^1 c_{k}^0 c_{q}^1 f_j^1  
\end{align}

Note that $\mathcal{P}_{\rm ph}^{(2)}$ and $\mathcal{P}_{\rm ph}^{(5)}$ are related by $ph$ symmetry.

The contribution to the interaction is
\begin{align}
    &\mathcal{P}_{S_0} \left[ \mathcal{V} \hat G^{0}(\varepsilon_0) \mathcal{V} \hat G^{0}(\varepsilon_0) \mathcal{V} \hat G^{0}(\varepsilon_0) \mathcal{V}   \right] \mathcal{P}_{S_0} = \nonumber \\
    &= \frac{1}{N^2}  \sum_{\bm r_i, \bm r_j,\bm k, \bm q}  
     e^{i(\bm k - \bm q)\cdot (\bm r_i - \bm r_j)} \ g_{\rm ph}^{f^2(5)} \  \mathcal{P}_{S_0} 
     \bm f^{\dag}_{\bm r_i} \mathbb{V}_{\bm{k}\mu}^{} \bm c_{\bm{k}\mu}^{}  \nonumber \\
    & \times  
    \bm c_{\bm{q}\mu'}^{\dag} \mathbb{V}_{\bm{q} \mu'}^{\dag} \bm f_{\bm r_i}^{}
    \bm c_{\bm{k}\mu}^{\dag} \mathbb{V}_{\bm{k} \mu}^{\dag} \bm f_{\bm r_j}^{}
    |\eta_{j\nu} \rangle \langle \eta_{j\nu} |
    \bm f^{\dag}_{\bm r_j} \mathbb{V}_{\bm{q}\mu'}^{} \bm c_{\bm{q}\mu'}^{} \mathcal{P}_{S_0} \nonumber \\
    & = -\frac{1}{N^2}  \sum_{\bm r_i, \bm r_j,\bm k, \bm q}  e^{i(\bm k - \bm q)\cdot (\bm r_i - \bm r_j)} \ g_{\rm ph}^{f^2(5)} f(\varepsilon_{\bm q \mu'}) [1-f(\varepsilon_{\bm k \mu})]   \nonumber \\
    & \times [\mathbb{V}_{\bm{k}\mu}^{} ]_{\sigma_1^{} s}
    [\mathbb{V}_{\bm{q} \mu'}^{*} ]_{\sigma_1' s'}  
    [\mathbb{V}_{\bm{k} \mu}^{*}]_{\alpha' s}
    [\mathbb{V}_{\bm{q}\mu'}^{}]_{\alpha s'}  \ 
    \gamma_{\sigma_2^{}\sigma_2',\alpha\alpha'}^{\nu}  \nonumber \\
    & \times  (  f_{\bm r_i \sigma_1^{}}^{\dag} f_{\bm r_i \sigma_1'}^{} \ f_{\bm r_j \sigma_2^{}}^{\dag} f_{\bm r_j \sigma_2'}^{} )
\end{align}
where the function of energy is
\begin{equation}
    g_{\rm ph}^{f^2(5)} = g_{\rm ph}^{f^2(2)} = \frac{1}{(\varepsilon_f - \varepsilon_{\bm k \mu} ) (\varepsilon_{\bm q \mu'}-\varepsilon_f - \varepsilon_{\nu} ) (\varepsilon_{\bm q \mu'} - \varepsilon_{\bm k \mu}) }
\end{equation}

The virtual process $\mathcal{P}_{\rm ph}^{(6)}$ is obtained by the order $bcad$ in Fig.~\ref{fig:app-ph-diag}(a),
\begin{align}
    \mathcal{P}_{\rm ph}^{f^2(6)} (bcad) & f_i^1 c_{k}^0 c_{q}^1 f_j^1 \rightarrow f_i^1 c_{k}^0 c_{q}^0 f_j^2 \rightarrow \nonumber \\
    & f_i^0 c_{k}^0 c_{q}^1 f_j^2 \rightarrow f_i^0 c_{k}^1 c_{q}^1 f_j^1 \rightarrow f_i^1 c_{k}^0 c_{q}^1 f_j^1   
\end{align}
($\mathcal{P}_{\rm ph}^{(3)}$ and $\mathcal{P}_{\rm ph}^{(6)}$ are related by $ph$ symmetry)

The contribution to the exchange interaction differs from $\mathcal{P}_{\rm ph}^{(5)}$ only in the function of energy, which is given by
\begin{equation}
    g_{\rm ph}^{f^2(6)} = g_{\rm ph}^{f^2(3)} = \frac{1}{(\varepsilon_{\bm q \mu'}-\varepsilon_f - \varepsilon_\nu ) (- \varepsilon_{\nu}) (\varepsilon_f - \varepsilon_{\bm k \mu}) } .
\end{equation}

By swapping the sites $i$ and $j$ we obtain the processes shown in Fig.~\ref{fig:app-ph-diag}(b). Then, we can get the contribution to the exchange interaction by exchanging $\bm r_i$ with $\bm r_j$, $\sigma_1$ with $\sigma_2$, and $\sigma_1'$ with $\sigma_2'$.

The total contribution of the processes with $f^0$ intermediate states is
\begin{align}
    & [\mathbb K^{(f^0:\text{ph})}(i,j)]_{\sigma_1^{}\sigma_1' \sigma_2^{}\sigma_2'}  = \nonumber \\
    & = - \frac{1}{N^2}  \sum_{\bm r_i, \bm r_j,\bm k, \bm q}   \frac{f(\varepsilon_{\bm q \mu'}) [1-f(\varepsilon_{\bm k \mu})]}{(\varepsilon_{\bm q \mu'}- \varepsilon_{\bm k \mu})(\varepsilon_f - \varepsilon_{\bm k \mu})^2 }  \nonumber \\
    & \times \big( [\mathbb{V}_{\bm{k}\mu}^{} ]_{\sigma_1^{} s}  
    [\mathbb{V}_{\bm{q} \mu'}^{*} ]_{\sigma_1' s'} 
    [\mathbb{V}_{\bm{q}\mu'}^{}]_{\sigma_2^{}s'} 
    [\mathbb{V}_{\bm{k} \mu}^{*}]_{\sigma_2' s} \  e^{i(\bm k - \bm q)\cdot (\bm r_i - \bm r_j)} \nonumber \\
    & \ \, + [\mathbb{V}_{\bm{k}\mu}^{} ]_{\sigma_2^{} s}  
    [\mathbb{V}_{\bm{q} \mu'}^{*} ]_{\sigma_2' s'} 
    [\mathbb{V}_{\bm{q}\mu'}^{}]_{\sigma_1^{}s'} 
    [\mathbb{V}_{\bm{k} \mu}^{*}]_{\sigma_1' s} \  e^{i(\bm q - \bm k)\cdot (\bm r_i - \bm r_j)} \big)
\end{align}

In momentum space,
\begin{align}\label{eq:app:f0ph}
    & [\mathbb K^{(f^0:\text{ph})}(\bm k)]_{\sigma_1^{}\sigma_1' \sigma_2^{}\sigma_2'}   = - \frac{1}{N}  \sum_{\bm q} g_{\varepsilon_{\bm k+\bm{q}\mu}^{},\varepsilon_{\bm{q}\mu'}}^{(f^0:\text{ph})}  \nonumber \\
    & \times [\mathbb{V}_{\bm k+\bm{q}\mu}^{}]_{\sigma_1^{} s} 
    [\mathbb{V}_{\bm{q} \mu'}^{*}]_{\sigma_1' s'}
    [\mathbb{V}_{\bm{q}\mu'}^{} ]_{\sigma_2^{} s'} 
    [\mathbb{V}_{\bm k+\bm{q} \mu}^{*} ]_{\sigma_2' s}   
\end{align}
where the function $g_{\varepsilon_{\bm k+\bm{q}\mu}^{},\varepsilon_{\bm{q}\mu'}}^{(f^0:\text{ph})}$ is given in Eq.~\eqref{eq:f0-ph}.

The total contribution of the processes with $f^2$ intermediate states can be separated as
\begin{equation}
    \mathbb K^{(f^2:\text{ph})}(i,j) = \mathbb K^{(f^{2:1}:\text{ph})}(i,j) + \mathbb K^{(f^{2:2}:\text{ph})}(i,j)
\end{equation}
where $\mathbb K^{(f^{2:1}:\text{ph})}$ contains the contribution from processes with $f^2$ intermediate in only one site, and reads as 
\begin{align}
    & [\mathbb K^{(f^{2:1}:\text{ph})}(i,j)]_{\sigma_1^{}\sigma_1' \sigma_2^{}\sigma_2'}  = \nonumber \\
    & =  -\frac{1}{N^2}  \sum_{\bm r_i, \bm r_j,\bm k, \bm q}    (g_{\rm ph}^{f^2(2)}+g_{\rm ph}^{f^2(3)}) f(\varepsilon_{\bm q \mu'}) [1-f(\varepsilon_{\bm k \mu})]   \nonumber \\
    &  \times  \Big[  \big( [\mathbb{V}_{\bm{q} \mu'}^{*} ]_{\alpha' s'} 
    [\mathbb{V}_{\bm{k}\mu}^{} ]_{\alpha^{} s}    
    [\mathbb{V}_{\bm{q}\mu'}^{}]_{\sigma_2^{}s'} 
    [\mathbb{V}_{\bm{k} \mu}^{*}]_{\sigma_2' s} e^{i(\bm k - \bm q)\cdot \bm r_{ij} } \nonumber \\
    & + [\mathbb{V}_{\bm{k}\mu}^{} ]_{\sigma_2^{} s}
    [\mathbb{V}_{\bm{q} \mu'}^{*} ]_{\sigma_2' s'}  
    [\mathbb{V}_{\bm{k} \mu}^{*}]_{\alpha' s}
    [\mathbb{V}_{\bm{q}\mu'}^{}]_{\alpha s'} e^{i(\bm q - \bm k)\cdot \bm r_{ij} } \big) \gamma_{\sigma_1^{}\sigma_1',\alpha\alpha'}^\nu \nonumber \\
    &  + \big( [\mathbb{V}_{\bm{q} \mu'}^{*} ]_{\alpha' s'} 
    [\mathbb{V}_{\bm{k}\mu}^{} ]_{\alpha^{} s}    
    [\mathbb{V}_{\bm{q}\mu'}^{}]_{\sigma_1^{}s'} 
    [\mathbb{V}_{\bm{k} \mu}^{*}]_{\sigma_1' s} e^{i(\bm q - \bm k)\cdot \bm r_{ij} } \nonumber \\
    & + [\mathbb{V}_{\bm{k}\mu}^{} ]_{\sigma_1^{} s}
    [\mathbb{V}_{\bm{q} \mu'}^{*} ]_{\sigma_1' s'}  
    [\mathbb{V}_{\bm{k} \mu}^{*}]_{\alpha' s}
    [\mathbb{V}_{\bm{q}\mu'}^{}]_{\alpha s'} e^{i(\bm k - \bm q)\cdot \bm r_{ij} } \big) \gamma_{\sigma_2^{}\sigma_2',\alpha\alpha'}^\nu \Big]
\end{align}
where $\bm r_{ij}=\bm r_i - \bm r_j$.

In momentum space,
\begin{align}\label{eq:app:f21ph}
    & [\mathbb K^{(f^{2:1}:\text{ph})}(\bm k)]_{\sigma_1^{}\sigma_1' \sigma_2^{}\sigma_2'}   =  \frac{1}{N}  \sum_{\bm q} g_{\varepsilon_{\bm k+\bm{q}\mu}^{},\varepsilon_{\bm{q}\mu'},\varepsilon_\nu}^{(f^{2:1}:\text{ph})}  \nonumber \\
    & \times \big( [\mathbb{V}_{\bm k+\bm{q}\mu}^{}]_{\alpha s} 
    [\mathbb{V}_{\bm{q} \mu'}^{*}]_{\alpha' s'}
    [\mathbb{V}_{\bm{q}\mu'}^{} ]_{\sigma_2^{} s'} 
    [\mathbb{V}_{\bm k+\bm{q} \mu}^{*} ]_{\sigma_2' s}   \gamma_{\sigma_1^{}\sigma_1',\alpha\alpha'}^\nu  \nonumber \\
    & \ \, +  [\mathbb{V}_{\bm k+\bm{q}\mu}^{}]_{\sigma_1^{} s} 
    [\mathbb{V}_{\bm{q} \mu'}^{*}]_{\sigma_1' s'}
    [\mathbb{V}_{\bm{q}\mu'}^{} ]_{\alpha^{} s'} 
    [\mathbb{V}_{\bm k+\bm{q} \mu}^{*} ]_{\alpha' s}   \gamma_{\sigma_2^{}\sigma_2',\alpha\alpha'}^\nu \big)
\end{align}
where the function $g_{\varepsilon_{\bm k+\bm{q}\mu}^{},\varepsilon_{\bm{q}\mu'},\varepsilon_\nu}^{(f^{2:1}:\text{ph})}$ is given in Eq.~\eqref{eq:f2:1-ph}.

The processes with $f^2$ intermediate states in both sites contribute to $\mathbb K^{(f^{2:2}:\text{ph})}$, and read as
\begin{align}
    & [\mathbb K^{(f^{2:2:}ph)}(i,j)]_{\sigma_1^{}\sigma_1' \sigma_2^{}\sigma_2'}  = \nonumber \\
    & =  \frac{1}{N^2} \! \!   \sum_{\bm r_i, \bm r_j,\bm k, \bm q} \! \!   g_{\rm ph}^{f^2(4)} f(\varepsilon_{\bm q \mu'}) [1-f(\varepsilon_{\bm k \mu})]  \gamma_{\sigma_1^{}\sigma_1',\alpha\alpha'}^{\nu_1} \gamma_{\sigma_2^{}\sigma_2',\beta\beta'}^{\nu_2} \nonumber \\
    &  \times  \big(  [\mathbb{V}_{\bm{q} \mu'}^{*} ]_{\alpha' s'} 
    [\mathbb{V}_{\bm{k}\mu}^{} ]_{\alpha^{} s} 
    [\mathbb{V}_{\bm{k} \mu}^{*}]_{\beta' s}
    [\mathbb{V}_{\bm{q}\mu'}^{}]_{\beta s'} e^{i(\bm k - \bm q)\cdot \bm r_{ij} } \nonumber \\
    & \ \, +  [\mathbb{V}_{\bm{q} \mu'}^{*} ]_{\beta' s'} 
    [\mathbb{V}_{\bm{k}\mu}^{} ]_{\beta^{} s} 
    [\mathbb{V}_{\bm{k} \mu}^{*}]_{\alpha' s}
    [\mathbb{V}_{\bm{q}\mu'}^{}]_{\alpha s'} e^{i(\bm q - \bm k)\cdot \bm r_{ij} } \big) 
\end{align}

In momentum space,
\begin{align}\label{eq:app:f22ph}
    & [\mathbb K^{(f^{2:2}:\text{ph})}(\bm k)]_{\sigma_1^{}\sigma_1' \sigma_2^{}\sigma_2'}   = \nonumber \\
    & \ \   = \frac{1}{N}  \sum_{\bm q} g_{\varepsilon_{\bm k+\bm{q}\mu}^{},\varepsilon_{\bm{q}\mu'},\varepsilon_{\nu_1},\varepsilon_{\nu_2}}^{(f^{2:2}:\text{ph})}   \gamma_{\sigma_1^{}\sigma_1',\alpha\alpha'}^{\nu_1}   \gamma_{\sigma_2^{}\sigma_2',\beta\beta'}^{\nu_2}   \nonumber \\
    & \ \ \ \ \ \times  [\mathbb{V}_{\bm k+\bm{q}\mu}^{}]_{\alpha s} 
    [\mathbb{V}_{\bm{q} \mu'}^{*}]_{\alpha' s'}
    [\mathbb{V}_{\bm{q}\mu'}^{} ]_{\sigma_2^{} s'} 
    [\mathbb{V}_{\bm k+\bm{q} \mu}^{*} ]_{\sigma_2' s} 
\end{align}
where the function $g_{\varepsilon_{\bm k+\bm{q}\mu}^{},\varepsilon_{\bm{q}\mu'},\varepsilon_{\nu_1},\varepsilon_{\nu_2}}^{(f^{2:2}:\text{ph})}$ is given in Eq.~\eqref{eq:f2:2-ph}.


\subsection{Hole-hole channel}
\begin{figure}[h!]
    \centering
    \includegraphics[width=0.4\textwidth]{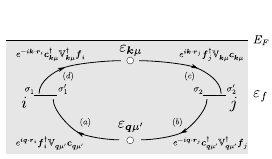}
    \caption{Diagram of virtual processes where the hopping between the $f$ orbital in sites $i$ and $j$ is mediated by the creation of a hole. The Fermi level is indicated by $E_F$ and the energy of the $f$ orbital by $\varepsilon_f$. The indices $\sigma_1^{}$ ($\sigma_2^{}$) and $\sigma_1'$ ($\sigma_2'$) indicate, respectively, the final and initial states of the site $i$ ($j$). The hole created in the $c$ bands are labeled by their energy, $\varepsilon_{\bm k \mu}$ and  $\varepsilon_{\bm q \mu'}$. Each arrow, labeled by a letter, represents a virtual process that is given by the perturbation written next to it. There are six different processes given by the orders $acbd$, $acdb$, $cabd$, $cadb$, $abcd$, $cdab$.}
    \label{fig:app-hh-diag}
\end{figure}

Here we consider the virtual processes that involve the creation of holes. All the processes of this kind are given by different orders of the processes depicted in Fig.~\ref{fig:app-hh-diag}. 

The virtual process $\mathcal{P}_{\rm hh}^{(1)}$ is given in the order $acbd$ of Fig.~\ref{fig:app-hh-diag},
\begin{align}
    \mathcal{P}_{\rm hh}^{f^2(1)}\ (acbd):& f_i^1 c_{k}^1 c_{q}^1 f_j^1 \rightarrow f_i^2 c_{k}^1 c_{q}^0 f_j^1 \rightarrow \nonumber \\
    & f_i^2 c_{k}^0 c_{q}^0 f_j^2 \rightarrow f_i^2 c_{k}^0 c_{q}^1 f_j^1 \rightarrow f_i^1 c_{k}^1 c_{q}^1 f_j^1
\end{align}
The contribution to the interaction is
\begin{align}
    &\mathcal{P}_{S_0} \left[ \mathcal{V} \hat G^{0}(\varepsilon_0) \mathcal{V} \hat G^{0}(\varepsilon_0) \mathcal{V} \hat G^{0}(\varepsilon_0) \mathcal{V}   \right] \mathcal{P}_{S_0} = \nonumber \\
    &= \frac{1}{N^2}  \sum_{\bm r_i, \bm r_j,\bm k, \bm q}  
     e^{i(\bm q - \bm k)\cdot (\bm r_i - \bm r_j)} \ g_{\rm hh}^{f^2(1)}     \nonumber \\
    & \times   \mathcal{P}_{S_0} 
    \bm c_{\bm{k}\mu}^{\dag} \mathbb{V}_{\bm{k} \mu}^{\dag} \bm f_{\bm r_i}^{}
    \bm c_{\bm{q}\mu'}^{\dag} \mathbb{V}_{\bm{q} \mu'}^{\dag} \bm f_{\bm r_j}^{}
    |\eta_{j\nu_2} \rangle \langle \eta_{j\nu_2} |
    \bm f^{\dag}_{\bm r_j} \mathbb{V}_{\bm{k}\mu}^{} \bm c_{\bm{k}\mu}^{}  \nonumber \\
    & \times 
    |\eta_{i\nu_1} \rangle \langle \eta_{i\nu_1} |
    \bm f^{\dag}_{\bm r_i} \mathbb{V}_{\bm{q}\mu'}^{} \bm c_{\bm{q}\mu'}^{} \mathcal{P}_{S_0} \nonumber \\
    & = -\frac{1}{N^2}  \sum_{\bm r_i, \bm r_j,\bm k, \bm q}  e^{i(\bm q - \bm k)\cdot (\bm r_i - \bm r_j)} \ g_{\rm hh}^{f^2(1)} f(\varepsilon_{\bm q \mu'}) f(\varepsilon_{\bm k \mu})   \nonumber \\
    & \times [\mathbb{V}_{\bm{k} \mu}^{*}]_{\alpha' s}
    [\mathbb{V}_{\bm{q} \mu'}^{*} ]_{\beta' s'}  
    [\mathbb{V}_{\bm{k}\mu}^{} ]_{\beta s}
    [\mathbb{V}_{\bm{q}\mu'}^{}]_{\alpha s'}  \ 
    \gamma_{\sigma_1^{}\sigma_1',\alpha\alpha'}^{\nu_1} \gamma_{\sigma_2^{}\sigma_2',\beta\beta'}^{\nu_2} \nonumber \\
    & \times  (  f_{\bm r_i \sigma_1^{}}^{\dag} f_{\bm r_i \sigma_1'}^{} \ f_{\bm r_j \sigma_2^{}}^{\dag} f_{\bm r_j \sigma_2'}^{} )
\end{align}
where the function of energy is
\begin{align}
    g_{\rm hh}^{f^2(1)} &= \frac{1}{(\varepsilon_{\bm q \mu'} - \varepsilon_f - \varepsilon_{\nu_1}) (\varepsilon_{\bm k \mu} - \varepsilon_f - \varepsilon_{\nu_1}) } \nonumber \\
    & \ \times \frac{1}{(\varepsilon_{\bm k \mu} + \varepsilon_{\bm q \mu'} - 2\varepsilon_f - \varepsilon_{\nu_1} - \varepsilon_{\nu_2}) } .
\end{align}

The virtual process $\mathcal{P}_{\rm hh}^{(2)}$ is given in the order $acdb$ of Fig.~\ref{fig:app-hh-diag},
\begin{align}
    \mathcal{P}_{\rm hh}^{f^2(2)} \ (acdb) : & f_i^1 c_{k}^1 c_{q}^1 f_j^1 \rightarrow  f_i^2 c_{k}^1 c_{q}^0 f_j^1 \rightarrow \nonumber \\
    & f_i^2 c_{k}^0 c_{q}^0 f_j^2 \rightarrow f_i^1 c_{k}^1 c_{q}^0 f_j^2 \rightarrow f_i^1 c_{k}^1 c_{q}^1 f_j^1
\end{align}
The contribution to the exchange interaction differs from $\mathcal{P}_{\rm hh}^{(1)}$ only in the function of energy, which is given by
\begin{align}
    g_{\rm hh}^{f^2(2)} &= \frac{1}{(\varepsilon_{\bm q \mu'} - \varepsilon_f - \varepsilon_{\nu_1}) (\varepsilon_{\bm q \mu'} - \varepsilon_f - \varepsilon_{\nu_2}) } \nonumber \\
    & \ \times \frac{1}{(\varepsilon_{\bm k \mu} + \varepsilon_{\bm q \mu'} - 2\varepsilon_f - \varepsilon_{\nu_1} - \varepsilon_{\nu_2}) } .
\end{align}

By swapping sites $i$ with $j$ in $\mathcal{P}_{\rm hh}^{(2)}$, we obtain the virtual process $\mathcal{P}_{\rm hh}^{(3)}$, which is given by the order $cabd$ in Fig.~\ref{fig:app-hh-diag},
\begin{align}
    \mathcal{P}_{\rm hh}^{f^2(3)}\ (cabd): & f_i^1 c_{k}^1 c_{q}^1 f_j^1 \rightarrow f_i^1 c_{k}^0 c_{q}^1 f_j^2 \rightarrow \nonumber \\
    & f_i^2 c_{k}^0 c_{q}^0 f_j^2 \rightarrow f_i^2 c_{k}^0 c_{q}^1 f_j^1 \rightarrow f_i^1 c_{k}^1 c_{q}^1 f_j^1 
\end{align}
The contribution is the same as for $\mathcal{P}_{\rm hh}^{(1)}$ but with the function of energy given by
\begin{align}
    g_{\rm hh}^{f^2(3)} &= \frac{1}{(\varepsilon_{\bm k \mu} - \varepsilon_f - \varepsilon_{\nu_1}) (\varepsilon_{\bm k \mu} - \varepsilon_f - \varepsilon_{\nu_2}) } \nonumber \\
    & \ \times \frac{1}{(\varepsilon_{\bm k \mu} + \varepsilon_{\bm q \mu'} - 2\varepsilon_f - \varepsilon_{\nu_1} - \varepsilon_{\nu_2}) } .
\end{align}

By swapping sites $i$ with $j$ in $\mathcal{P}_{\rm hh}^{(1)}$, we obtain the virtual process $\mathcal{P}_{\rm hh}^{(4)}$, which is given by the order $cadb$ in Fig.~\ref{fig:app-hh-diag},
\begin{align}
    \mathcal{P}_{\rm hh}^{f^2(4)}\ (cadb): & f_i^1 c_{k}^1 c_{q}^1 f_j^1 \rightarrow  f_i^1 c_{k}^0 c_{q}^1 f_j^2 \rightarrow \nonumber \\
    & f_i^2 c_{k}^0 c_{q}^0 f_j^2 \rightarrow f_i^1 c_{k}^1 c_{q}^0 f_j^2 \rightarrow f_i^1 c_{k}^1 c_{q}^1 f_j^1
\end{align}
The contribution is the same as for $\mathcal{P}_{\rm hh}^{(1)}$ with the function of energy given by
\begin{align}
    g_{\rm hh}^{f^2(4)} &= \frac{1}{(\varepsilon_{\bm q \mu'} - \varepsilon_f - \varepsilon_{\nu_2}) (\varepsilon_{\bm k \mu} - \varepsilon_f - \varepsilon_{\nu_2}) } \nonumber \\
    & \ \times \frac{1}{(\varepsilon_{\bm k \mu} + \varepsilon_{\bm q \mu'} - 2\varepsilon_f - \varepsilon_{\nu_1} - \varepsilon_{\nu_2}) } .
\end{align}

The following processes contain $f^2$ intermediate states for only one site.
The virtual process $\mathcal{P}_{\rm hh}^{(5)}$, is given by the order $abcd$ in Fig.~\ref{fig:app-hh-diag},
\begin{align}
    \mathcal{P}_{\rm hh}^{f^2(5)}\ (abcd): & f_i^1 c_{k}^1 c_{q}^1 f_j^1 \rightarrow  f_i^2 c_{k}^1 c_{q}^0 f_j^1 \rightarrow \nonumber \\
    & f_i^2 c_{k}^1 c_{q}^1 f_j^0 \rightarrow f_i^2 c_{k}^0 c_{q}^1 f_j^1 \rightarrow f_i^1 c_{k}^1 c_{q}^1 f_j^1 .
\end{align}
The contribution to the interaction is
\begin{align}
    &\mathcal{P}_{S_0} \left[ \mathcal{V} \hat G^{0}(\varepsilon_0) \mathcal{V} \hat G^{0}(\varepsilon_0) \mathcal{V} \hat G^{0}(\varepsilon_0) \mathcal{V}   \right] \mathcal{P}_{S_0} = \nonumber \\
    &= \frac{1}{N^2}  \sum_{\bm r_i, \bm r_j,\bm k, \bm q}  
     e^{i(\bm q - \bm k)\cdot (\bm r_i - \bm r_j)} \ g_{\rm hh}^{f^2(5)}   \mathcal{P}_{S_0} 
    \bm c_{\bm{k}\mu}^{\dag} \mathbb{V}_{\bm{k} \mu}^{\dag} \bm f_{\bm r_i}^{}    \nonumber \\
    & \ \ \ \times  \bm f^{\dag}_{\bm r_j} \mathbb{V}_{\bm{k}\mu}^{} \bm c_{\bm{k}\mu}^{} 
    \bm c_{\bm{q}\mu'}^{\dag} \mathbb{V}_{\bm{q} \mu'}^{\dag} \bm f_{\bm r_j}^{}
    |\eta_{i\nu} \rangle \langle \eta_{i\nu} |
    \bm f^{\dag}_{\bm r_i} \mathbb{V}_{\bm{q}\mu'}^{} \bm c_{\bm{q}\mu'}^{} \mathcal{P}_{S_0} \nonumber \\
    & = -\frac{1}{N^2}  \sum_{\bm r_i, \bm r_j,\bm k, \bm q}  e^{i(\bm q - \bm k)\cdot (\bm r_i - \bm r_j)} \ g_{\rm hh}^{f^2(5)} f(\varepsilon_{\bm q \mu'}) f(\varepsilon_{\bm k \mu})   \nonumber \\
    & \times [\mathbb{V}_{\bm{k} \mu}^{*}]_{\alpha' s}
    [\mathbb{V}_{\bm{k}\mu}^{} ]_{\sigma_2 s}
    [\mathbb{V}_{\bm{q} \mu'}^{*} ]_{\sigma_2' s'}  
    [\mathbb{V}_{\bm{q}\mu'}^{}]_{\alpha s'}  \ 
    \gamma_{\sigma_1^{}\sigma_1',\alpha\alpha'}^{\nu}  \nonumber \\
    & \times  (  f_{\bm r_i \sigma_1^{}}^{\dag} f_{\bm r_i \sigma_1'}^{} \ f_{\bm r_j \sigma_2^{}}^{\dag} f_{\bm r_j \sigma_2'}^{} )
\end{align}
where the function of energy is
\begin{align}
    g_{\rm hh}^{f^2(5)} &= \frac{1}{(\varepsilon_{\bm k \mu} - \varepsilon_f - \varepsilon_{\nu}) (\varepsilon_{\bm q \mu'} - \varepsilon_f - \varepsilon_{\nu}) \varepsilon_{\nu} } .
\end{align}

By swapping sites $i$ with $j$ in $\mathcal{P}_{\rm hh}^{(5)}$, we obtain the virtual process $\mathcal{P}_{\rm hh}^{(6)}$, which is given by the order $cdab$ in Fig.~\ref{fig:app-hh-diag},
\begin{align}
    \mathcal{P}_{\rm hh}^{f^2 (6)} \ (cdab) : & f_i^1 c_{k}^1 c_{q}^1 f_j^1 \rightarrow  f_i^1 c_{k}^0 c_{q}^1 f_j^2 \rightarrow \nonumber \\
    & f_i^0 c_{k}^1 c_{q}^1 f_j^2 \rightarrow f_i^1 c_{k}^1 c_{q}^0 f_j^2 \rightarrow f_i^1 c_{k}^1 c_{q}^1 f_j^1 .
\end{align}

The total contribution of the processes of the hole-hole channel can be split into 
\begin{equation}
    \mathbb K^{(f^2:hh)}(i,j) = \mathbb K^{(f^{2:1}:\text{hh})}(i,j) + \mathbb K^{(f^{2:2}:\text{hh})}(i,j)
\end{equation}
where $\mathbb K^{(f^{2:1}:\text{hh})}$ contains the contribution from processes with $f^2$ intermediate in only one site, and reads as 
\begin{align}
    & [\mathbb K^{(f^{2:1}:\text{hh})}(i,j)]_{\sigma_1^{}\sigma_1' \sigma_2^{}\sigma_2'}  = \nonumber \\
    & =  -\frac{1}{N^2}  \sum_{\bm r_i, \bm r_j,\bm k, \bm q}    g_{\rm hh}^{f^2(5)} f(\varepsilon_{\bm q \mu'}) f(\varepsilon_{\bm k \mu}) \ e^{i(\bm q - \bm k)\cdot \bm r_{ij} }  \nonumber \\
    &  \times  \Big( [\mathbb{V}_{\bm{k} \mu}^{*}]_{\alpha' s}
    [\mathbb{V}_{\bm{k}\mu}^{} ]_{\sigma_2 s}
    [\mathbb{V}_{\bm{q} \mu'}^{*} ]_{\sigma_2' s'}  
    [\mathbb{V}_{\bm{q}\mu'}^{}]_{\alpha s'} \  \gamma_{\sigma_1^{}\sigma_1',\alpha\alpha'}^\nu \nonumber \\
    & \ \, + [\mathbb{V}_{\bm{k} \mu}^{*}]_{\sigma_1' s}
    [\mathbb{V}_{\bm{k}\mu}^{} ]_{\alpha s}
    [\mathbb{V}_{\bm{q} \mu'}^{*} ]_{\alpha' s'}  
    [\mathbb{V}_{\bm{q}\mu'}^{}]_{\sigma_1^{} s'} \   \gamma_{\sigma_2^{}\sigma_2',\alpha\alpha'}^\nu \Big)
\end{align}

In momentum space,
\begin{align}\label{eq:app:f21hh}
    & [\mathbb K^{(f^{2:1}:\text{hh})}(\bm k)]_{\sigma_1^{}\sigma_1' \sigma_2^{}\sigma_2'}   = \frac{1}{N}  \sum_{\bm q} g_{\varepsilon_{\bm k+\bm{q}\mu}^{},\varepsilon_{\bm{q}\mu'},\varepsilon_{\nu}}^{(f^{2:1}:\text{hh})}      \nonumber \\
    &  \times  \Big( [\mathbb{V}_{\bm k+\bm{q}\mu}^{}]_{\alpha s'}
    [\mathbb{V}_{\bm{q} \mu'}^{*}]_{\alpha' s}
    [\mathbb{V}_{\bm{q}\mu'}^{} ]_{\sigma_2 s}
    [\mathbb{V}_{\bm k+\bm{q} \mu}^{*} ]_{\sigma_2' s'}   \  \gamma_{\sigma_1^{}\sigma_1',\alpha\alpha'}^\nu \nonumber \\
    & \ \, + [\mathbb{V}_{\bm k+\bm{q}\mu}^{}]_{\sigma_1^{} s'}
    [\mathbb{V}_{\bm{q} \mu'}^{*}]_{\sigma_1' s}
    [\mathbb{V}_{\bm{q}\mu'}^{} ]_{\alpha s}
    [\mathbb{V}_{\bm k+\bm{q} \mu}^{*} ]_{\alpha' s'}   \   \gamma_{\sigma_2^{}\sigma_2',\alpha\alpha'}^\nu \Big)
\end{align}
where the function $g_{\varepsilon_{\bm k+\bm{q}\mu}^{},\varepsilon_{\bm{q}\mu'},\varepsilon_{\nu_1},\varepsilon_{\nu_2}}^{(f^{2:1}:\text{hh})}$ is given in Eq.~\eqref{eq:f2:1-hh}.

The processes with $f^2$ intermediate states in both sites contribute to $\mathbb K^{(f^{2:2}:\text{hh})}$, and read as
\begin{align}
    & [\mathbb K^{(f^{2:2}:\text{hh})}(i,j)]_{\sigma_1^{}\sigma_1' \sigma_2^{}\sigma_2'}  =  - \frac{1}{N^2} \! \!   \sum_{\bm r_i, \bm r_j,\bm k, \bm q} \! \! e^{i(\bm q - \bm k)\cdot \bm r_{ij} }   g_{\rm hh}^{f^2(1-4)} \nonumber \\
    & \times f(\varepsilon_{\bm q \mu'}) f(\varepsilon_{\bm k \mu})  \gamma_{\sigma_1^{}\sigma_1',\alpha\alpha'}^{\nu_1} \gamma_{\sigma_2^{}\sigma_2',\beta\beta'}^{\nu_2} \nonumber \\
    &  \times   [\mathbb{V}_{\bm{k} \mu}^{*}]_{\alpha' s}
    [\mathbb{V}_{\bm{q} \mu'}^{*} ]_{\beta' s'}  
    [\mathbb{V}_{\bm{k}\mu}^{} ]_{\beta s}
    [\mathbb{V}_{\bm{q}\mu'}^{}]_{\alpha s'}
\end{align}
with the function of energy,
\begin{align}
    g_{\rm hh}^{f^2(1-4)} &= \frac{( \varepsilon_{\bm k \mu} + \varepsilon_{\bm q \mu'} - 2 \varepsilon_f - \varepsilon_{\nu_1} - \varepsilon_{\nu_2})}{ ( \varepsilon_{\bm q \mu'} -   \varepsilon_f - \varepsilon_{\nu_1} ) ( \varepsilon_{\bm k \mu} -   \varepsilon_f - \varepsilon_{\nu_1} )  }  \nonumber \\
    & \times \frac{ 1 }{ ( \varepsilon_{\bm q \mu'} -   \varepsilon_f - \varepsilon_{\nu_2} ) ( \varepsilon_{\bm k \mu} -   \varepsilon_f - \varepsilon_{\nu_2} )  } .
\end{align}
In momentum space,
\begin{align}\label{eq:app:f22hh}
    & [\mathbb K^{(f^{2:2}:\text{hh})}(\bm k)]_{\sigma_1^{}\sigma_1' \sigma_2^{}\sigma_2'}   = \nonumber \\
    & \ \   =  \frac{1}{N}  \sum_{\bm q} g_{\varepsilon_{\bm k+\bm{q}\mu}^{},\varepsilon_{\bm{q}\mu'},\varepsilon_{\nu_1},\varepsilon_{\nu_2}}^{(f^{2:2}:\text{hh})} \  \gamma_{\sigma_1^{}\sigma_1',\alpha\alpha'}^{\nu_1}   \gamma_{\sigma_2^{}\sigma_2',\beta\beta'}^{\nu_2}   \nonumber \\
    & \ \ \ \ \ \times  [\mathbb{V}_{\bm k+\bm{q}\mu}^{}]_{\alpha s} 
    [\mathbb{V}_{\bm{q} \mu'}^{*}]_{\alpha' s'}
    [\mathbb{V}_{\bm{q}\mu'}^{} ]_{\beta s'} 
    [\mathbb{V}_{\bm k+\bm{q} \mu}^{*} ]_{\beta' s} 
\end{align}
where the function $g_{\varepsilon_{\bm k+\bm{q}\mu}^{},\varepsilon_{\bm{q}\mu'},\varepsilon_{\nu_1},\varepsilon_{\nu_2}}^{(f^{2:2}:\text{hh})}$ is given in Eq.~\eqref{eq:f2:2-hh}.

By using the effective hopping defined in the main text, $\mathbb{\tilde T}_{\bm k \mu}^{} = \mathbb{\tilde V}_{\bm k \mu}^{} \mathbb{\tilde V}_{\bm k \mu}^{\dag}$, and summing up Eq.~\eqref{eq:app:f0pp} for the $pp$ channel and Eq.~\eqref{eq:app:f0ph} for the $ph$ channel, we obtain the final expression given in Eq.~\eqref{eq:f0}.

The addition of the $pp$, $ph$ and $hh$ channels given, respectively, in Eqs.~\eqref{eq:app:f2pp}, \eqref{eq:app:f21ph}, and (\ref{eq:app:f21hh}), and the effective hopping $\mathbb{\hat T}_{\bm k \mu}^{} = \mathbb{\tilde V}_{\bm k \mu}^{} \mathbb{V}_{\bm k \mu}^{\dag}$, gives rise to Eq.  (\ref{eq:Kf2}).

The final expression in Eq.~\eqref{eq:Kf2f2} is obtained by adding the $ph$ and $hh$ channels given in Eqs.~\eqref{eq:app:f22ph} and \eqref{eq:app:f22hh}, and by using the effective hopping $\mathbb{T}_{\bm k \mu}^{} = \mathbb{V}_{\bm k \mu}^{} \mathbb{V}_{\bm k \mu}^{\dag}$.
\\~\\

Notice that for the $pp$ and $hh$ channels, some contributions must be removed due to the Pauli exclusion principle. Specifically, in the processes $\mathcal{P}^{f^0(n)}_{\rm pp}$ and $\mathcal{P}^{f^2(n)}_{\rm hh}$ with $n=1,2,3,4$, the two particles at the $c$-bands cannot be created in the same state, meaning that if $\bm k=\bm q$ and $\mu=\mu'$, the spin must be different ($s\neq s'$). Nonetheless, this particular contribution, which must be eliminated, is compensated by a group of diagrams that we have neglected, as they solely contribute to a density-density interaction. These are the diagrams that do not exchange particles, which means that the second order hybridization takes place between the $c$ bands and the same $f$ site. After clarifying this, we only consider these diagrams when the Pauli exclusion principle acts, which is equivalent to ignoring the Pauli exclusion principle in the processes mentioned above. This is important for the magnetic interaction to be continuous at the $\Gamma$ point.

\section{Derivation of Equation~\eqref{eq:hopp-TR}} \label{app:aux-calculation}

In this appendix, we present the demonstration of Eq.~\eqref{eq:hopp-TR}, which stems from time-reversal and spatial-inversion symmetry.  

The time-reversal symmetry implies that the effective hopping
\begin{equation}
    [\mathbb{T}_{\bm k \mu}]_{\alpha \beta}^{} = {\rm sgn}(\alpha) {\rm sgn}(\beta) \ [\mathbb{T}_{-\bm k \mu}]_{\bar\alpha \bar\beta}^{*} \ .\label{eq:T-TR} 
\end{equation}
Due to the spatial-inversion symmetry, the dispersion of the conduction bands
\begin{equation}
    \varepsilon_{\bm k\mu} = \varepsilon_{-\bm k\mu} \ .   \label{eq:g-sign}
\end{equation}
The function of energy for the $f^{2:1}$ channel given in Eq.~\eqref{eq:f2:1} is invariant under the exchange between $\varepsilon_{\bm k+ \bm q \mu}$ and $\varepsilon_{\bm q \mu'}$; that is,
\begin{equation}
    g_{\varepsilon_{\bm k+ \bm q \mu},\varepsilon_{\bm q \mu'},F^0}^{(f^{2:1})} = g_{\varepsilon_{\bm q \mu'},\varepsilon_{\bm k+ \bm q \mu},F^0}^{(f^{2:1})} \ . \label{eq:g-swap}
\end{equation}
By means of these properties, we obtain
\begin{align}
    & \sum_{ \bm q }  [\mathbb{T}_{\bm k+\bm q \mu}^{*}]_{\sigma \alpha }^{} [\mathbb{T}_{\bm q \mu'}^{}]_{\sigma' \alpha}^{} \   g_{\varepsilon_{\bm k+\bm q \mu}, \varepsilon_{\bm q \mu'} , F^0}^{(f^{2:1})}  \nonumber \\
    & \overset{\ref{eq:T-TR}}{=} {\rm sgn}(\sigma) {\rm sgn}(\sigma') \sum_{ \bm q} [\mathbb{T}_{-\bm k-\bm q \mu}^{}]_{\bar \sigma \bar \alpha}^{} [\mathbb{T}_{-\bm q \mu'}^{*}]_{\bar \sigma' \bar \alpha }^{}  \  g_{\varepsilon_{\bm k+\bm q \mu}, \varepsilon_{\bm q \mu'} , F^0}^{(f^{2:1})}  \nonumber \\
    & \overset{\bm q \rightarrow \text{-}\bm q \text{-}\bm k}{=} \!\!  {\rm sgn}(\sigma) {\rm sgn}(\sigma') \sum_{ \bm q } [\mathbb{T}_{\bm q \mu}^{}]_{\bar \sigma \bar \alpha }^{} [\mathbb{T}_{\bm k+\bm q \mu'}^{*}]_{\bar \sigma' \bar \alpha}^{} \   g_{\varepsilon_{\text{-}\bm q \mu}, \varepsilon_{\text{-}\bm k\text{-}\bm q \mu'} , F^0}^{(f^{2:1})}   \nonumber \\
    & \overset{\ref{eq:g-sign}}{=}  {\rm sgn}(\sigma) {\rm sgn}(\sigma') \sum_{ \bm q } [\mathbb{T}_{\bm k+\bm q \mu'}^{*}]_{\bar \sigma' \bar \alpha }^{} [\mathbb{T}_{\bm q \mu}^{}]_{\bar \sigma \bar \alpha }^{}  \   g_{\varepsilon_{\bm q \mu}, \varepsilon_{\bm k+\bm q \mu'} , F^0}^{(f^{2:1})}  \nonumber \\
    & \overset{\ref{eq:g-swap}}{=}  {\rm sgn}(\sigma) {\rm sgn}(\sigma') \sum_{ \bm q } [\mathbb{T}_{\bm k+\bm q \mu'}^{*}]_{\bar \sigma' \bar \alpha }^{} [\mathbb{T}_{\bm q \mu}^{}]_{\bar \sigma \bar \alpha}^{}  \   g_{\varepsilon_{\bm k+\bm q \mu}, \varepsilon_{\bm q \mu'} , F^0}^{(f^{2:1})}  .
\end{align}
In this equation there is no summation over the repeated indices $\mu$, $\mu'$ and $\alpha$. However, for the cancellation in Eq.~\eqref{eq:a-b} it is necessary to sum over the Kramers doublet ${\alpha,\bar\alpha}$.


\section{Energy spectrum of the ion Ce$^{2+}$} \label{app:Ce-ion}

To derive the superexchange contribution that involves $f^2$ intermediate states, we need to compute the energies and states of the Ce$^{2+}$ ion~\cite{ZhangQ2014_thesis}. The two-particle states of the $f^2$ configuration are constructed in the basis
\begin{equation}
    |m_1\sigma_1,m_2\sigma_2 \rangle = f_{m_1 \sigma_1}^{\dag} f_{m_2  \sigma_2}^{\dag} | 0 \rangle.
\end{equation}
The single-ion Hamiltonian is the sum of the Coulomb interaction and the spin-orbit coupling,
\begin{equation}\label{eq:Hion}
    H_{\rm ion} = H_{\rm soc} + H_{\rm Coulomb},
\end{equation}
where 
\begin{widetext}
\begin{align}
    H_{\rm Coulomb} = \frac{1}{2} \sum_{\substack{m_1, m_2, \sigma \\ m'_1, m'_2, \sigma'}} U_{m_1, m_2,m_1',m_2'} 
 f_{m_1\sigma}^\dag f_{m_2\sigma'}^\dag f_{m_2'\sigma'}^{} f_{m_1'\sigma}^{}
\end{align}
with $U_{m_1, m_2,m_1',m_2'}$ coming from the Coulomb integral
\begin{align}
    U_{m_1, m_2,m_1',m_2'} ={ \frac{e^2}{4\pi\epsilon_0} } \int d\bm r_1 \int d \bm r_2 \psi_{m_1}^*(\bm r_1)  \psi_{m_2}^*(\bm r_2)  
    \frac{1}{|\bm r_1 - \bm r_2|} \psi_{m_2'}^{} (\bm r_2) \psi_{m_1'}^{}(\bm r_1),
\end{align}
{where $e$ is the electron charge and $\epsilon_0$ the vacuum permittivity.}
The electron wave function {$\psi_{m}^{} (\bm r)$} is well approximated by the atomic 4$f$ orbitals giving rise to the Slater integral $F^{2k}$ with $k\in\{0,1,2,3\}$. Then, the Coulomb integral can be written as
    \begin{align}
    U_{m_1, m_2,m_1',m_2'} = \delta_{m_1^{}+m_2^{},m_1'+m_2'} (-1)^{m_1^{}-m_1'} 
     \sum_{k=0}^{3} c^{2k}(m_1^{},m_1') c^{2k}(m_2^{},m_2') F^{2k}
\end{align}
\end{widetext}
By using the atomic wave function $\psi_{m}^{} (\bm r) = R_{nl}(r)Y_{l}^{m}(\theta,\phi)$, the Slater integral yields
\begin{equation}\label{eq:F2k}
    F^{2k} = { \frac{e^2}{4\pi\epsilon_0} } \int_0^\infty r_1^2 dr_1  \int_0^\infty r_2^2 dr_2 R_{43}^2(r_1) R_{43}^2(r_2) \frac{r_<^{2k}}{r_>^{2k+1}}
\end{equation}
where $r_>={\rm max}(r_1,r_2)$ and $r_<={\rm min}(r_1,r_2)$. The Guant coefficients are defined in terms of the spherical harmonics
\begin{align}
    c^{2k}&(m,m') = \sqrt{\frac{4\pi}{4k+1}} (-1)^m  \nonumber \\
    & \times \int d\Omega \ Y_{3}^{-m}(\theta,\phi) Y_{2k}^{m-m'}(\theta,\phi) Y_{3}^{m'}(\theta,\phi),
\end{align}
where $\int d\Omega \equiv \int_{0}^\pi\sin(\theta)d\theta \int_0^{2\pi}d\phi$. We adopted
the following values for the Coulomb integrals~\cite{Jang2023}:
\begin{equation}
\label{eq:Hion-param}
\begin{split}
F^0 &= \SI{6.4}{eV}, \\ 
F^2 &= \SI{8.3436}{eV}, \\
F^4 &= \SI{5.57482}{eV}, \\ 
F^6 &= \SI{4.12446}{eV}.
\end{split}
\end{equation}
    
The spin-orbit coupling Hamiltonian reads as
\begin{equation}
  H_{\rm soc} =  \zeta_{\rm SO} \sum_{mm'} \sum_{\sigma \sigma'} \bm L_{mm'} \cdot \bm S_{\sigma \sigma'} \ f_{m\sigma}^\dag f_{m'\sigma'} 
\end{equation}
where $\bm L=(L^x,L^y,L^z)$ is the angular momentum for $l=3$, and $\bm S= \bm \sigma/2$ with $\bm \sigma$ being the Pauli matrices.
For the spin-orbit coupling we used $\zeta_{\rm SO} = \SI{0.08}{eV}$, which is equivalent to a spin-orbit splitting of \SI{280}{meV} between the $J=5/2$ and $J=7/2$ states. 




\section{Exchange from direct $f-f$ hopping}\label{app:Hff}

An exchange interaction between the $f$-orbital can be derived by considering the direct hopping between them.

The tight-binding Hamiltonian obtained from DFT is
\begin{align}
    \mathcal{\hat H}^{f-f}_{\rm TB} =  \sum_{{\bm R}, {\bm r}}    {\bm f}_{\bm R + \bm r, m}^{\dag} \mathbb{t}^{00}_{m,m'}{(\bm r)} {\bm f}_{\bm R, m'}^{}
\end{align} 
where ${\bm r}$ runs over the six vectors $\{ \pm a \hat{x}, \pm a \hat{y}, \pm a \hat{z} \}$ connecting a given Ce ion with its nearest neighbors. The indices $m$ and $m'$ run over the projection of the orbital angular momentum $(-3\leq m \leq 3)$ of the $f$-doublet-Wannier orbitals.

\begin{figure}[h!]
    \centering
    \includegraphics[width=0.48\textwidth]{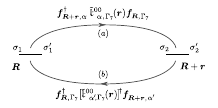}
    \caption{Diagram of the virtual processes involved in the derivation of the effective Hamiltonian due to $f$-$f$ hopping. The final and initial states in site $\bm R$ ($\bm R+\bm r$) are given by $\sigma_1^{}$ and $\sigma_1'$ ($\sigma_2^{}$ and $\sigma_2'$), respectively. }
    \label{fig:ff-hopping-diag}
\end{figure}

By unitary transformation, the hopping matrix can be written in the basis of the CF eigenstates for the states with $J=5/2$ and the basis of $J_z$ for the states with $J=7/2$,
\begin{eqnarray}
    \mathbb{\tilde t}_{\alpha \alpha'}^{00}(\bm r) = \mathbb{U}_{\alpha m} \mathbb{\tilde t}_{m m'}^{00}(\bm r) [\mathbb{U}_{\alpha' m'}]^\dag 
\end{eqnarray}
where the indices $\alpha$ and $\alpha'$ run over the set of doublets $\{ \Gamma_7^{}, \ \Gamma_8^a, \ \Gamma_8^b, \ \tfrac{1}{2}, \ \tfrac{3}{2}, \ \tfrac{5}{2}, \ \tfrac{7}{2}  \}$
. The virtual processes involved in the direct hopping between $f$ orbitals are depicted in Fig.~\ref{fig:ff-hopping-diag}.
By using perturbation theory,
\begin{equation}
    \mathcal{H}_{\rm eff} = \mathcal{P}_{\rm S}  V  G(\varepsilon_0)  V  \mathcal{P}_{\rm S}
\end{equation}
The low-energy Hamiltonian is given by
\begin{equation}
    \mathcal{H}_{\rm eff}^{\rm ff} = \frac{1}{2} \sum_{\bm R, \bm r}  K_{\sigma_1^{} \sigma_1' \sigma_2^{} \sigma_2'}^{\rm ff}(\bm r) \ f_{\bm r, \sigma_1^{}}^{\dag} f_{\bm r, \sigma_1'}^{}  f_{\bm R + \bm r,\sigma_2^{}}^{\dag} f_{\bm R +\bm r, \sigma_2'}^{}
\end{equation}
where the exchange interaction reads as 
\begin{align}\label{eq:Kff-real}
    K_{\sigma_1^{} \sigma_1' \sigma_2^{} \sigma_2'}^{\rm ff} ( \bm r & )  =  - \sum_{\nu} \frac{1}{\varepsilon_\nu} \nonumber  \\ 
    \times \sum_{ \alpha \alpha',ss'} \big\{ & [\mathbb{\tilde t}_{\alpha',\Gamma_7}^{00}(\bm r)]_{s' \sigma_1 }^{*}    [ \mathbb{\tilde t}_{\alpha,\Gamma_7}^{00}(\bm r)]_{s \sigma_1'}^{} \  \gamma_{\sigma_2^{} \sigma_2', \alpha_s^{} \alpha_{s'}'}^{\nu} \nonumber  \\ 
    + & [\mathbb{\tilde t}_{\Gamma_7, \alpha'}^{00}(\bm r)]_{\sigma_2^{} s' }^{} [\mathbb{\tilde t}_{\Gamma_7, \alpha}^{00}(\bm r)]_{\sigma_2' s}^{*} \  \gamma_{\sigma_1^{} \sigma_1', \alpha_s^{} \alpha_{s'}'}^{\nu} \big\}
\end{align}
where the indices $\sigma$ and $s$ run over the spin $\{ \uparrow, \downarrow \}$, and the index $\alpha_s$ stands for the state in the doublet $\alpha$ and spin $s$.

Finally, by mean of Eqs.~(\ref{eq:mag-int-total}) we express the spin Hamiltonian $\mathcal{H}_{\rm eff}^{\rm ff}$ with spin operators as
\begin{equation}
    \mathcal{H}_{\rm eff}^{\rm ff} = \frac{1}{2} \sum_{\bm R, \bm r} I_{\bm r}^{\alpha\beta} S_{\bm R}^\alpha S_{\bm R+\bm r}^\beta 
\end{equation}
For the nearest neighbors, the magnetic interaction is very anisotropic and compass-like. For instance, for the neighbor $\bm r=(1,0,0)$ it yields
\begin{align}
I_{\bm{r}}^{xx} = \SI{2.33}{\mu eV}, && I_{\bm{r}}^{yy} = \SI{-1.67}{\mu eV}, &&  I_{\bm r}^{zz} = \SI{-1.67}{\mu eV}. \nonumber 
\end{align}
Due to its short-range nature, interactions with second neighbors are approximately ten times weaker.

The strong anisotropy can be understood by observing that the elements of the $f$-$f$ intra-doublet hopping, $\mathbb{\tilde t}_{\Gamma_7, \Gamma_7}^{00}(\bm r) = -0.76 \mathbb{1}_{2\times2}$ meV, are one order of magnitude smaller than the largest element of the out-of-doublet hopping, $[\mathbb{\tilde t}_{\Gamma_7, \alpha}^{00}(\bm r)]_{\downarrow\uparrow}^{} =$ \SI{-9.85}{meV} with $\alpha=(J=7/2,J_z=3/2)$. The smallness of the hopping between $J=5/2$-states, in comparison with the hopping between $J=5/2$- and $J=7/2$-states, is related with the cancellation that occurs in presence of axial symmetry under rotation around the bond, as described in Ref.~\cite{BatistaCD2009}.



\subsection{Degenerate case}
Here we show that for the direct $f$-$f$ hopping, the $f^2$ \emph{out-of-doublet} contributions cancel out when the $f^2$ configurations are fully degenerate. The remaining $f^2$ \emph{intradoublet} contributions generate an isotropic magnetic interaction.

When all the $f^2$ states degenerate, the function that projects the creation and annihilation of a particle with state $\alpha_s$ and $\alpha_{s'}'$, respectively, in a site with initial state $\sigma'$ and final states $\sigma$, is given by Eq.~\eqref{eq:gamma-degen},
\begin{equation}
    \sum_{\nu} \gamma_{\sigma \sigma',\alpha\alpha'}^{\nu} = \delta_{\sigma \sigma'} \delta_{\alpha_s \alpha_{s}'} - \delta_{\sigma \alpha_s} \delta_{\sigma' \alpha_{s}'}
\end{equation}
Then, the interaction due to the direct hopping in Eq.~\eqref{eq:Kff-real} becomes,
\begin{widetext}
    \begin{align}\label{eq:Kff-real-deg}
    K_{\sigma_1^{} \sigma_1' \sigma_2^{} \sigma_2'}^{\rm ff} ( \bm r  )  =  - \frac{1}{F^0}  \big\{ & [\mathbb{\tilde t}_{\alpha,\Gamma_7}^{00}(\bm r)]_{s \sigma_1 }^{*}    [ \mathbb{\tilde t}_{\alpha,\Gamma_7}^{00}(\bm r)]_{s \sigma_1'}^{} \delta_{\sigma_2^{} \sigma_2'} 
     +  [\mathbb{\tilde t}_{\Gamma_7, \alpha}^{00}(\bm r)]_{\sigma_2^{} s }^{} [\mathbb{\tilde t}_{\Gamma_7, \alpha}^{00}(\bm r)]_{\sigma_2' s}^{*} \delta_{\sigma_1^{} \sigma_1'} \nonumber \\
     - & [\mathbb{\tilde t}_{\Gamma_7,\Gamma_7}^{00}(\bm r)]_{\sigma_2' \sigma_1 }^{*}    [ \mathbb{\tilde t}_{\Gamma_7,\Gamma_7}^{00}(\bm r)]_{\sigma_2 \sigma_1'}^{}  
    - [\mathbb{\tilde t}_{\Gamma_7, \Gamma_7}^{00}(\bm r)]_{\sigma_2^{} \sigma_1' }^{} [\mathbb{\tilde t}_{\Gamma_7, \Gamma_7}^{00}(\bm r)]_{\sigma_2' \sigma_1^{}}^{*}  \big\}
\end{align}
\end{widetext}

In the following calculations, we resort to the property
\begin{align}\label{eq:app:TR}
    & \quad \sum_{\alpha,s}  [\mathbb{\tilde t}_{\Gamma_7, \alpha}^{00}(\bm r)]_{\sigma^{} s }^{} [\mathbb{\tilde t}_{\Gamma_7, \alpha}^{00}(\bm r)]_{\sigma' s}^{*}  \nonumber \\
    & = {\rm sgn}(\sigma) {\rm sgn}(\sigma') \sum_{\alpha,s} [\mathbb{\tilde t}_{\Gamma_7, \alpha}^{00}(\bm r)]_{\bar\sigma^{} \bar s }^{*} [\mathbb{\tilde t}_{\Gamma_7, \alpha}^{00}(\bm r)]_{\bar \sigma' \bar s}^{},
\end{align}
where we have used the time-reversal symmetry.

The $zz$ component of the magnetic interaction is proportional to
\begin{align}
    & \quad  K_{\uparrow\uparrow\uparrow\uparrow}^{\rm ff}(\bm r) - K_{\uparrow\uparrow\downarrow\downarrow}^{\rm ff}(\bm r)  \nonumber \\
    & = - \frac{1}{F^0} \big\{ [\mathbb{\tilde t}_{\Gamma_7, \alpha}^{00}(\bm r)]_{\uparrow s }^{} [\mathbb{\tilde t}_{\Gamma_7, \alpha}^{00}(\bm r)]_{\uparrow s}^{*} - [\mathbb{\tilde t}_{\Gamma_7, \alpha}^{00}(\bm r)]_{\downarrow  s }^{} [\mathbb{\tilde t}_{\Gamma_7, \alpha}^{00}(\bm r)]_{\downarrow s}^{*}   \nonumber \\
    & \quad \quad  -   2 |[\mathbb{\tilde t}_{\Gamma_7, \Gamma_7}^{00}(\bm r)]_{\uparrow \uparrow }^{}|^2 - 2 |[\mathbb{\tilde t}_{\Gamma_7, \Gamma_7}^{00}(\bm r)]_{\downarrow \uparrow }^{}|^2  \big\}
\end{align}

Using Eq.~\eqref{eq:app:TR}, the terms in the second line cancel, showing that the contributions from out-of-doublet $\Gamma_7$ are zero. Finally, the time-reversal symmetry implies that $[\mathbb{\tilde t}_{\Gamma_7, \Gamma_7}^{00}(\bm r)]_{\downarrow \uparrow }^{} = 0$, and the spatial-inversion symmetry combined with the Hermiticity ($[\mathbb{\tilde t}_{\Gamma_7, \Gamma_7}^{00}(\bm r)]_{\downarrow \uparrow }^{} = [\mathbb{\tilde t}_{\Gamma_7, \Gamma_7}^{00}(\bm r)]_{\downarrow \uparrow }^{} $) shows that $[\mathbb{\tilde t}_{\Gamma_7, \Gamma_7}^{00}(\bm r)]_{\sigma \sigma }^{} $ is real. Then,
\begin{align}
    & K_{\uparrow\uparrow\uparrow\uparrow}^{\rm ff}(\bm r) - K_{\uparrow\uparrow\downarrow\downarrow}^{\rm ff}(\bm r) = \frac{2}{F^0} ([\mathbb{\tilde t}_{\Gamma_7, \Gamma_7}^{00}(\bm r)]_{\uparrow \uparrow }^{})^2 
\end{align}

The interaction $ K_{\uparrow\downarrow\downarrow\uparrow}^{\rm ff}$ is obtained from Eq.~\eqref{eq:Kff-real-deg}:
\begin{equation}
    K_{\uparrow\downarrow\downarrow\uparrow}^{\rm ff}(\bm r)  =  \frac{2}{F^0} ([\mathbb{\tilde t}_{\Gamma_7, \Gamma_7}^{00}(\bm r)]_{\uparrow \uparrow }^{})^2  .
\end{equation}

The interaction $ K_{\uparrow\downarrow\uparrow\downarrow}^{\rm ff}$ yields,
\begin{equation}
    K_{\uparrow\downarrow\uparrow\downarrow}^{\rm ff}(\bm r)  =  \frac{2}{F^0}   [\mathbb{\tilde t}_{\Gamma_7, \Gamma_7}^{00}(\bm r)]_{\downarrow \uparrow }^{*}  [\mathbb{\tilde t}_{\Gamma_7, \Gamma_7}^{00}(\bm r)]_{\uparrow \downarrow }^{}  = 0.
\end{equation}
And the interaction $ K_{\uparrow\uparrow\uparrow\downarrow}^{\rm ff}$ yields
\begin{align}
    K_{\uparrow\uparrow\uparrow\downarrow}^{\rm ff}(\bm r)  =  - \frac{1}{F^0}  & [\mathbb{\tilde t}_{\Gamma_7, \alpha}^{00}(\bm r)]_{\uparrow s }^{}  [\mathbb{\tilde t}_{\Gamma_7, \alpha}^{00}(\bm r)]_{\downarrow s }^{*}  \nonumber \\
    + \frac{2}{F^0} & [\mathbb{\tilde t}_{\Gamma_7, \alpha}^{00}(\bm r)]_{\downarrow \uparrow }^{*}  [\mathbb{\tilde t}_{\Gamma_7, \alpha}^{00}(\bm r)]_{\uparrow \uparrow }^{} = 0.
\end{align}
The contributions of the term in the first line are zero according to Eq.~\eqref{eq:app:TR}, whereas the term in the second line is zero since  $[\mathbb{\tilde t}_{\Gamma_7, \Gamma_7}^{00}(\bm r)]_{\downarrow \uparrow }^{} = 0$.

This procedure demonstrates the cancellation of the remaining interaction, $ K_{\uparrow\downarrow\uparrow\uparrow}^{\rm ff}=0$. 

Finally, from Eqs.~\eqref{eq:mag-int-total} we obtain the 
magnetic interaction tensor, which is isotropic
\begin{subequations}
\begin{align}
I_{\bm{r}}^{xx} = I_{\bm{r}}^{yy} = I_{\bm r}^{zz}  &= \frac{4}{F^0} ([\mathbb{\tilde t}_{\Gamma_7, \Gamma_7}^{00}(\bm r)]_{\uparrow \uparrow }^{})^2 ,\\
I_{\bm{r}}^{xy} = I_{\bm{r}}^{yz} = I_{\bm r}^{zx}  &=  0 . 
\end{align}
\end{subequations}

\bibliography{refs_CeIn3}

\begin{thebibliography}{63}%
\makeatletter
\providecommand \@ifxundefined [1]{%
 \@ifx{#1\undefined}
}%
\providecommand \@ifnum [1]{%
 \ifnum #1\expandafter \@firstoftwo
 \else \expandafter \@secondoftwo
 \fi
}%
\providecommand \@ifx [1]{%
 \ifx #1\expandafter \@firstoftwo
 \else \expandafter \@secondoftwo
 \fi
}%
\providecommand \natexlab [1]{#1}%
\providecommand \enquote  [1]{``#1''}%
\providecommand \bibnamefont  [1]{#1}%
\providecommand \bibfnamefont [1]{#1}%
\providecommand \citenamefont [1]{#1}%
\providecommand \href@noop [0]{\@secondoftwo}%
\providecommand \href [0]{\begingroup \@sanitize@url \@href}%
\providecommand \@href[1]{\@@startlink{#1}\@@href}%
\providecommand \@@href[1]{\endgroup#1\@@endlink}%
\providecommand \@sanitize@url [0]{\catcode `\\12\catcode `\$12\catcode
  `\&12\catcode `\#12\catcode `\^12\catcode `\_12\catcode `\%12\relax}%
\providecommand \@@startlink[1]{}%
\providecommand \@@endlink[0]{}%
\providecommand \url  [0]{\begingroup\@sanitize@url \@url }%
\providecommand \@url [1]{\endgroup\@href {#1}{\urlprefix }}%
\providecommand \urlprefix  [0]{URL }%
\providecommand \Eprint [0]{\href }%
\providecommand \doibase [0]{https://doi.org/}%
\providecommand \selectlanguage [0]{\@gobble}%
\providecommand \bibinfo  [0]{\@secondoftwo}%
\providecommand \bibfield  [0]{\@secondoftwo}%
\providecommand \translation [1]{[#1]}%
\providecommand \BibitemOpen [0]{}%
\providecommand \bibitemStop [0]{}%
\providecommand \bibitemNoStop [0]{.\EOS\space}%
\providecommand \EOS [0]{\spacefactor3000\relax}%
\providecommand \BibitemShut  [1]{\csname bibitem#1\endcsname}%
\let\auto@bib@innerbib\@empty
\bibitem [{\citenamefont {Andres}\ \emph {et~al.}(1975)\citenamefont {Andres},
  \citenamefont {Graebner},\ and\ \citenamefont {Ott}}]{AndresK1975}%
  \BibitemOpen
  \bibfield  {author} {\bibinfo {author} {\bibfnamefont {K.}~\bibnamefont
  {Andres}}, \bibinfo {author} {\bibfnamefont {J.~E.}\ \bibnamefont
  {Graebner}},\ and\ \bibinfo {author} {\bibfnamefont {H.~R.}\ \bibnamefont
  {Ott}},\ }\bibfield  {title} {\bibinfo {title}
  {4{\emph{f}}-{{Virtual-Bound-State Formation}} in {{CeAl}}{$_3$} at {{Low
  Temperatures}}},\ }\href {https://doi.org/10.1103/PhysRevLett.35.1779}
  {\bibfield  {journal} {\bibinfo  {journal} {Phys. Rev. Lett.}\ }\textbf
  {\bibinfo {volume} {35}},\ \bibinfo {pages} {1779} (\bibinfo {year}
  {1975})}\BibitemShut {NoStop}%
\bibitem [{\citenamefont {Steglich}\ \emph {et~al.}(1979)\citenamefont
  {Steglich}, \citenamefont {Aarts}, \citenamefont {Bredl}, \citenamefont
  {Lieke}, \citenamefont {Meschede}, \citenamefont {Franz},\ and\ \citenamefont
  {Sch{\"a}fer}}]{SteglichF1979}%
  \BibitemOpen
  \bibfield  {author} {\bibinfo {author} {\bibfnamefont {F.}~\bibnamefont
  {Steglich}}, \bibinfo {author} {\bibfnamefont {J.}~\bibnamefont {Aarts}},
  \bibinfo {author} {\bibfnamefont {C.~D.}\ \bibnamefont {Bredl}}, \bibinfo
  {author} {\bibfnamefont {W.}~\bibnamefont {Lieke}}, \bibinfo {author}
  {\bibfnamefont {D.}~\bibnamefont {Meschede}}, \bibinfo {author}
  {\bibfnamefont {W.}~\bibnamefont {Franz}},\ and\ \bibinfo {author}
  {\bibfnamefont {H.}~\bibnamefont {Sch{\"a}fer}},\ }\bibfield  {title}
  {\bibinfo {title} {Superconductivity in the {{Presence}} of {{Strong Pauli
  Paramagnetism}}: {{CeCu}}{$_{2}$}{{Si}}{$_{2}$}},\ }\href
  {https://doi.org/10.1103/PhysRevLett.43.1892} {\bibfield  {journal} {\bibinfo
   {journal} {Phys. Rev. Lett.}\ }\textbf {\bibinfo {volume} {43}},\ \bibinfo
  {pages} {1892} (\bibinfo {year} {1979})}\BibitemShut {NoStop}%
\bibitem [{\citenamefont {Ott}\ \emph {et~al.}(1983)\citenamefont {Ott},
  \citenamefont {Rudigier}, \citenamefont {Fisk},\ and\ \citenamefont
  {Smith}}]{OttHR1983}%
  \BibitemOpen
  \bibfield  {author} {\bibinfo {author} {\bibfnamefont {H.~R.}\ \bibnamefont
  {Ott}}, \bibinfo {author} {\bibfnamefont {H.}~\bibnamefont {Rudigier}},
  \bibinfo {author} {\bibfnamefont {Z.}~\bibnamefont {Fisk}},\ and\ \bibinfo
  {author} {\bibfnamefont {J.~L.}\ \bibnamefont {Smith}},\ }\bibfield  {title}
  {\bibinfo {title} {{{UBe}}{$_{13}$}: {{An Unconventional Actinide
  Superconductor}}},\ }\href {https://doi.org/10.1103/PhysRevLett.50.1595}
  {\bibfield  {journal} {\bibinfo  {journal} {Phys. Rev. Lett.}\ }\textbf
  {\bibinfo {volume} {50}},\ \bibinfo {pages} {1595} (\bibinfo {year}
  {1983})}\BibitemShut {NoStop}%
\bibitem [{\citenamefont {Stewart}\ \emph {et~al.}(1984)\citenamefont
  {Stewart}, \citenamefont {Fisk}, \citenamefont {Willis},\ and\ \citenamefont
  {Smith}}]{StewartGR1984_UPt3}%
  \BibitemOpen
  \bibfield  {author} {\bibinfo {author} {\bibfnamefont {G.~R.}\ \bibnamefont
  {Stewart}}, \bibinfo {author} {\bibfnamefont {Z.}~\bibnamefont {Fisk}},
  \bibinfo {author} {\bibfnamefont {J.~O.}\ \bibnamefont {Willis}},\ and\
  \bibinfo {author} {\bibfnamefont {J.~L.}\ \bibnamefont {Smith}},\ }\bibfield
  {title} {\bibinfo {title} {Possibility of {{Coexistence}} of {{Bulk
  Superconductivity}} and {{Spin Fluctuations}} in {{UPt}}{$_{3}$}},\ }\href
  {https://doi.org/10.1103/PhysRevLett.52.679} {\bibfield  {journal} {\bibinfo
  {journal} {Phys. Rev. Lett.}\ }\textbf {\bibinfo {volume} {52}},\ \bibinfo
  {pages} {679} (\bibinfo {year} {1984})}\BibitemShut {NoStop}%
\bibitem [{\citenamefont {Fisher}\ \emph {et~al.}(1989)\citenamefont {Fisher},
  \citenamefont {Kim}, \citenamefont {Woodfield}, \citenamefont {Phillips},
  \citenamefont {Taillefer}, \citenamefont {Hasselbach}, \citenamefont
  {Flouquet}, \citenamefont {Giorgi},\ and\ \citenamefont
  {Smith}}]{FisherRA1989}%
  \BibitemOpen
  \bibfield  {author} {\bibinfo {author} {\bibfnamefont {R.~A.}\ \bibnamefont
  {Fisher}}, \bibinfo {author} {\bibfnamefont {S.}~\bibnamefont {Kim}},
  \bibinfo {author} {\bibfnamefont {B.~F.}\ \bibnamefont {Woodfield}}, \bibinfo
  {author} {\bibfnamefont {N.~E.}\ \bibnamefont {Phillips}}, \bibinfo {author}
  {\bibfnamefont {L.}~\bibnamefont {Taillefer}}, \bibinfo {author}
  {\bibfnamefont {K.}~\bibnamefont {Hasselbach}}, \bibinfo {author}
  {\bibfnamefont {J.}~\bibnamefont {Flouquet}}, \bibinfo {author}
  {\bibfnamefont {A.~L.}\ \bibnamefont {Giorgi}},\ and\ \bibinfo {author}
  {\bibfnamefont {J.~L.}\ \bibnamefont {Smith}},\ }\bibfield  {title} {\bibinfo
  {title} {Specific heat of {{UPt}}{$_3$}: {{Evidence}} for unconventional
  superconductivity},\ }\href {https://doi.org/10.1103/PhysRevLett.62.1411}
  {\bibfield  {journal} {\bibinfo  {journal} {Phys. Rev. Lett.}\ }\textbf
  {\bibinfo {volume} {62}},\ \bibinfo {pages} {1411} (\bibinfo {year}
  {1989})}\BibitemShut {NoStop}%
\bibitem [{\citenamefont {Bruls}\ \emph {et~al.}(1990)\citenamefont {Bruls},
  \citenamefont {Weber}, \citenamefont {Wolf}, \citenamefont {Thalmeier},
  \citenamefont {L{\"u}thi}, \citenamefont {de~Visser},\ and\ \citenamefont
  {Menovsky}}]{BrulsG1990}%
  \BibitemOpen
  \bibfield  {author} {\bibinfo {author} {\bibfnamefont {G.}~\bibnamefont
  {Bruls}}, \bibinfo {author} {\bibfnamefont {D.}~\bibnamefont {Weber}},
  \bibinfo {author} {\bibfnamefont {B.}~\bibnamefont {Wolf}}, \bibinfo {author}
  {\bibfnamefont {P.}~\bibnamefont {Thalmeier}}, \bibinfo {author}
  {\bibfnamefont {B.}~\bibnamefont {L{\"u}thi}}, \bibinfo {author}
  {\bibfnamefont {A.}~\bibnamefont {de~Visser}},\ and\ \bibinfo {author}
  {\bibfnamefont {A.}~\bibnamefont {Menovsky}},\ }\bibfield  {title} {\bibinfo
  {title} {Strain--order-parameter coupling and phase diagrams in
  superconducting {{UPt}}{$_{3}$}},\ }\href
  {https://doi.org/10.1103/PhysRevLett.65.2294} {\bibfield  {journal} {\bibinfo
   {journal} {Phys. Rev. Lett.}\ }\textbf {\bibinfo {volume} {65}},\ \bibinfo
  {pages} {2294} (\bibinfo {year} {1990})}\BibitemShut {NoStop}%
\bibitem [{\citenamefont {Adenwalla}\ \emph {et~al.}(1990)\citenamefont
  {Adenwalla}, \citenamefont {Lin}, \citenamefont {Ran}, \citenamefont {Zhao},
  \citenamefont {Ketterson}, \citenamefont {Sauls}, \citenamefont {Taillefer},
  \citenamefont {Hinks}, \citenamefont {Levy},\ and\ \citenamefont
  {Sarma}}]{AdenwallaS1990}%
  \BibitemOpen
  \bibfield  {author} {\bibinfo {author} {\bibfnamefont {S.}~\bibnamefont
  {Adenwalla}}, \bibinfo {author} {\bibfnamefont {S.~W.}\ \bibnamefont {Lin}},
  \bibinfo {author} {\bibfnamefont {Q.~Z.}\ \bibnamefont {Ran}}, \bibinfo
  {author} {\bibfnamefont {Z.}~\bibnamefont {Zhao}}, \bibinfo {author}
  {\bibfnamefont {J.~B.}\ \bibnamefont {Ketterson}}, \bibinfo {author}
  {\bibfnamefont {J.~A.}\ \bibnamefont {Sauls}}, \bibinfo {author}
  {\bibfnamefont {L.}~\bibnamefont {Taillefer}}, \bibinfo {author}
  {\bibfnamefont {D.~G.}\ \bibnamefont {Hinks}}, \bibinfo {author}
  {\bibfnamefont {M.}~\bibnamefont {Levy}},\ and\ \bibinfo {author}
  {\bibfnamefont {B.~K.}\ \bibnamefont {Sarma}},\ }\bibfield  {title} {\bibinfo
  {title} {Phase diagram of {{UPt}}{$_3$} from ultrasonic velocity
  measurements},\ }\href {https://doi.org/10.1103/PhysRevLett.65.2298}
  {\bibfield  {journal} {\bibinfo  {journal} {Phys. Rev. Lett.}\ }\textbf
  {\bibinfo {volume} {65}},\ \bibinfo {pages} {2298} (\bibinfo {year}
  {1990})}\BibitemShut {NoStop}%
\bibitem [{\citenamefont {Aeppli}\ and\ \citenamefont
  {Fisk}(1992)}]{AeppliG1992}%
  \BibitemOpen
  \bibfield  {author} {\bibinfo {author} {\bibfnamefont {G.}~\bibnamefont
  {Aeppli}}\ and\ \bibinfo {author} {\bibfnamefont {Z.}~\bibnamefont {Fisk}},\
  }\bibfield  {title} {\bibinfo {title} {Kondo insulators},\ }\href@noop {}
  {\bibfield  {journal} {\bibinfo  {journal} {Comm. Condens. Matter Phys.}\
  }\textbf {\bibinfo {volume} {16}},\ \bibinfo {pages} {155} (\bibinfo {year}
  {1992})}\BibitemShut {NoStop}%
\bibitem [{\citenamefont {Jaccard}\ \emph {et~al.}(1992)\citenamefont
  {Jaccard}, \citenamefont {Behnia},\ and\ \citenamefont
  {Sierro}}]{JaccardD1992}%
  \BibitemOpen
  \bibfield  {author} {\bibinfo {author} {\bibfnamefont {D.}~\bibnamefont
  {Jaccard}}, \bibinfo {author} {\bibfnamefont {K.}~\bibnamefont {Behnia}},\
  and\ \bibinfo {author} {\bibfnamefont {J.}~\bibnamefont {Sierro}},\
  }\bibfield  {title} {\bibinfo {title} {Pressure induced heavy fermion
  superconductivity of {{CeCu}}{$_{2}$}{{Ge}}{$_{2}$}},\ }\href
  {https://doi.org/10.1016/0375-9601(92)90860-O} {\bibfield  {journal}
  {\bibinfo  {journal} {Physics Letters A}\ }\textbf {\bibinfo {volume}
  {163}},\ \bibinfo {pages} {475} (\bibinfo {year} {1992})}\BibitemShut
  {NoStop}%
\bibitem [{\citenamefont {Movshovich}\ \emph {et~al.}(1996)\citenamefont
  {Movshovich}, \citenamefont {Graf}, \citenamefont {Mandrus}, \citenamefont
  {Thompson}, \citenamefont {Smith},\ and\ \citenamefont
  {Fisk}}]{MovshovichR1996}%
  \BibitemOpen
  \bibfield  {author} {\bibinfo {author} {\bibfnamefont {R.}~\bibnamefont
  {Movshovich}}, \bibinfo {author} {\bibfnamefont {T.}~\bibnamefont {Graf}},
  \bibinfo {author} {\bibfnamefont {D.}~\bibnamefont {Mandrus}}, \bibinfo
  {author} {\bibfnamefont {J.~D.}\ \bibnamefont {Thompson}}, \bibinfo {author}
  {\bibfnamefont {J.~L.}\ \bibnamefont {Smith}},\ and\ \bibinfo {author}
  {\bibfnamefont {Z.}~\bibnamefont {Fisk}},\ }\bibfield  {title} {\bibinfo
  {title} {Superconductivity in heavy-fermion {{CeRh}}{$_{2}$}{{Si}}{$_{2}$}},\
  }\href {https://doi.org/10.1103/PhysRevB.53.8241} {\bibfield  {journal}
  {\bibinfo  {journal} {Phys. Rev. B}\ }\textbf {\bibinfo {volume} {53}},\
  \bibinfo {pages} {8241} (\bibinfo {year} {1996})}\BibitemShut {NoStop}%
\bibitem [{\citenamefont {Mathur}\ \emph {et~al.}(1998)\citenamefont {Mathur},
  \citenamefont {Grosche}, \citenamefont {Julian}, \citenamefont {Walker},
  \citenamefont {Freye}, \citenamefont {Haselwimmer},\ and\ \citenamefont
  {Lonzarich}}]{MathurND1998}%
  \BibitemOpen
  \bibfield  {author} {\bibinfo {author} {\bibfnamefont {N.~D.}\ \bibnamefont
  {Mathur}}, \bibinfo {author} {\bibfnamefont {F.~M.}\ \bibnamefont {Grosche}},
  \bibinfo {author} {\bibfnamefont {S.~R.}\ \bibnamefont {Julian}}, \bibinfo
  {author} {\bibfnamefont {I.~R.}\ \bibnamefont {Walker}}, \bibinfo {author}
  {\bibfnamefont {D.~M.}\ \bibnamefont {Freye}}, \bibinfo {author}
  {\bibfnamefont {R.~K.~W.}\ \bibnamefont {Haselwimmer}},\ and\ \bibinfo
  {author} {\bibfnamefont {G.~G.}\ \bibnamefont {Lonzarich}},\ }\bibfield
  {title} {\bibinfo {title} {Magnetically mediated superconductivity in heavy
  fermion compounds},\ }\href {https://doi.org/10.1038/27838} {\bibfield
  {journal} {\bibinfo  {journal} {Nature}\ }\textbf {\bibinfo {volume} {394}},\
  \bibinfo {pages} {39} (\bibinfo {year} {1998})}\BibitemShut {NoStop}%
\bibitem [{\citenamefont {Schr{\"o}der}\ \emph {et~al.}(2000)\citenamefont
  {Schr{\"o}der}, \citenamefont {Aeppli}, \citenamefont {Coldea}, \citenamefont
  {Adams}, \citenamefont {Stockert}, \citenamefont {v~L{\"o}hneysen},
  \citenamefont {Bucher}, \citenamefont {Ramazashvili},\ and\ \citenamefont
  {Coleman}}]{SchroderA2000}%
  \BibitemOpen
  \bibfield  {author} {\bibinfo {author} {\bibfnamefont {A.}~\bibnamefont
  {Schr{\"o}der}}, \bibinfo {author} {\bibfnamefont {G.}~\bibnamefont
  {Aeppli}}, \bibinfo {author} {\bibfnamefont {R.}~\bibnamefont {Coldea}},
  \bibinfo {author} {\bibfnamefont {M.}~\bibnamefont {Adams}}, \bibinfo
  {author} {\bibfnamefont {O.}~\bibnamefont {Stockert}}, \bibinfo {author}
  {\bibfnamefont {H.}~\bibnamefont {v~L{\"o}hneysen}}, \bibinfo {author}
  {\bibfnamefont {E.}~\bibnamefont {Bucher}}, \bibinfo {author} {\bibfnamefont
  {R.}~\bibnamefont {Ramazashvili}},\ and\ \bibinfo {author} {\bibfnamefont
  {P.}~\bibnamefont {Coleman}},\ }\bibfield  {title} {\bibinfo {title} {Onset
  of antiferromagnetism in heavy-fermion metals},\ }\href
  {https://doi.org/10.1038/35030039} {\bibfield  {journal} {\bibinfo  {journal}
  {Nature}\ }\textbf {\bibinfo {volume} {407}},\ \bibinfo {pages} {351}
  (\bibinfo {year} {2000})}\BibitemShut {NoStop}%
\bibitem [{\citenamefont {Paglione}\ \emph {et~al.}(2003)\citenamefont
  {Paglione}, \citenamefont {Tanatar}, \citenamefont {Hawthorn}, \citenamefont
  {Boaknin}, \citenamefont {Hill}, \citenamefont {Ronning}, \citenamefont
  {Sutherland}, \citenamefont {Taillefer}, \citenamefont {Petrovic},\ and\
  \citenamefont {Canfield}}]{PaglioneJ2003}%
  \BibitemOpen
  \bibfield  {author} {\bibinfo {author} {\bibfnamefont {J.}~\bibnamefont
  {Paglione}}, \bibinfo {author} {\bibfnamefont {M.~A.}\ \bibnamefont
  {Tanatar}}, \bibinfo {author} {\bibfnamefont {D.~G.}\ \bibnamefont
  {Hawthorn}}, \bibinfo {author} {\bibfnamefont {E.}~\bibnamefont {Boaknin}},
  \bibinfo {author} {\bibfnamefont {R.~W.}\ \bibnamefont {Hill}}, \bibinfo
  {author} {\bibfnamefont {F.}~\bibnamefont {Ronning}}, \bibinfo {author}
  {\bibfnamefont {M.}~\bibnamefont {Sutherland}}, \bibinfo {author}
  {\bibfnamefont {L.}~\bibnamefont {Taillefer}}, \bibinfo {author}
  {\bibfnamefont {C.}~\bibnamefont {Petrovic}},\ and\ \bibinfo {author}
  {\bibfnamefont {P.~C.}\ \bibnamefont {Canfield}},\ }\bibfield  {title}
  {\bibinfo {title} {Field-{{Induced Quantum Critical Point}} in
  {{CeCoIn}}{$_{5}$}},\ }\href {https://doi.org/10.1103/PhysRevLett.91.246405}
  {\bibfield  {journal} {\bibinfo  {journal} {Phys. Rev. Lett.}\ }\textbf
  {\bibinfo {volume} {91}},\ \bibinfo {pages} {246405} (\bibinfo {year}
  {2003})}\BibitemShut {NoStop}%
\bibitem [{\citenamefont {Bianchi}\ \emph {et~al.}(2003)\citenamefont
  {Bianchi}, \citenamefont {Movshovich}, \citenamefont {Vekhter}, \citenamefont
  {Pagliuso},\ and\ \citenamefont {Sarrao}}]{BianchiA2003}%
  \BibitemOpen
  \bibfield  {author} {\bibinfo {author} {\bibfnamefont {A.}~\bibnamefont
  {Bianchi}}, \bibinfo {author} {\bibfnamefont {R.}~\bibnamefont {Movshovich}},
  \bibinfo {author} {\bibfnamefont {I.}~\bibnamefont {Vekhter}}, \bibinfo
  {author} {\bibfnamefont {P.~G.}\ \bibnamefont {Pagliuso}},\ and\ \bibinfo
  {author} {\bibfnamefont {J.~L.}\ \bibnamefont {Sarrao}},\ }\bibfield  {title}
  {\bibinfo {title} {Avoided {{Antiferromagnetic Order}} and {{Quantum Critical
  Point}} in {{CeCoIn}}{$_{5}$}},\ }\href
  {https://doi.org/10.1103/PhysRevLett.91.257001} {\bibfield  {journal}
  {\bibinfo  {journal} {Phys. Rev. Lett.}\ }\textbf {\bibinfo {volume} {91}},\
  \bibinfo {pages} {257001} (\bibinfo {year} {2003})}\BibitemShut {NoStop}%
\bibitem [{\citenamefont {Paschen}\ \emph {et~al.}(2004)\citenamefont
  {Paschen}, \citenamefont {L{\"u}hmann}, \citenamefont {Wirth}, \citenamefont
  {Gegenwart}, \citenamefont {Trovarelli}, \citenamefont {Geibel},
  \citenamefont {Steglich}, \citenamefont {Coleman},\ and\ \citenamefont
  {Si}}]{PaschenS2004}%
  \BibitemOpen
  \bibfield  {author} {\bibinfo {author} {\bibfnamefont {S.}~\bibnamefont
  {Paschen}}, \bibinfo {author} {\bibfnamefont {T.}~\bibnamefont
  {L{\"u}hmann}}, \bibinfo {author} {\bibfnamefont {S.}~\bibnamefont {Wirth}},
  \bibinfo {author} {\bibfnamefont {P.}~\bibnamefont {Gegenwart}}, \bibinfo
  {author} {\bibfnamefont {O.}~\bibnamefont {Trovarelli}}, \bibinfo {author}
  {\bibfnamefont {C.}~\bibnamefont {Geibel}}, \bibinfo {author} {\bibfnamefont
  {F.}~\bibnamefont {Steglich}}, \bibinfo {author} {\bibfnamefont
  {P.}~\bibnamefont {Coleman}},\ and\ \bibinfo {author} {\bibfnamefont
  {Q.}~\bibnamefont {Si}},\ }\bibfield  {title} {\bibinfo {title} {Hall-effect
  evolution across a heavy-fermion quantum critical point},\ }\href
  {https://doi.org/10.1038/nature03129} {\bibfield  {journal} {\bibinfo
  {journal} {Nature}\ }\textbf {\bibinfo {volume} {432}},\ \bibinfo {pages}
  {881} (\bibinfo {year} {2004})}\BibitemShut {NoStop}%
\bibitem [{\citenamefont {Shishido}\ \emph {et~al.}(2005)\citenamefont
  {Shishido}, \citenamefont {Settai}, \citenamefont {Harima},\ and\
  \citenamefont {{\=O}nuki}}]{ShishidoH2005}%
  \BibitemOpen
  \bibfield  {author} {\bibinfo {author} {\bibfnamefont {H.}~\bibnamefont
  {Shishido}}, \bibinfo {author} {\bibfnamefont {R.}~\bibnamefont {Settai}},
  \bibinfo {author} {\bibfnamefont {H.}~\bibnamefont {Harima}},\ and\ \bibinfo
  {author} {\bibfnamefont {Y.}~\bibnamefont {{\=O}nuki}},\ }\bibfield  {title}
  {\bibinfo {title} {A {{Drastic Change}} of the {{Fermi Surface}} at a
  {{Critical Pressure}} in {{CeRhIn5}}: {{dHvA Study}} under {{Pressure}}},\
  }\href {https://doi.org/10.1143/JPSJ.74.1103} {\bibfield  {journal} {\bibinfo
   {journal} {J. Phys. Soc. Jpn.}\ }\textbf {\bibinfo {volume} {74}},\ \bibinfo
  {pages} {1103} (\bibinfo {year} {2005})}\BibitemShut {NoStop}%
\bibitem [{\citenamefont {Dzero}\ \emph {et~al.}(2010)\citenamefont {Dzero},
  \citenamefont {Sun}, \citenamefont {Galitski},\ and\ \citenamefont
  {Coleman}}]{DzeroM2010}%
  \BibitemOpen
  \bibfield  {author} {\bibinfo {author} {\bibfnamefont {M.}~\bibnamefont
  {Dzero}}, \bibinfo {author} {\bibfnamefont {K.}~\bibnamefont {Sun}}, \bibinfo
  {author} {\bibfnamefont {V.}~\bibnamefont {Galitski}},\ and\ \bibinfo
  {author} {\bibfnamefont {P.}~\bibnamefont {Coleman}},\ }\bibfield  {title}
  {\bibinfo {title} {Topological {{Kondo Insulators}}},\ }\href
  {https://doi.org/10.1103/PhysRevLett.104.106408} {\bibfield  {journal}
  {\bibinfo  {journal} {Phys. Rev. Lett.}\ }\textbf {\bibinfo {volume} {104}},\
  \bibinfo {pages} {106408} (\bibinfo {year} {2010})}\BibitemShut {NoStop}%
\bibitem [{\citenamefont {Ronning}\ \emph {et~al.}(2017)\citenamefont
  {Ronning}, \citenamefont {Helm}, \citenamefont {Shirer}, \citenamefont
  {Bachmann}, \citenamefont {Balicas}, \citenamefont {Chan}, \citenamefont
  {Ramshaw}, \citenamefont {McDonald}, \citenamefont {Balakirev}, \citenamefont
  {Jaime}, \citenamefont {Bauer},\ and\ \citenamefont {Moll}}]{RonningF2017}%
  \BibitemOpen
  \bibfield  {author} {\bibinfo {author} {\bibfnamefont {F.}~\bibnamefont
  {Ronning}}, \bibinfo {author} {\bibfnamefont {T.}~\bibnamefont {Helm}},
  \bibinfo {author} {\bibfnamefont {K.~R.}\ \bibnamefont {Shirer}}, \bibinfo
  {author} {\bibfnamefont {M.~D.}\ \bibnamefont {Bachmann}}, \bibinfo {author}
  {\bibfnamefont {L.}~\bibnamefont {Balicas}}, \bibinfo {author} {\bibfnamefont
  {M.~K.}\ \bibnamefont {Chan}}, \bibinfo {author} {\bibfnamefont {B.~J.}\
  \bibnamefont {Ramshaw}}, \bibinfo {author} {\bibfnamefont {R.~D.}\
  \bibnamefont {McDonald}}, \bibinfo {author} {\bibfnamefont {F.~F.}\
  \bibnamefont {Balakirev}}, \bibinfo {author} {\bibfnamefont {M.}~\bibnamefont
  {Jaime}}, \bibinfo {author} {\bibfnamefont {E.~D.}\ \bibnamefont {Bauer}},\
  and\ \bibinfo {author} {\bibfnamefont {P.~J.~W.}\ \bibnamefont {Moll}},\
  }\bibfield  {title} {\bibinfo {title} {Electronic in-plane symmetry breaking
  at field-tuned quantum criticality in {{CeRhIn}}{$_{5}$}},\ }\href
  {https://doi.org/10.1038/nature23315} {\bibfield  {journal} {\bibinfo
  {journal} {Nature}\ }\textbf {\bibinfo {volume} {548}},\ \bibinfo {pages}
  {313} (\bibinfo {year} {2017})}\BibitemShut {NoStop}%
\bibitem [{\citenamefont {Coleman}(2007)}]{ColemanP2007}%
  \BibitemOpen
  \bibfield  {author} {\bibinfo {author} {\bibfnamefont {P.}~\bibnamefont
  {Coleman}},\ }\bibfield  {title} {\bibinfo {title} {Heavy {{Fermions}}:
  {{Electrons}} at the {{Edge}} of {{Magnetism}}},\ }in\ \href
  {https://doi.org/10.1002/9780470022184.hmm105} {\emph {\bibinfo {booktitle}
  {Handbook of {{Magnetism}} and {{Advanced Magnetic Materials}}}}}\ (\bibinfo
  {publisher} {John Wiley \& Sons, Ltd},\ \bibinfo {year} {2007})\BibitemShut
  {NoStop}%
\bibitem [{\citenamefont {Jackeli}\ and\ \citenamefont
  {Khaliullin}(2009)}]{JackeliG2009}%
  \BibitemOpen
  \bibfield  {author} {\bibinfo {author} {\bibfnamefont {G.}~\bibnamefont
  {Jackeli}}\ and\ \bibinfo {author} {\bibfnamefont {G.}~\bibnamefont
  {Khaliullin}},\ }\bibfield  {title} {\bibinfo {title} {Mott {{Insulators}} in
  the {{Strong Spin-Orbit Coupling Limit}}: {{From Heisenberg}} to a {{Quantum
  Compass}} and {{Kitaev Models}}},\ }\href
  {https://doi.org/10.1103/PhysRevLett.102.017205} {\bibfield  {journal}
  {\bibinfo  {journal} {Phys. Rev. Lett.}\ }\textbf {\bibinfo {volume} {102}},\
  \bibinfo {pages} {017205} (\bibinfo {year} {2009})}\BibitemShut {NoStop}%
\bibitem [{\citenamefont {{Witczak-Krempa}}\ \emph {et~al.}(2014)\citenamefont
  {{Witczak-Krempa}}, \citenamefont {Chen}, \citenamefont {Kim},\ and\
  \citenamefont {Balents}}]{Witczak-KrempaW2014_review}%
  \BibitemOpen
  \bibfield  {author} {\bibinfo {author} {\bibfnamefont {W.}~\bibnamefont
  {{Witczak-Krempa}}}, \bibinfo {author} {\bibfnamefont {G.}~\bibnamefont
  {Chen}}, \bibinfo {author} {\bibfnamefont {Y.~B.}\ \bibnamefont {Kim}},\ and\
  \bibinfo {author} {\bibfnamefont {L.}~\bibnamefont {Balents}},\ }\bibfield
  {title} {\bibinfo {title} {Correlated {{Quantum Phenomena}} in the {{Strong
  Spin-Orbit Regime}}},\ }\href
  {https://doi.org/10.1146/annurev-conmatphys-020911-125138} {\bibfield
  {journal} {\bibinfo  {journal} {Annu. Rev. Condens. Matter Phys.}\ }\textbf
  {\bibinfo {volume} {5}},\ \bibinfo {pages} {57} (\bibinfo {year}
  {2014})}\BibitemShut {NoStop}%
\bibitem [{\citenamefont {Goodenough}(1955)}]{GoodenoughJB1955}%
  \BibitemOpen
  \bibfield  {author} {\bibinfo {author} {\bibfnamefont {J.~B.}\ \bibnamefont
  {Goodenough}},\ }\bibfield  {title} {\bibinfo {title} {Theory of the {{Role}}
  of {{Covalence}} in the {{Perovskite-Type Manganites}}
  [{{La}},{{{\emph{M}}}}({{II}})]{{MnO}}{$_{3}$}},\ }\href
  {https://doi.org/10.1103/PhysRev.100.564} {\bibfield  {journal} {\bibinfo
  {journal} {Phys. Rev.}\ }\textbf {\bibinfo {volume} {100}},\ \bibinfo {pages}
  {564} (\bibinfo {year} {1955})}\BibitemShut {NoStop}%
\bibitem [{\citenamefont {Kanamori}(1957)}]{KanamoriJ1957}%
  \BibitemOpen
  \bibfield  {author} {\bibinfo {author} {\bibfnamefont {J.}~\bibnamefont
  {Kanamori}},\ }\bibfield  {title} {\bibinfo {title} {Theory of the {{Magnetic
  Properties}} of {{Ferrous}} and {{Cobaltous Oxides}}, {{I}}},\ }\href
  {https://doi.org/10.1143/PTP.17.177} {\bibfield  {journal} {\bibinfo
  {journal} {Prog. Theor. Phys.}\ }\textbf {\bibinfo {volume} {17}},\ \bibinfo
  {pages} {177} (\bibinfo {year} {1957})}\BibitemShut {NoStop}%
\bibitem [{\citenamefont {Goodenough}(1958)}]{GoodenoughJB1958}%
  \BibitemOpen
  \bibfield  {author} {\bibinfo {author} {\bibfnamefont {J.~B.}\ \bibnamefont
  {Goodenough}},\ }\bibfield  {title} {\bibinfo {title} {An interpretation of
  the magnetic properties of the perovskite-type mixed crystals
  {{La}}{\textsubscript{1-x}}{{Sr}}{\textsubscript{x}}{{CoO}}{\textsubscript{3-{$\lambda$}}}},\
  }\href {https://doi.org/10.1016/0022-3697(58)90107-0} {\bibfield  {journal}
  {\bibinfo  {journal} {J. Phys. Chem. Solids}\ }\textbf {\bibinfo {volume}
  {6}},\ \bibinfo {pages} {287} (\bibinfo {year} {1958})}\BibitemShut {NoStop}%
\bibitem [{\citenamefont {Moriya}(1960)}]{MoriyaT1960b}%
  \BibitemOpen
  \bibfield  {author} {\bibinfo {author} {\bibfnamefont {T.}~\bibnamefont
  {Moriya}},\ }\bibfield  {title} {\bibinfo {title} {Anisotropic
  {{Superexchange Interaction}} and {{Weak Ferromagnetism}}},\ }\href
  {https://doi.org/10.1103/PhysRev.120.91} {\bibfield  {journal} {\bibinfo
  {journal} {Phys. Rev.}\ }\textbf {\bibinfo {volume} {120}},\ \bibinfo {pages}
  {91} (\bibinfo {year} {1960})}\BibitemShut {NoStop}%
\bibitem [{\citenamefont {Russell}\ and\ \citenamefont
  {Saunders}(1925)}]{RussellHN1925}%
  \BibitemOpen
  \bibfield  {author} {\bibinfo {author} {\bibfnamefont {H.~N.}\ \bibnamefont
  {Russell}}\ and\ \bibinfo {author} {\bibfnamefont {F.~A.}\ \bibnamefont
  {Saunders}},\ }\bibfield  {title} {\bibinfo {title} {New {{Regularities}} in
  the {{Spectra}} of the {{Alkaline Earths}}},\ }\href
  {https://doi.org/10.1086/142872} {\bibfield  {journal} {\bibinfo  {journal}
  {Astrophys. J.}\ }\textbf {\bibinfo {volume} {61}},\ \bibinfo {pages} {38}
  (\bibinfo {year} {1925})}\BibitemShut {NoStop}%
\bibitem [{\citenamefont {Jang}\ \emph {et~al.}(2019)\citenamefont {Jang},
  \citenamefont {Sano}, \citenamefont {Kato},\ and\ \citenamefont
  {Motome}}]{JangSH2019}%
  \BibitemOpen
  \bibfield  {author} {\bibinfo {author} {\bibfnamefont {S.-H.}\ \bibnamefont
  {Jang}}, \bibinfo {author} {\bibfnamefont {R.}~\bibnamefont {Sano}}, \bibinfo
  {author} {\bibfnamefont {Y.}~\bibnamefont {Kato}},\ and\ \bibinfo {author}
  {\bibfnamefont {Y.}~\bibnamefont {Motome}},\ }\bibfield  {title} {\bibinfo
  {title} {Antiferromagnetic {{Kitaev}} interaction in {\emph{f}}-electron
  based honeycomb magnets},\ }\href
  {https://doi.org/10.1103/PhysRevB.99.241106} {\bibfield  {journal} {\bibinfo
  {journal} {Phys. Rev. B}\ }\textbf {\bibinfo {volume} {99}},\ \bibinfo
  {pages} {241106} (\bibinfo {year} {2019})}\BibitemShut {NoStop}%
\bibitem [{\citenamefont {Jang}\ \emph {et~al.}(2020)\citenamefont {Jang},
  \citenamefont {Sano}, \citenamefont {Kato},\ and\ \citenamefont
  {Motome}}]{JangSH2020}%
  \BibitemOpen
  \bibfield  {author} {\bibinfo {author} {\bibfnamefont {S.-H.}\ \bibnamefont
  {Jang}}, \bibinfo {author} {\bibfnamefont {R.}~\bibnamefont {Sano}}, \bibinfo
  {author} {\bibfnamefont {Y.}~\bibnamefont {Kato}},\ and\ \bibinfo {author}
  {\bibfnamefont {Y.}~\bibnamefont {Motome}},\ }\bibfield  {title} {\bibinfo
  {title} {Computational design of {\emph{f}}-electron {{Kitaev}} magnets:
  {{Honeycomb}} and hyperhoneycomb compounds
  {{{\emph{A}}}}{$_{2}$}{{PrO}}{$_3$} ({{{\emph{A}}}}=alkali metals)},\ }\href
  {https://doi.org/10.1103/PhysRevMaterials.4.104420} {\bibfield  {journal}
  {\bibinfo  {journal} {Phys. Rev. Mater.}\ }\textbf {\bibinfo {volume} {4}},\
  \bibinfo {pages} {104420} (\bibinfo {year} {2020})}\BibitemShut {NoStop}%
\bibitem [{\citenamefont {Daum}\ \emph {et~al.}(2021)\citenamefont {Daum},
  \citenamefont {Ramanathan}, \citenamefont {Kolesnikov}, \citenamefont
  {Calder}, \citenamefont {Mourigal},\ and\ \citenamefont
  {La~Pierre}}]{DaumMJ2021}%
  \BibitemOpen
  \bibfield  {author} {\bibinfo {author} {\bibfnamefont {M.~J.}\ \bibnamefont
  {Daum}}, \bibinfo {author} {\bibfnamefont {A.}~\bibnamefont {Ramanathan}},
  \bibinfo {author} {\bibfnamefont {A.~I.}\ \bibnamefont {Kolesnikov}},
  \bibinfo {author} {\bibfnamefont {S.}~\bibnamefont {Calder}}, \bibinfo
  {author} {\bibfnamefont {M.}~\bibnamefont {Mourigal}},\ and\ \bibinfo
  {author} {\bibfnamefont {H.~S.}\ \bibnamefont {La~Pierre}},\ }\bibfield
  {title} {\bibinfo {title} {Collective excitations in the tetravalent
  lanthanide honeycomb antiferromagnet {{Na}}{$_{2}$}{{PrO}}{$_{3}$}},\ }\href
  {https://doi.org/10.1103/PhysRevB.103.L121109} {\bibfield  {journal}
  {\bibinfo  {journal} {Phys. Rev. B}\ }\textbf {\bibinfo {volume} {103}},\
  \bibinfo {pages} {L121109} (\bibinfo {year} {2021})}\BibitemShut {NoStop}%
\bibitem [{\citenamefont {Ramanathan}\ \emph
  {et~al.}(2023{\natexlab{a}})\citenamefont {Ramanathan}, \citenamefont
  {Kaplan}, \citenamefont {Sergentu}, \citenamefont {Branson}, \citenamefont
  {Ozerov}, \citenamefont {Kolesnikov}, \citenamefont {Minasian}, \citenamefont
  {Autschbach}, \citenamefont {Freeland}, \citenamefont {Jiang}, \citenamefont
  {Mourigal},\ and\ \citenamefont {La~Pierre}}]{RamanathanA2023a}%
  \BibitemOpen
  \bibfield  {author} {\bibinfo {author} {\bibfnamefont {A.}~\bibnamefont
  {Ramanathan}}, \bibinfo {author} {\bibfnamefont {J.}~\bibnamefont {Kaplan}},
  \bibinfo {author} {\bibfnamefont {D.-C.}\ \bibnamefont {Sergentu}}, \bibinfo
  {author} {\bibfnamefont {J.~A.}\ \bibnamefont {Branson}}, \bibinfo {author}
  {\bibfnamefont {M.}~\bibnamefont {Ozerov}}, \bibinfo {author} {\bibfnamefont
  {A.~I.}\ \bibnamefont {Kolesnikov}}, \bibinfo {author} {\bibfnamefont
  {S.~G.}\ \bibnamefont {Minasian}}, \bibinfo {author} {\bibfnamefont
  {J.}~\bibnamefont {Autschbach}}, \bibinfo {author} {\bibfnamefont {J.~W.}\
  \bibnamefont {Freeland}}, \bibinfo {author} {\bibfnamefont {Z.}~\bibnamefont
  {Jiang}}, \bibinfo {author} {\bibfnamefont {M.}~\bibnamefont {Mourigal}},\
  and\ \bibinfo {author} {\bibfnamefont {H.~S.}\ \bibnamefont {La~Pierre}},\
  }\bibfield  {title} {\bibinfo {title} {Chemical design of electronic and
  magnetic energy scales of tetravalent praseodymium materials},\ }\href
  {https://doi.org/10.1038/s41467-023-38431-7} {\bibfield  {journal} {\bibinfo
  {journal} {Nat. Commun.}\ }\textbf {\bibinfo {volume} {14}},\ \bibinfo
  {pages} {3134} (\bibinfo {year} {2023}{\natexlab{a}})}\BibitemShut {NoStop}%
\bibitem [{\citenamefont {Ramanathan}\ \emph
  {et~al.}(2023{\natexlab{b}})\citenamefont {Ramanathan}, \citenamefont
  {Walter}, \citenamefont {Mourigal},\ and\ \citenamefont
  {La~Pierre}}]{RamanathanA2023b}%
  \BibitemOpen
  \bibfield  {author} {\bibinfo {author} {\bibfnamefont {A.}~\bibnamefont
  {Ramanathan}}, \bibinfo {author} {\bibfnamefont {E.~D.}\ \bibnamefont
  {Walter}}, \bibinfo {author} {\bibfnamefont {M.}~\bibnamefont {Mourigal}},\
  and\ \bibinfo {author} {\bibfnamefont {H.~S.}\ \bibnamefont {La~Pierre}},\
  }\bibfield  {title} {\bibinfo {title} {Increased {{Crystal Field Drives
  Intermediate Coupling}} and {{Minimizes Decoherence}} in {{Tetravalent
  Praseodymium Qubits}}},\ }\href {https://doi.org/10.1021/jacs.3c02820}
  {\bibfield  {journal} {\bibinfo  {journal} {J. Am. Chem. Soc.}\ }\textbf
  {\bibinfo {volume} {145}},\ \bibinfo {pages} {17603} (\bibinfo {year}
  {2023}{\natexlab{b}})}\BibitemShut {NoStop}%
\bibitem [{\citenamefont {Simeth}\ \emph {et~al.}(2023)\citenamefont {Simeth},
  \citenamefont {Wang}, \citenamefont {Ghioldi}, \citenamefont {Fobes},
  \citenamefont {Podlesnyak}, \citenamefont {Sung}, \citenamefont {Bauer},
  \citenamefont {Lass}, \citenamefont {Flury}, \citenamefont {Vonka},
  \citenamefont {Mazzone}, \citenamefont {Niedermayer}, \citenamefont {Nomura},
  \citenamefont {Arita}, \citenamefont {Batista}, \citenamefont {Ronning},\
  and\ \citenamefont {Janoschek}}]{SimethW2023_CeIn3}%
  \BibitemOpen
  \bibfield  {author} {\bibinfo {author} {\bibfnamefont {W.}~\bibnamefont
  {Simeth}}, \bibinfo {author} {\bibfnamefont {Z.}~\bibnamefont {Wang}},
  \bibinfo {author} {\bibfnamefont {E.~A.}\ \bibnamefont {Ghioldi}}, \bibinfo
  {author} {\bibfnamefont {D.~M.}\ \bibnamefont {Fobes}}, \bibinfo {author}
  {\bibfnamefont {A.}~\bibnamefont {Podlesnyak}}, \bibinfo {author}
  {\bibfnamefont {N.~H.}\ \bibnamefont {Sung}}, \bibinfo {author}
  {\bibfnamefont {E.~D.}\ \bibnamefont {Bauer}}, \bibinfo {author}
  {\bibfnamefont {J.}~\bibnamefont {Lass}}, \bibinfo {author} {\bibfnamefont
  {S.}~\bibnamefont {Flury}}, \bibinfo {author} {\bibfnamefont
  {J.}~\bibnamefont {Vonka}}, \bibinfo {author} {\bibfnamefont {D.~G.}\
  \bibnamefont {Mazzone}}, \bibinfo {author} {\bibfnamefont {C.}~\bibnamefont
  {Niedermayer}}, \bibinfo {author} {\bibfnamefont {Y.}~\bibnamefont {Nomura}},
  \bibinfo {author} {\bibfnamefont {R.}~\bibnamefont {Arita}}, \bibinfo
  {author} {\bibfnamefont {C.~D.}\ \bibnamefont {Batista}}, \bibinfo {author}
  {\bibfnamefont {F.}~\bibnamefont {Ronning}},\ and\ \bibinfo {author}
  {\bibfnamefont {M.}~\bibnamefont {Janoschek}},\ }\bibfield  {title} {\bibinfo
  {title} {A microscopic {{Kondo}} lattice model for the heavy fermion
  antiferromagnet {{CeIn}}{$_{3}$}},\ }\href
  {https://doi.org/10.1038/s41467-023-43947-z} {\bibfield  {journal} {\bibinfo
  {journal} {Nat. Commun.}\ }\textbf {\bibinfo {volume} {14}},\ \bibinfo
  {pages} {8239} (\bibinfo {year} {2023})}\BibitemShut {NoStop}%
\bibitem [{\citenamefont {Kim}\ \emph {et~al.}(1997)\citenamefont {Kim},
  \citenamefont {Tjernberg}, \citenamefont {Chiaia}, \citenamefont
  {Kumigashira}, \citenamefont {Takahashi}, \citenamefont {Du{\`o}},
  \citenamefont {Sakai}, \citenamefont {Kasaya},\ and\ \citenamefont
  {Lindau}}]{KimHD1997}%
  \BibitemOpen
  \bibfield  {author} {\bibinfo {author} {\bibfnamefont {H.-D.}\ \bibnamefont
  {Kim}}, \bibinfo {author} {\bibfnamefont {O.}~\bibnamefont {Tjernberg}},
  \bibinfo {author} {\bibfnamefont {G.}~\bibnamefont {Chiaia}}, \bibinfo
  {author} {\bibfnamefont {H.}~\bibnamefont {Kumigashira}}, \bibinfo {author}
  {\bibfnamefont {T.}~\bibnamefont {Takahashi}}, \bibinfo {author}
  {\bibfnamefont {L.}~\bibnamefont {Du{\`o}}}, \bibinfo {author} {\bibfnamefont
  {O.}~\bibnamefont {Sakai}}, \bibinfo {author} {\bibfnamefont
  {M.}~\bibnamefont {Kasaya}},\ and\ \bibinfo {author} {\bibfnamefont
  {I.}~\bibnamefont {Lindau}},\ }\bibfield  {title} {\bibinfo {title} {Surface
  and bulk 4{\emph{f}}-photoemission spectra of {{CeIn}}{$_3$} and
  {{CeSn}}{$_{3}$}},\ }\href {https://doi.org/10.1103/PhysRevB.56.1620}
  {\bibfield  {journal} {\bibinfo  {journal} {Phys. Rev. B}\ }\textbf {\bibinfo
  {volume} {56}},\ \bibinfo {pages} {1620} (\bibinfo {year}
  {1997})}\BibitemShut {NoStop}%
\bibitem [{\citenamefont {Zhang}\ \emph {et~al.}(2016)\citenamefont {Zhang},
  \citenamefont {Lu}, \citenamefont {Zhu}, \citenamefont {Tan}, \citenamefont
  {Liu}, \citenamefont {Chen}, \citenamefont {Feng}, \citenamefont {Xie},
  \citenamefont {Luo}, \citenamefont {Liu}, \citenamefont {Song}, \citenamefont
  {Zhang},\ and\ \citenamefont {Lai}}]{ZhangY2016}%
  \BibitemOpen
  \bibfield  {author} {\bibinfo {author} {\bibfnamefont {Y.}~\bibnamefont
  {Zhang}}, \bibinfo {author} {\bibfnamefont {H.}~\bibnamefont {Lu}}, \bibinfo
  {author} {\bibfnamefont {X.}~\bibnamefont {Zhu}}, \bibinfo {author}
  {\bibfnamefont {S.}~\bibnamefont {Tan}}, \bibinfo {author} {\bibfnamefont
  {Q.}~\bibnamefont {Liu}}, \bibinfo {author} {\bibfnamefont {Q.}~\bibnamefont
  {Chen}}, \bibinfo {author} {\bibfnamefont {W.}~\bibnamefont {Feng}}, \bibinfo
  {author} {\bibfnamefont {D.}~\bibnamefont {Xie}}, \bibinfo {author}
  {\bibfnamefont {L.}~\bibnamefont {Luo}}, \bibinfo {author} {\bibfnamefont
  {Y.}~\bibnamefont {Liu}}, \bibinfo {author} {\bibfnamefont {H.}~\bibnamefont
  {Song}}, \bibinfo {author} {\bibfnamefont {Z.}~\bibnamefont {Zhang}},\ and\
  \bibinfo {author} {\bibfnamefont {X.}~\bibnamefont {Lai}},\ }\bibfield
  {title} {\bibinfo {title} {Three-dimensional bulk electronic structure of the
  {{Kondo}} lattice {{CeIn}}{$_3$} revealed by photoemission},\ }\href
  {https://doi.org/10.1038/srep33613} {\bibfield  {journal} {\bibinfo
  {journal} {Sci. Rep.}\ }\textbf {\bibinfo {volume} {6}},\ \bibinfo {pages}
  {33613} (\bibinfo {year} {2016})}\BibitemShut {NoStop}%
\bibitem [{\citenamefont {Sundermann}\ \emph {et~al.}(2016)\citenamefont
  {Sundermann}, \citenamefont {Strigari}, \citenamefont {Willers},
  \citenamefont {Weinen}, \citenamefont {Liao}, \citenamefont {Tsuei},
  \citenamefont {Hiraoka}, \citenamefont {Ishii}, \citenamefont {Yamaoka},
  \citenamefont {Mizuki}, \citenamefont {Zekko}, \citenamefont {Bauer},
  \citenamefont {Sarrao}, \citenamefont {Thompson}, \citenamefont {Lejay},
  \citenamefont {Muro}, \citenamefont {Yutani}, \citenamefont {Takabatake},
  \citenamefont {Tanaka}, \citenamefont {Hollmann}, \citenamefont {Tjeng},\
  and\ \citenamefont {Severing}}]{SundermannM2016}%
  \BibitemOpen
  \bibfield  {author} {\bibinfo {author} {\bibfnamefont {M.}~\bibnamefont
  {Sundermann}}, \bibinfo {author} {\bibfnamefont {F.}~\bibnamefont
  {Strigari}}, \bibinfo {author} {\bibfnamefont {T.}~\bibnamefont {Willers}},
  \bibinfo {author} {\bibfnamefont {J.}~\bibnamefont {Weinen}}, \bibinfo
  {author} {\bibfnamefont {Y.~F.}\ \bibnamefont {Liao}}, \bibinfo {author}
  {\bibfnamefont {K.~D.}\ \bibnamefont {Tsuei}}, \bibinfo {author}
  {\bibfnamefont {N.}~\bibnamefont {Hiraoka}}, \bibinfo {author} {\bibfnamefont
  {H.}~\bibnamefont {Ishii}}, \bibinfo {author} {\bibfnamefont
  {H.}~\bibnamefont {Yamaoka}}, \bibinfo {author} {\bibfnamefont
  {J.}~\bibnamefont {Mizuki}}, \bibinfo {author} {\bibfnamefont
  {Y.}~\bibnamefont {Zekko}}, \bibinfo {author} {\bibfnamefont {E.~D.}\
  \bibnamefont {Bauer}}, \bibinfo {author} {\bibfnamefont {J.~L.}\ \bibnamefont
  {Sarrao}}, \bibinfo {author} {\bibfnamefont {J.~D.}\ \bibnamefont
  {Thompson}}, \bibinfo {author} {\bibfnamefont {P.}~\bibnamefont {Lejay}},
  \bibinfo {author} {\bibfnamefont {Y.}~\bibnamefont {Muro}}, \bibinfo {author}
  {\bibfnamefont {K.}~\bibnamefont {Yutani}}, \bibinfo {author} {\bibfnamefont
  {T.}~\bibnamefont {Takabatake}}, \bibinfo {author} {\bibfnamefont
  {A.}~\bibnamefont {Tanaka}}, \bibinfo {author} {\bibfnamefont
  {N.}~\bibnamefont {Hollmann}}, \bibinfo {author} {\bibfnamefont {L.~H.}\
  \bibnamefont {Tjeng}},\ and\ \bibinfo {author} {\bibfnamefont
  {A.}~\bibnamefont {Severing}},\ }\bibfield  {title} {\bibinfo {title}
  {Quantitative study of the {\emph{f}} occupation in
  {{Ce}}{{{\emph{M}}}}{{In}}{$_5$} and other cerium compounds with hard
  {{X-rays}}},\ }\href {https://doi.org/10.1016/j.elspec.2016.02.002}
  {\bibfield  {journal} {\bibinfo  {journal} {J. Electron Spectros. Relat.
  Phenomena}\ }\textbf {\bibinfo {volume} {209}},\ \bibinfo {pages} {1}
  (\bibinfo {year} {2016})}\BibitemShut {NoStop}%
\bibitem [{\citenamefont {Braicovich}\ \emph {et~al.}(2007)\citenamefont
  {Braicovich}, \citenamefont {Tagliaferri}, \citenamefont {Annese},
  \citenamefont {Ghiringhelli}, \citenamefont {Dallera}, \citenamefont
  {Fracassi}, \citenamefont {Palenzona},\ and\ \citenamefont
  {Brookes}}]{BraicovichL2007}%
  \BibitemOpen
  \bibfield  {author} {\bibinfo {author} {\bibfnamefont {L.}~\bibnamefont
  {Braicovich}}, \bibinfo {author} {\bibfnamefont {A.}~\bibnamefont
  {Tagliaferri}}, \bibinfo {author} {\bibfnamefont {E.}~\bibnamefont {Annese}},
  \bibinfo {author} {\bibfnamefont {G.}~\bibnamefont {Ghiringhelli}}, \bibinfo
  {author} {\bibfnamefont {C.}~\bibnamefont {Dallera}}, \bibinfo {author}
  {\bibfnamefont {F.}~\bibnamefont {Fracassi}}, \bibinfo {author}
  {\bibfnamefont {A.}~\bibnamefont {Palenzona}},\ and\ \bibinfo {author}
  {\bibfnamefont {N.~B.}\ \bibnamefont {Brookes}},\ }\bibfield  {title}
  {\bibinfo {title} {Spectroscopy of strongly correlated systems: {{Resonant}}
  x-ray scattering without energy resolution in the scattered beam},\ }\href
  {https://doi.org/10.1103/PhysRevB.75.073104} {\bibfield  {journal} {\bibinfo
  {journal} {Phys. Rev. B}\ }\textbf {\bibinfo {volume} {75}},\ \bibinfo
  {pages} {073104} (\bibinfo {year} {2007})}\BibitemShut {NoStop}%
\bibitem [{\citenamefont {Lawrence}\ and\ \citenamefont
  {Shapiro}(1980)}]{LawrenceJM1980}%
  \BibitemOpen
  \bibfield  {author} {\bibinfo {author} {\bibfnamefont {J.~M.}\ \bibnamefont
  {Lawrence}}\ and\ \bibinfo {author} {\bibfnamefont {S.~M.}\ \bibnamefont
  {Shapiro}},\ }\bibfield  {title} {\bibinfo {title} {Magnetic ordering in the
  presence of fast spin fluctuations: {{A}} neutron scattering study of
  {{CeIn}}{$_{3}$}},\ }\href {https://doi.org/10.1103/PhysRevB.22.4379}
  {\bibfield  {journal} {\bibinfo  {journal} {Phys. Rev. B}\ }\textbf {\bibinfo
  {volume} {22}},\ \bibinfo {pages} {4379} (\bibinfo {year}
  {1980})}\BibitemShut {NoStop}%
\bibitem [{\citenamefont {Knafo}\ \emph {et~al.}(2003)\citenamefont {Knafo},
  \citenamefont {Raymond}, \citenamefont {F{\aa}k}, \citenamefont {Lapertot},
  \citenamefont {Canfield},\ and\ \citenamefont {Flouquet}}]{KnafoW2003}%
  \BibitemOpen
  \bibfield  {author} {\bibinfo {author} {\bibfnamefont {W.}~\bibnamefont
  {Knafo}}, \bibinfo {author} {\bibfnamefont {S.}~\bibnamefont {Raymond}},
  \bibinfo {author} {\bibfnamefont {B.}~\bibnamefont {F{\aa}k}}, \bibinfo
  {author} {\bibfnamefont {G.}~\bibnamefont {Lapertot}}, \bibinfo {author}
  {\bibfnamefont {P.~C.}\ \bibnamefont {Canfield}},\ and\ \bibinfo {author}
  {\bibfnamefont {J.}~\bibnamefont {Flouquet}},\ }\bibfield  {title} {\bibinfo
  {title} {Study of low-energy magnetic excitations in single-crystalline
  {{CeIn}}{$_3$} by inelastic neutron scattering},\ }\href
  {https://doi.org/10.1088/0953-8984/15/22/308} {\bibfield  {journal} {\bibinfo
   {journal} {J. Phys.: Condens. Matter}\ }\textbf {\bibinfo {volume} {15}},\
  \bibinfo {pages} {3741} (\bibinfo {year} {2003})}\BibitemShut {NoStop}%
\bibitem [{\citenamefont {Lindgren}(1974)}]{LindgrenI1974}%
  \BibitemOpen
  \bibfield  {author} {\bibinfo {author} {\bibfnamefont {I.}~\bibnamefont
  {Lindgren}},\ }\bibfield  {title} {\bibinfo {title} {The
  {{Rayleigh-Schrodinger}} perturbation and the linked-diagram theorem for a
  multi-configurational model space},\ }\href
  {https://doi.org/10.1088/0022-3700/7/18/010} {\bibfield  {journal} {\bibinfo
  {journal} {J. Phys. B: Atom. Mol. Phys.}\ }\textbf {\bibinfo {volume} {7}},\
  \bibinfo {pages} {2441} (\bibinfo {year} {1974})}\BibitemShut {NoStop}%
\bibitem [{\citenamefont {Doniach}(1977)}]{DoniachS1977}%
  \BibitemOpen
  \bibfield  {author} {\bibinfo {author} {\bibfnamefont {S.}~\bibnamefont
  {Doniach}},\ }\bibfield  {title} {\bibinfo {title} {The {{Kondo}} lattice and
  weak antiferromagnetism},\ }\href
  {https://doi.org/10.1016/0378-4363(77)90190-5} {\bibfield  {journal}
  {\bibinfo  {journal} {Physica B+C}\ }\textbf {\bibinfo {volume} {91}},\
  \bibinfo {pages} {231} (\bibinfo {year} {1977})}\BibitemShut {NoStop}%
\bibitem [{\citenamefont {Ruderman}\ and\ \citenamefont
  {Kittel}(1954)}]{RudermanMA1954}%
  \BibitemOpen
  \bibfield  {author} {\bibinfo {author} {\bibfnamefont {M.~A.}\ \bibnamefont
  {Ruderman}}\ and\ \bibinfo {author} {\bibfnamefont {C.}~\bibnamefont
  {Kittel}},\ }\bibfield  {title} {\bibinfo {title} {Indirect {{Exchange
  Coupling}} of {{Nuclear Magnetic Moments}} by {{Conduction Electrons}}},\
  }\href {https://doi.org/10.1103/PhysRev.96.99} {\bibfield  {journal}
  {\bibinfo  {journal} {Phys. Rev.}\ }\textbf {\bibinfo {volume} {96}},\
  \bibinfo {pages} {99} (\bibinfo {year} {1954})}\BibitemShut {NoStop}%
\bibitem [{\citenamefont {Kasuya}(1956)}]{KasuyaT1956}%
  \BibitemOpen
  \bibfield  {author} {\bibinfo {author} {\bibfnamefont {T.}~\bibnamefont
  {Kasuya}},\ }\bibfield  {title} {\bibinfo {title} {A {{Theory}} of {{Metallic
  Ferro-}} and {{Antiferromagnetism}} on {{Zener}}'s {{Model}}},\ }\href
  {https://doi.org/10.1143/PTP.16.45} {\bibfield  {journal} {\bibinfo
  {journal} {Prog. Theor. Phys.}\ }\textbf {\bibinfo {volume} {16}},\ \bibinfo
  {pages} {45} (\bibinfo {year} {1956})}\BibitemShut {NoStop}%
\bibitem [{\citenamefont {Yosida}(1957)}]{YosidaK1957}%
  \BibitemOpen
  \bibfield  {author} {\bibinfo {author} {\bibfnamefont {K.}~\bibnamefont
  {Yosida}},\ }\bibfield  {title} {\bibinfo {title} {Magnetic {{Properties}} of
  {{Cu-Mn Alloys}}},\ }\href {https://doi.org/10.1103/PhysRev.106.893}
  {\bibfield  {journal} {\bibinfo  {journal} {Phys. Rev.}\ }\textbf {\bibinfo
  {volume} {106}},\ \bibinfo {pages} {893} (\bibinfo {year}
  {1957})}\BibitemShut {NoStop}%
\bibitem [{\citenamefont {Batista}(2009)}]{BatistaCD2009}%
  \BibitemOpen
  \bibfield  {author} {\bibinfo {author} {\bibfnamefont {C.~D.}\ \bibnamefont
  {Batista}},\ }\bibfield  {title} {\bibinfo {title} {Effective {{Hamiltonian}}
  for metallic {{Pu}}},\ }\href {https://doi.org/10.1016/j.jnucmat.2008.10.018}
  {\bibfield  {journal} {\bibinfo  {journal} {J. Nucl. Mater.}\ }\textbf
  {\bibinfo {volume} {385}},\ \bibinfo {pages} {60} (\bibinfo {year}
  {2009})}\BibitemShut {NoStop}%
\bibitem [{\citenamefont {Giannozzi}\ \emph {et~al.}(2017)\citenamefont
  {Giannozzi}, \citenamefont {Andreussi}, \citenamefont {Brumme}, \citenamefont
  {Bunau}, \citenamefont {Nardelli}, \citenamefont {Calandra}, \citenamefont
  {Car}, \citenamefont {Cavazzoni}, \citenamefont {Ceresoli}, \citenamefont
  {Cococcioni}, \citenamefont {Colonna}, \citenamefont {Carnimeo},
  \citenamefont {Corso}, \citenamefont {de~Gironcoli}, \citenamefont {Delugas},
  \citenamefont {DiStasio}, \citenamefont {Ferretti}, \citenamefont {Floris},
  \citenamefont {Fratesi}, \citenamefont {Fugallo}, \citenamefont {Gebauer},
  \citenamefont {Gerstmann}, \citenamefont {Giustino}, \citenamefont {Gorni},
  \citenamefont {Jia}, \citenamefont {Kawamura}, \citenamefont {Ko},
  \citenamefont {Kokalj}, \citenamefont {K{\"u}{\c c}{\"u}kbenli},
  \citenamefont {Lazzeri}, \citenamefont {Marsili}, \citenamefont {Marzari},
  \citenamefont {Mauri}, \citenamefont {Nguyen}, \citenamefont {Nguyen},
  \citenamefont {{Otero-de-la-Roza}}, \citenamefont {Paulatto}, \citenamefont
  {Ponc{\'e}}, \citenamefont {Rocca}, \citenamefont {Sabatini}, \citenamefont
  {Santra}, \citenamefont {Schlipf}, \citenamefont {Seitsonen}, \citenamefont
  {Smogunov}, \citenamefont {Timrov}, \citenamefont {Thonhauser}, \citenamefont
  {Umari}, \citenamefont {Vast}, \citenamefont {Wu},\ and\ \citenamefont
  {Baroni}}]{GiannozziP2017}%
  \BibitemOpen
  \bibfield  {author} {\bibinfo {author} {\bibfnamefont {P.}~\bibnamefont
  {Giannozzi}}, \bibinfo {author} {\bibfnamefont {O.}~\bibnamefont
  {Andreussi}}, \bibinfo {author} {\bibfnamefont {T.}~\bibnamefont {Brumme}},
  \bibinfo {author} {\bibfnamefont {O.}~\bibnamefont {Bunau}}, \bibinfo
  {author} {\bibfnamefont {M.~B.}\ \bibnamefont {Nardelli}}, \bibinfo {author}
  {\bibfnamefont {M.}~\bibnamefont {Calandra}}, \bibinfo {author}
  {\bibfnamefont {R.}~\bibnamefont {Car}}, \bibinfo {author} {\bibfnamefont
  {C.}~\bibnamefont {Cavazzoni}}, \bibinfo {author} {\bibfnamefont
  {D.}~\bibnamefont {Ceresoli}}, \bibinfo {author} {\bibfnamefont
  {M.}~\bibnamefont {Cococcioni}}, \bibinfo {author} {\bibfnamefont
  {N.}~\bibnamefont {Colonna}}, \bibinfo {author} {\bibfnamefont
  {I.}~\bibnamefont {Carnimeo}}, \bibinfo {author} {\bibfnamefont {A.~D.}\
  \bibnamefont {Corso}}, \bibinfo {author} {\bibfnamefont {S.}~\bibnamefont
  {de~Gironcoli}}, \bibinfo {author} {\bibfnamefont {P.}~\bibnamefont
  {Delugas}}, \bibinfo {author} {\bibfnamefont {R.~A.}\ \bibnamefont
  {DiStasio}}, \bibinfo {author} {\bibfnamefont {A.}~\bibnamefont {Ferretti}},
  \bibinfo {author} {\bibfnamefont {A.}~\bibnamefont {Floris}}, \bibinfo
  {author} {\bibfnamefont {G.}~\bibnamefont {Fratesi}}, \bibinfo {author}
  {\bibfnamefont {G.}~\bibnamefont {Fugallo}}, \bibinfo {author} {\bibfnamefont
  {R.}~\bibnamefont {Gebauer}}, \bibinfo {author} {\bibfnamefont
  {U.}~\bibnamefont {Gerstmann}}, \bibinfo {author} {\bibfnamefont
  {F.}~\bibnamefont {Giustino}}, \bibinfo {author} {\bibfnamefont
  {T.}~\bibnamefont {Gorni}}, \bibinfo {author} {\bibfnamefont
  {J.}~\bibnamefont {Jia}}, \bibinfo {author} {\bibfnamefont {M.}~\bibnamefont
  {Kawamura}}, \bibinfo {author} {\bibfnamefont {H.-Y.}\ \bibnamefont {Ko}},
  \bibinfo {author} {\bibfnamefont {A.}~\bibnamefont {Kokalj}}, \bibinfo
  {author} {\bibfnamefont {E.}~\bibnamefont {K{\"u}{\c c}{\"u}kbenli}},
  \bibinfo {author} {\bibfnamefont {M.}~\bibnamefont {Lazzeri}}, \bibinfo
  {author} {\bibfnamefont {M.}~\bibnamefont {Marsili}}, \bibinfo {author}
  {\bibfnamefont {N.}~\bibnamefont {Marzari}}, \bibinfo {author} {\bibfnamefont
  {F.}~\bibnamefont {Mauri}}, \bibinfo {author} {\bibfnamefont {N.~L.}\
  \bibnamefont {Nguyen}}, \bibinfo {author} {\bibfnamefont {H.-V.}\
  \bibnamefont {Nguyen}}, \bibinfo {author} {\bibfnamefont {A.}~\bibnamefont
  {{Otero-de-la-Roza}}}, \bibinfo {author} {\bibfnamefont {L.}~\bibnamefont
  {Paulatto}}, \bibinfo {author} {\bibfnamefont {S.}~\bibnamefont {Ponc{\'e}}},
  \bibinfo {author} {\bibfnamefont {D.}~\bibnamefont {Rocca}}, \bibinfo
  {author} {\bibfnamefont {R.}~\bibnamefont {Sabatini}}, \bibinfo {author}
  {\bibfnamefont {B.}~\bibnamefont {Santra}}, \bibinfo {author} {\bibfnamefont
  {M.}~\bibnamefont {Schlipf}}, \bibinfo {author} {\bibfnamefont {A.~P.}\
  \bibnamefont {Seitsonen}}, \bibinfo {author} {\bibfnamefont {A.}~\bibnamefont
  {Smogunov}}, \bibinfo {author} {\bibfnamefont {I.}~\bibnamefont {Timrov}},
  \bibinfo {author} {\bibfnamefont {T.}~\bibnamefont {Thonhauser}}, \bibinfo
  {author} {\bibfnamefont {P.}~\bibnamefont {Umari}}, \bibinfo {author}
  {\bibfnamefont {N.}~\bibnamefont {Vast}}, \bibinfo {author} {\bibfnamefont
  {X.}~\bibnamefont {Wu}},\ and\ \bibinfo {author} {\bibfnamefont
  {S.}~\bibnamefont {Baroni}},\ }\bibfield  {title} {\bibinfo {title} {Advanced
  capabilities for materials modelling with {{Quantum ESPRESSO}}},\ }\href
  {https://doi.org/10.1088/1361-648X/aa8f79} {\bibfield  {journal} {\bibinfo
  {journal} {J. Phys.: Condens. Matter}\ }\textbf {\bibinfo {volume} {29}},\
  \bibinfo {pages} {465901} (\bibinfo {year} {2017})}\BibitemShut {NoStop}%
\bibitem [{\citenamefont {Dal~Corso}(2014)}]{DalCorsoA2014}%
  \BibitemOpen
  \bibfield  {author} {\bibinfo {author} {\bibfnamefont {A.}~\bibnamefont
  {Dal~Corso}},\ }\bibfield  {title} {\bibinfo {title} {Pseudopotentials
  periodic table: {{From H}} to {{Pu}}},\ }\href
  {https://doi.org/10.1016/j.commatsci.2014.07.043} {\bibfield  {journal}
  {\bibinfo  {journal} {Comput. Mater. Sci.}\ }\textbf {\bibinfo {volume}
  {95}},\ \bibinfo {pages} {337} (\bibinfo {year} {2014})}\BibitemShut
  {NoStop}%
\bibitem [{\citenamefont {Pizzi}\ \emph {et~al.}(2020)\citenamefont {Pizzi},
  \citenamefont {Vitale}, \citenamefont {Arita}, \citenamefont {Bl{\"u}gel},
  \citenamefont {Freimuth}, \citenamefont {G{\'e}ranton}, \citenamefont
  {Gibertini}, \citenamefont {Gresch}, \citenamefont {Johnson}, \citenamefont
  {Koretsune}, \citenamefont {{Iba{\~n}ez-Azpiroz}}, \citenamefont {Lee},
  \citenamefont {Lihm}, \citenamefont {Marchand}, \citenamefont {Marrazzo},
  \citenamefont {Mokrousov}, \citenamefont {Mustafa}, \citenamefont {Nohara},
  \citenamefont {Nomura}, \citenamefont {Paulatto}, \citenamefont {Ponc{\'e}},
  \citenamefont {Ponweiser}, \citenamefont {Qiao}, \citenamefont {Th{\"o}le},
  \citenamefont {Tsirkin}, \citenamefont {Wierzbowska}, \citenamefont
  {Marzari}, \citenamefont {Vanderbilt}, \citenamefont {Souza}, \citenamefont
  {Mostofi},\ and\ \citenamefont {Yates}}]{PizziG2020}%
  \BibitemOpen
  \bibfield  {author} {\bibinfo {author} {\bibfnamefont {G.}~\bibnamefont
  {Pizzi}}, \bibinfo {author} {\bibfnamefont {V.}~\bibnamefont {Vitale}},
  \bibinfo {author} {\bibfnamefont {R.}~\bibnamefont {Arita}}, \bibinfo
  {author} {\bibfnamefont {S.}~\bibnamefont {Bl{\"u}gel}}, \bibinfo {author}
  {\bibfnamefont {F.}~\bibnamefont {Freimuth}}, \bibinfo {author}
  {\bibfnamefont {G.}~\bibnamefont {G{\'e}ranton}}, \bibinfo {author}
  {\bibfnamefont {M.}~\bibnamefont {Gibertini}}, \bibinfo {author}
  {\bibfnamefont {D.}~\bibnamefont {Gresch}}, \bibinfo {author} {\bibfnamefont
  {C.}~\bibnamefont {Johnson}}, \bibinfo {author} {\bibfnamefont
  {T.}~\bibnamefont {Koretsune}}, \bibinfo {author} {\bibfnamefont
  {J.}~\bibnamefont {{Iba{\~n}ez-Azpiroz}}}, \bibinfo {author} {\bibfnamefont
  {H.}~\bibnamefont {Lee}}, \bibinfo {author} {\bibfnamefont {J.-M.}\
  \bibnamefont {Lihm}}, \bibinfo {author} {\bibfnamefont {D.}~\bibnamefont
  {Marchand}}, \bibinfo {author} {\bibfnamefont {A.}~\bibnamefont {Marrazzo}},
  \bibinfo {author} {\bibfnamefont {Y.}~\bibnamefont {Mokrousov}}, \bibinfo
  {author} {\bibfnamefont {J.~I.}\ \bibnamefont {Mustafa}}, \bibinfo {author}
  {\bibfnamefont {Y.}~\bibnamefont {Nohara}}, \bibinfo {author} {\bibfnamefont
  {Y.}~\bibnamefont {Nomura}}, \bibinfo {author} {\bibfnamefont
  {L.}~\bibnamefont {Paulatto}}, \bibinfo {author} {\bibfnamefont
  {S.}~\bibnamefont {Ponc{\'e}}}, \bibinfo {author} {\bibfnamefont
  {T.}~\bibnamefont {Ponweiser}}, \bibinfo {author} {\bibfnamefont
  {J.}~\bibnamefont {Qiao}}, \bibinfo {author} {\bibfnamefont {F.}~\bibnamefont
  {Th{\"o}le}}, \bibinfo {author} {\bibfnamefont {S.~S.}\ \bibnamefont
  {Tsirkin}}, \bibinfo {author} {\bibfnamefont {M.}~\bibnamefont
  {Wierzbowska}}, \bibinfo {author} {\bibfnamefont {N.}~\bibnamefont
  {Marzari}}, \bibinfo {author} {\bibfnamefont {D.}~\bibnamefont {Vanderbilt}},
  \bibinfo {author} {\bibfnamefont {I.}~\bibnamefont {Souza}}, \bibinfo
  {author} {\bibfnamefont {A.~A.}\ \bibnamefont {Mostofi}},\ and\ \bibinfo
  {author} {\bibfnamefont {J.~R.}\ \bibnamefont {Yates}},\ }\bibfield  {title}
  {\bibinfo {title} {Wannier90 as a community code: New features and
  applications},\ }\href {https://doi.org/10.1088/1361-648X/ab51ff} {\bibfield
  {journal} {\bibinfo  {journal} {J. Phys.: Condens. Matter}\ }\textbf
  {\bibinfo {volume} {32}},\ \bibinfo {pages} {165902} (\bibinfo {year}
  {2020})}\BibitemShut {NoStop}%
\bibitem [{\citenamefont {Schneider}\ \emph {et~al.}(1985)\citenamefont
  {Schneider}, \citenamefont {Delley}, \citenamefont {Wuilloud}, \citenamefont
  {Imer},\ and\ \citenamefont {Baer}}]{SchneiderWD1985}%
  \BibitemOpen
  \bibfield  {author} {\bibinfo {author} {\bibfnamefont {W.-D.}\ \bibnamefont
  {Schneider}}, \bibinfo {author} {\bibfnamefont {B.}~\bibnamefont {Delley}},
  \bibinfo {author} {\bibfnamefont {E.}~\bibnamefont {Wuilloud}}, \bibinfo
  {author} {\bibfnamefont {J.-M.}\ \bibnamefont {Imer}},\ and\ \bibinfo
  {author} {\bibfnamefont {Y.}~\bibnamefont {Baer}},\ }\bibfield  {title}
  {\bibinfo {title} {Electron-spectroscopic manifestations of the 4f states in
  light rare-earth solids},\ }\href {https://doi.org/10.1103/PhysRevB.32.6819}
  {\bibfield  {journal} {\bibinfo  {journal} {Phys. Rev. B}\ }\textbf {\bibinfo
  {volume} {32}},\ \bibinfo {pages} {6819} (\bibinfo {year}
  {1985})}\BibitemShut {NoStop}%
\bibitem [{\citenamefont {Gam{\.z}a}\ \emph {et~al.}(2008)\citenamefont
  {Gam{\.z}a}, \citenamefont {{\'S}lebarski},\ and\ \citenamefont
  {Deniszczyk}}]{GamzaM2008}%
  \BibitemOpen
  \bibfield  {author} {\bibinfo {author} {\bibfnamefont {M.}~\bibnamefont
  {Gam{\.z}a}}, \bibinfo {author} {\bibfnamefont {A.}~\bibnamefont
  {{\'S}lebarski}},\ and\ \bibinfo {author} {\bibfnamefont {J.}~\bibnamefont
  {Deniszczyk}},\ }\bibfield  {title} {\bibinfo {title} {Electronic structure
  of
  {{Ce}}{\textsubscript{n}}{{M}}{\textsubscript{m}}{{In}}{\textsubscript{2m+3n}},
  where {\emph{n}} = 1, 2; {\emph{m}} = 0, 1;{{M}} = {{Co}}, {{Rh}} or {{Ir}}:
  Experiment and calculations},\ }\href
  {https://doi.org/10.1088/0953-8984/20/11/115202} {\bibfield  {journal}
  {\bibinfo  {journal} {J. Phys.: Condens. Matter}\ }\textbf {\bibinfo {volume}
  {20}},\ \bibinfo {pages} {115202} (\bibinfo {year} {2008})}\BibitemShut
  {NoStop}%
\bibitem [{\citenamefont {Jang}\ \emph {et~al.}(2023)\citenamefont {Jang},
  \citenamefont {O'Neal}, \citenamefont {Lane}, \citenamefont {B\"ohm},
  \citenamefont {Sirica}, \citenamefont {Yarotski}, \citenamefont {Bauer},
  \citenamefont {Ronning}, \citenamefont {Prasankumar},\ and\ \citenamefont
  {Zhu}}]{Jang2023}%
  \BibitemOpen
  \bibfield  {author} {\bibinfo {author} {\bibfnamefont {B.~G.}\ \bibnamefont
  {Jang}}, \bibinfo {author} {\bibfnamefont {K.~R.}\ \bibnamefont {O'Neal}},
  \bibinfo {author} {\bibfnamefont {C.}~\bibnamefont {Lane}}, \bibinfo {author}
  {\bibfnamefont {T.~U.}\ \bibnamefont {B\"ohm}}, \bibinfo {author}
  {\bibfnamefont {N.}~\bibnamefont {Sirica}}, \bibinfo {author} {\bibfnamefont
  {D.}~\bibnamefont {Yarotski}}, \bibinfo {author} {\bibfnamefont {E.~D.}\
  \bibnamefont {Bauer}}, \bibinfo {author} {\bibfnamefont {F.}~\bibnamefont
  {Ronning}}, \bibinfo {author} {\bibfnamefont {R.}~\bibnamefont
  {Prasankumar}},\ and\ \bibinfo {author} {\bibfnamefont {J.-X.}\ \bibnamefont
  {Zhu}},\ }\bibfield  {title} {\bibinfo {title} {One-dimensionality signature
  in optical conductivity of heavy-fermion
  ${\mathrm{ceir}}_{3}{\mathrm{b}}_{2}$},\ }\href
  {https://doi.org/10.1103/PhysRevB.107.205116} {\bibfield  {journal} {\bibinfo
   {journal} {Phys. Rev. B}\ }\textbf {\bibinfo {volume} {107}},\ \bibinfo
  {pages} {205116} (\bibinfo {year} {2023})}\BibitemShut {NoStop}%
\bibitem [{\citenamefont {Lawrence}(1979)}]{LawrenceJ1979}%
  \BibitemOpen
  \bibfield  {author} {\bibinfo {author} {\bibfnamefont {J.}~\bibnamefont
  {Lawrence}},\ }\bibfield  {title} {\bibinfo {title} {Scaling behavior near a
  valence instability: {{The}} magnetic susceptibility of
  {{CeIn}}{\textsubscript{3-x}}{{Sn}}{\textsubscript{x}}},\ }\href
  {https://doi.org/10.1103/PhysRevB.20.3770} {\bibfield  {journal} {\bibinfo
  {journal} {Phys. Rev. B}\ }\textbf {\bibinfo {volume} {20}},\ \bibinfo
  {pages} {3770} (\bibinfo {year} {1979})}\BibitemShut {NoStop}%
\bibitem [{\citenamefont {Fuggle}\ \emph {et~al.}(1983)\citenamefont {Fuggle},
  \citenamefont {Hillebrecht}, \citenamefont {Zo{\l}nierek}, \citenamefont
  {L{\"a}sser}, \citenamefont {Freiburg}, \citenamefont {Gunnarsson},\ and\
  \citenamefont {Sch{\"o}nhammer}}]{FuggleJC1983}%
  \BibitemOpen
  \bibfield  {author} {\bibinfo {author} {\bibfnamefont {J.~C.}\ \bibnamefont
  {Fuggle}}, \bibinfo {author} {\bibfnamefont {F.~U.}\ \bibnamefont
  {Hillebrecht}}, \bibinfo {author} {\bibfnamefont {Z.}~\bibnamefont
  {Zo{\l}nierek}}, \bibinfo {author} {\bibfnamefont {R.}~\bibnamefont
  {L{\"a}sser}}, \bibinfo {author} {\bibfnamefont {{\relax Ch}.}~\bibnamefont
  {Freiburg}}, \bibinfo {author} {\bibfnamefont {O.}~\bibnamefont
  {Gunnarsson}},\ and\ \bibinfo {author} {\bibfnamefont {K.}~\bibnamefont
  {Sch{\"o}nhammer}},\ }\bibfield  {title} {\bibinfo {title} {Electronic
  structure of {{Ce}} and its intermetallic compounds},\ }\href
  {https://doi.org/10.1103/PhysRevB.27.7330} {\bibfield  {journal} {\bibinfo
  {journal} {Phys. Rev. B}\ }\textbf {\bibinfo {volume} {27}},\ \bibinfo
  {pages} {7330} (\bibinfo {year} {1983})}\BibitemShut {NoStop}%
\bibitem [{\citenamefont {Sala}\ \emph {et~al.}(2021)\citenamefont {Sala},
  \citenamefont {Stone}, \citenamefont {Rai}, \citenamefont {May},
  \citenamefont {Laurell}, \citenamefont {Garlea}, \citenamefont {Butch},
  \citenamefont {Lumsden}, \citenamefont {Ehlers}, \citenamefont {Pokharel},
  \citenamefont {Podlesnyak}, \citenamefont {Mandrus}, \citenamefont {Parker},
  \citenamefont {Okamoto}, \citenamefont {Hal{\'a}sz},\ and\ \citenamefont
  {Christianson}}]{SalaG2021}%
  \BibitemOpen
  \bibfield  {author} {\bibinfo {author} {\bibfnamefont {G.}~\bibnamefont
  {Sala}}, \bibinfo {author} {\bibfnamefont {M.~B.}\ \bibnamefont {Stone}},
  \bibinfo {author} {\bibfnamefont {B.~K.}\ \bibnamefont {Rai}}, \bibinfo
  {author} {\bibfnamefont {A.~F.}\ \bibnamefont {May}}, \bibinfo {author}
  {\bibfnamefont {P.}~\bibnamefont {Laurell}}, \bibinfo {author} {\bibfnamefont
  {V.~O.}\ \bibnamefont {Garlea}}, \bibinfo {author} {\bibfnamefont {N.~P.}\
  \bibnamefont {Butch}}, \bibinfo {author} {\bibfnamefont {M.~D.}\ \bibnamefont
  {Lumsden}}, \bibinfo {author} {\bibfnamefont {G.}~\bibnamefont {Ehlers}},
  \bibinfo {author} {\bibfnamefont {G.}~\bibnamefont {Pokharel}}, \bibinfo
  {author} {\bibfnamefont {A.}~\bibnamefont {Podlesnyak}}, \bibinfo {author}
  {\bibfnamefont {D.}~\bibnamefont {Mandrus}}, \bibinfo {author} {\bibfnamefont
  {D.~S.}\ \bibnamefont {Parker}}, \bibinfo {author} {\bibfnamefont
  {S.}~\bibnamefont {Okamoto}}, \bibinfo {author} {\bibfnamefont {G.~B.}\
  \bibnamefont {Hal{\'a}sz}},\ and\ \bibinfo {author} {\bibfnamefont {A.~D.}\
  \bibnamefont {Christianson}},\ }\bibfield  {title} {\bibinfo {title} {Van
  {{Hove}} singularity in the magnon spectrum of the antiferromagnetic quantum
  honeycomb lattice},\ }\href {https://doi.org/10.1038/s41467-020-20335-5}
  {\bibfield  {journal} {\bibinfo  {journal} {Nat. Commun.}\ }\textbf {\bibinfo
  {volume} {12}},\ \bibinfo {pages} {171} (\bibinfo {year} {2021})}\BibitemShut
  {NoStop}%
\bibitem [{\citenamefont {Ding}\ \emph {et~al.}(2019)\citenamefont {Ding},
  \citenamefont {Manuel}, \citenamefont {Bachus}, \citenamefont {Gru{\ss}ler},
  \citenamefont {Gegenwart}, \citenamefont {Singleton}, \citenamefont
  {Johnson}, \citenamefont {Walker}, \citenamefont {Adroja}, \citenamefont
  {Hillier},\ and\ \citenamefont {Tsirlin}}]{DingL2019}%
  \BibitemOpen
  \bibfield  {author} {\bibinfo {author} {\bibfnamefont {L.}~\bibnamefont
  {Ding}}, \bibinfo {author} {\bibfnamefont {P.}~\bibnamefont {Manuel}},
  \bibinfo {author} {\bibfnamefont {S.}~\bibnamefont {Bachus}}, \bibinfo
  {author} {\bibfnamefont {F.}~\bibnamefont {Gru{\ss}ler}}, \bibinfo {author}
  {\bibfnamefont {P.}~\bibnamefont {Gegenwart}}, \bibinfo {author}
  {\bibfnamefont {J.}~\bibnamefont {Singleton}}, \bibinfo {author}
  {\bibfnamefont {R.~D.}\ \bibnamefont {Johnson}}, \bibinfo {author}
  {\bibfnamefont {H.~C.}\ \bibnamefont {Walker}}, \bibinfo {author}
  {\bibfnamefont {D.~T.}\ \bibnamefont {Adroja}}, \bibinfo {author}
  {\bibfnamefont {A.~D.}\ \bibnamefont {Hillier}},\ and\ \bibinfo {author}
  {\bibfnamefont {A.~A.}\ \bibnamefont {Tsirlin}},\ }\bibfield  {title}
  {\bibinfo {title} {Gapless spin-liquid state in the structurally
  disorder-free triangular antiferromagnet {{NaYbO}}{$_{2}$}},\ }\href
  {https://doi.org/10.1103/PhysRevB.100.144432} {\bibfield  {journal} {\bibinfo
   {journal} {Phys. Rev. B}\ }\textbf {\bibinfo {volume} {100}},\ \bibinfo
  {pages} {144432} (\bibinfo {year} {2019})}\BibitemShut {NoStop}%
\bibitem [{\citenamefont {Bordelon}\ \emph {et~al.}(2019)\citenamefont
  {Bordelon}, \citenamefont {Kenney}, \citenamefont {Liu}, \citenamefont
  {Hogan}, \citenamefont {Posthuma}, \citenamefont {Kavand}, \citenamefont
  {Lyu}, \citenamefont {Sherwin}, \citenamefont {Butch}, \citenamefont {Brown},
  \citenamefont {Graf}, \citenamefont {Balents},\ and\ \citenamefont
  {Wilson}}]{BordelonMM2019}%
  \BibitemOpen
  \bibfield  {author} {\bibinfo {author} {\bibfnamefont {M.~M.}\ \bibnamefont
  {Bordelon}}, \bibinfo {author} {\bibfnamefont {E.}~\bibnamefont {Kenney}},
  \bibinfo {author} {\bibfnamefont {C.}~\bibnamefont {Liu}}, \bibinfo {author}
  {\bibfnamefont {T.}~\bibnamefont {Hogan}}, \bibinfo {author} {\bibfnamefont
  {L.}~\bibnamefont {Posthuma}}, \bibinfo {author} {\bibfnamefont
  {M.}~\bibnamefont {Kavand}}, \bibinfo {author} {\bibfnamefont
  {Y.}~\bibnamefont {Lyu}}, \bibinfo {author} {\bibfnamefont {M.}~\bibnamefont
  {Sherwin}}, \bibinfo {author} {\bibfnamefont {N.~P.}\ \bibnamefont {Butch}},
  \bibinfo {author} {\bibfnamefont {C.}~\bibnamefont {Brown}}, \bibinfo
  {author} {\bibfnamefont {M.~J.}\ \bibnamefont {Graf}}, \bibinfo {author}
  {\bibfnamefont {L.}~\bibnamefont {Balents}},\ and\ \bibinfo {author}
  {\bibfnamefont {S.~D.}\ \bibnamefont {Wilson}},\ }\bibfield  {title}
  {\bibinfo {title} {Field-tunable quantum disordered ground state in the
  triangular-lattice antiferromagnet {{NaYbO}}{$_{2}$}},\ }\href
  {https://doi.org/10.1038/s41567-019-0594-5} {\bibfield  {journal} {\bibinfo
  {journal} {Nat. Phys.}\ }\textbf {\bibinfo {volume} {15}},\ \bibinfo {pages}
  {1058} (\bibinfo {year} {2019})}\BibitemShut {NoStop}%
\bibitem [{\citenamefont {Bordelon}\ \emph {et~al.}(2020)\citenamefont
  {Bordelon}, \citenamefont {Liu}, \citenamefont {Posthuma}, \citenamefont
  {Sarte}, \citenamefont {Butch}, \citenamefont {Pajerowski}, \citenamefont
  {Banerjee}, \citenamefont {Balents},\ and\ \citenamefont
  {Wilson}}]{BordelonMM2020}%
  \BibitemOpen
  \bibfield  {author} {\bibinfo {author} {\bibfnamefont {M.~M.}\ \bibnamefont
  {Bordelon}}, \bibinfo {author} {\bibfnamefont {C.}~\bibnamefont {Liu}},
  \bibinfo {author} {\bibfnamefont {L.}~\bibnamefont {Posthuma}}, \bibinfo
  {author} {\bibfnamefont {P.~M.}\ \bibnamefont {Sarte}}, \bibinfo {author}
  {\bibfnamefont {N.~P.}\ \bibnamefont {Butch}}, \bibinfo {author}
  {\bibfnamefont {D.~M.}\ \bibnamefont {Pajerowski}}, \bibinfo {author}
  {\bibfnamefont {A.}~\bibnamefont {Banerjee}}, \bibinfo {author}
  {\bibfnamefont {L.}~\bibnamefont {Balents}},\ and\ \bibinfo {author}
  {\bibfnamefont {S.~D.}\ \bibnamefont {Wilson}},\ }\bibfield  {title}
  {\bibinfo {title} {Spin excitations in the frustrated triangular lattice
  antiferromagnet {{NaYbO}}{$_{2}$}},\ }\href
  {https://doi.org/10.1103/PhysRevB.101.224427} {\bibfield  {journal} {\bibinfo
   {journal} {Phys. Rev. B}\ }\textbf {\bibinfo {volume} {101}},\ \bibinfo
  {pages} {224427} (\bibinfo {year} {2020})}\BibitemShut {NoStop}%
\bibitem [{\citenamefont {Xing}\ \emph {et~al.}(2019)\citenamefont {Xing},
  \citenamefont {Sanjeewa}, \citenamefont {Kim}, \citenamefont {Stewart},
  \citenamefont {Podlesnyak},\ and\ \citenamefont {Sefat}}]{XingJ2019_CsYbSe2}%
  \BibitemOpen
  \bibfield  {author} {\bibinfo {author} {\bibfnamefont {J.}~\bibnamefont
  {Xing}}, \bibinfo {author} {\bibfnamefont {L.~D.}\ \bibnamefont {Sanjeewa}},
  \bibinfo {author} {\bibfnamefont {J.}~\bibnamefont {Kim}}, \bibinfo {author}
  {\bibfnamefont {G.~R.}\ \bibnamefont {Stewart}}, \bibinfo {author}
  {\bibfnamefont {A.}~\bibnamefont {Podlesnyak}},\ and\ \bibinfo {author}
  {\bibfnamefont {A.~S.}\ \bibnamefont {Sefat}},\ }\bibfield  {title} {\bibinfo
  {title} {Field-induced magnetic transition and spin fluctuations in the
  quantum spin-liquid candidate {{CsYbSe}}{$_{2}$}},\ }\href
  {https://doi.org/10.1103/PhysRevB.100.220407} {\bibfield  {journal} {\bibinfo
   {journal} {Phys. Rev. B}\ }\textbf {\bibinfo {volume} {100}},\ \bibinfo
  {pages} {220407} (\bibinfo {year} {2019})}\BibitemShut {NoStop}%
\bibitem [{\citenamefont {Xie}\ \emph {et~al.}(2023)\citenamefont {Xie},
  \citenamefont {Eberharter}, \citenamefont {Xing}, \citenamefont {Nishimoto},
  \citenamefont {Brando}, \citenamefont {Khanenko}, \citenamefont
  {Sichelschmidt}, \citenamefont {Turrini}, \citenamefont {Mazzone},
  \citenamefont {Naumov}, \citenamefont {Sanjeewa}, \citenamefont {Harrison},
  \citenamefont {Sefat}, \citenamefont {Normand}, \citenamefont {L{\"a}uchli},
  \citenamefont {Podlesnyak},\ and\ \citenamefont {Nikitin}}]{XieT2023}%
  \BibitemOpen
  \bibfield  {author} {\bibinfo {author} {\bibfnamefont {T.}~\bibnamefont
  {Xie}}, \bibinfo {author} {\bibfnamefont {A.~A.}\ \bibnamefont {Eberharter}},
  \bibinfo {author} {\bibfnamefont {J.}~\bibnamefont {Xing}}, \bibinfo {author}
  {\bibfnamefont {S.}~\bibnamefont {Nishimoto}}, \bibinfo {author}
  {\bibfnamefont {M.}~\bibnamefont {Brando}}, \bibinfo {author} {\bibfnamefont
  {P.}~\bibnamefont {Khanenko}}, \bibinfo {author} {\bibfnamefont
  {J.}~\bibnamefont {Sichelschmidt}}, \bibinfo {author} {\bibfnamefont {A.~A.}\
  \bibnamefont {Turrini}}, \bibinfo {author} {\bibfnamefont {D.~G.}\
  \bibnamefont {Mazzone}}, \bibinfo {author} {\bibfnamefont {P.~G.}\
  \bibnamefont {Naumov}}, \bibinfo {author} {\bibfnamefont {L.~D.}\
  \bibnamefont {Sanjeewa}}, \bibinfo {author} {\bibfnamefont {N.}~\bibnamefont
  {Harrison}}, \bibinfo {author} {\bibfnamefont {A.~S.}\ \bibnamefont {Sefat}},
  \bibinfo {author} {\bibfnamefont {B.}~\bibnamefont {Normand}}, \bibinfo
  {author} {\bibfnamefont {A.~M.}\ \bibnamefont {L{\"a}uchli}}, \bibinfo
  {author} {\bibfnamefont {A.}~\bibnamefont {Podlesnyak}},\ and\ \bibinfo
  {author} {\bibfnamefont {S.~E.}\ \bibnamefont {Nikitin}},\ }\bibfield
  {title} {\bibinfo {title} {Complete field-induced spectral response of the
  spin-1/2 triangular-lattice antiferromagnet {{CsYbSe}}{$_{2}$}},\ }\href
  {https://doi.org/10.1038/s41535-023-00580-9} {\bibfield  {journal} {\bibinfo
  {journal} {npj Quantum Mater.}\ }\textbf {\bibinfo {volume} {8}},\ \bibinfo
  {pages} {1} (\bibinfo {year} {2023})}\BibitemShut {NoStop}%
\bibitem [{\citenamefont {Villanova}\ \emph {et~al.}(2023)\citenamefont
  {Villanova}, \citenamefont {Scheie}, \citenamefont {Tennant}, \citenamefont
  {Okamoto},\ and\ \citenamefont {Berlijn}}]{VillanovaJW2023}%
  \BibitemOpen
  \bibfield  {author} {\bibinfo {author} {\bibfnamefont {J.~W.}\ \bibnamefont
  {Villanova}}, \bibinfo {author} {\bibfnamefont {A.~O.}\ \bibnamefont
  {Scheie}}, \bibinfo {author} {\bibfnamefont {D.~A.}\ \bibnamefont {Tennant}},
  \bibinfo {author} {\bibfnamefont {S.}~\bibnamefont {Okamoto}},\ and\ \bibinfo
  {author} {\bibfnamefont {T.}~\bibnamefont {Berlijn}},\ }\bibfield  {title}
  {\bibinfo {title} {First-principles derivation of magnetic interactions in
  the triangular quantum spin liquid candidates {{KYb}}{{{\emph{Ch}}}}{$_2$}
  ({{{\emph{Ch}}}}{$_2$}={{S}}, {{Se}}, {{Te}}) and
  {{{\emph{A}}}}{{YbSe}}{$_2$} ({{{\emph{A}}}} = {{Na}}, {{Rb}})},\ }\href
  {https://doi.org/10.1103/PhysRevResearch.5.033050} {\bibfield  {journal}
  {\bibinfo  {journal} {Phys. Rev. Res.}\ }\textbf {\bibinfo {volume} {5}},\
  \bibinfo {pages} {033050} (\bibinfo {year} {2023})}\BibitemShut {NoStop}%
\bibitem [{\citenamefont {Scheie}\ \emph {et~al.}(2024)\citenamefont {Scheie},
  \citenamefont {Kamiya}, \citenamefont {Zhang}, \citenamefont {Lee},
  \citenamefont {Woods}, \citenamefont {Ajeesh}, \citenamefont {Gonzalez},
  \citenamefont {Bernu}, \citenamefont {Villanova}, \citenamefont {Xing},
  \citenamefont {Huang}, \citenamefont {Zhang}, \citenamefont {Ma},
  \citenamefont {Choi}, \citenamefont {Pajerowski}, \citenamefont {Zhou},
  \citenamefont {Sefat}, \citenamefont {Okamoto}, \citenamefont {Berlijn},
  \citenamefont {Messio}, \citenamefont {Movshovich}, \citenamefont {Batista},\
  and\ \citenamefont {Tennant}}]{ScheieAO2024}%
  \BibitemOpen
  \bibfield  {author} {\bibinfo {author} {\bibfnamefont {A.~O.}\ \bibnamefont
  {Scheie}}, \bibinfo {author} {\bibfnamefont {Y.}~\bibnamefont {Kamiya}},
  \bibinfo {author} {\bibfnamefont {H.}~\bibnamefont {Zhang}}, \bibinfo
  {author} {\bibfnamefont {S.}~\bibnamefont {Lee}}, \bibinfo {author}
  {\bibfnamefont {A.~J.}\ \bibnamefont {Woods}}, \bibinfo {author}
  {\bibfnamefont {M.~O.}\ \bibnamefont {Ajeesh}}, \bibinfo {author}
  {\bibfnamefont {M.~G.}\ \bibnamefont {Gonzalez}}, \bibinfo {author}
  {\bibfnamefont {B.}~\bibnamefont {Bernu}}, \bibinfo {author} {\bibfnamefont
  {J.~W.}\ \bibnamefont {Villanova}}, \bibinfo {author} {\bibfnamefont
  {J.}~\bibnamefont {Xing}}, \bibinfo {author} {\bibfnamefont {Q.}~\bibnamefont
  {Huang}}, \bibinfo {author} {\bibfnamefont {Q.}~\bibnamefont {Zhang}},
  \bibinfo {author} {\bibfnamefont {J.}~\bibnamefont {Ma}}, \bibinfo {author}
  {\bibfnamefont {E.~S.}\ \bibnamefont {Choi}}, \bibinfo {author}
  {\bibfnamefont {D.~M.}\ \bibnamefont {Pajerowski}}, \bibinfo {author}
  {\bibfnamefont {H.}~\bibnamefont {Zhou}}, \bibinfo {author} {\bibfnamefont
  {A.~S.}\ \bibnamefont {Sefat}}, \bibinfo {author} {\bibfnamefont
  {S.}~\bibnamefont {Okamoto}}, \bibinfo {author} {\bibfnamefont
  {T.}~\bibnamefont {Berlijn}}, \bibinfo {author} {\bibfnamefont
  {L.}~\bibnamefont {Messio}}, \bibinfo {author} {\bibfnamefont
  {R.}~\bibnamefont {Movshovich}}, \bibinfo {author} {\bibfnamefont {C.~D.}\
  \bibnamefont {Batista}},\ and\ \bibinfo {author} {\bibfnamefont {D.~A.}\
  \bibnamefont {Tennant}},\ }\bibfield  {title} {\bibinfo {title} {Nonlinear
  magnons and exchange {{Hamiltonians}} of the delafossite proximate quantum
  spin liquid candidates {{KYbSe}}{$_2$} and {{NaYbSe}}{$_{2}$}},\ }\href
  {https://doi.org/10.1103/PhysRevB.109.014425} {\bibfield  {journal} {\bibinfo
   {journal} {Phys. Rev. B}\ }\textbf {\bibinfo {volume} {109}},\ \bibinfo
  {pages} {014425} (\bibinfo {year} {2024})}\BibitemShut {NoStop}%
\bibitem [{\citenamefont {Wang}\ \emph {et~al.}(2020)\citenamefont {Wang},
  \citenamefont {Su}, \citenamefont {Lin},\ and\ \citenamefont
  {Batista}}]{WangZ2020_RKKY}%
  \BibitemOpen
  \bibfield  {author} {\bibinfo {author} {\bibfnamefont {Z.}~\bibnamefont
  {Wang}}, \bibinfo {author} {\bibfnamefont {Y.}~\bibnamefont {Su}}, \bibinfo
  {author} {\bibfnamefont {S.-Z.}\ \bibnamefont {Lin}},\ and\ \bibinfo {author}
  {\bibfnamefont {C.~D.}\ \bibnamefont {Batista}},\ }\bibfield  {title}
  {\bibinfo {title} {Skyrmion {{Crystal}} from {{RKKY Interaction Mediated}} by
  {{2D Electron Gas}}},\ }\href
  {https://doi.org/10.1103/PhysRevLett.124.207201} {\bibfield  {journal}
  {\bibinfo  {journal} {Phys. Rev. Lett.}\ }\textbf {\bibinfo {volume} {124}},\
  \bibinfo {pages} {207201} (\bibinfo {year} {2020})}\BibitemShut {NoStop}%
\bibitem [{\citenamefont {Kitaev}(2006)}]{KitaevA2006}%
  \BibitemOpen
  \bibfield  {author} {\bibinfo {author} {\bibfnamefont {A.}~\bibnamefont
  {Kitaev}},\ }\bibfield  {title} {\bibinfo {title} {Anyons in an exactly
  solved model and beyond},\ }\href {https://doi.org/10.1016/j.aop.2005.10.005}
  {\bibfield  {journal} {\bibinfo  {journal} {Ann. Phys.}\ }\textbf {\bibinfo
  {volume} {321}},\ \bibinfo {pages} {2} (\bibinfo {year} {2006})}\BibitemShut
  {NoStop}%
\bibitem [{\citenamefont {Zhang}(2014)}]{ZhangQ2014_thesis}%
  \BibitemOpen
  \bibfield  {author} {\bibinfo {author} {\bibfnamefont {Q.}~\bibnamefont
  {Zhang}},\ }\emph {\bibinfo {title} {Calculations of Atomic Multiplets across
  the Periodic Table}},\ \href@noop {} {Master's thesis},\ \bibinfo  {school}
  {RWTH Aachen}, \bibinfo {address} {J{\"u}lich} (\bibinfo {year}
  {2014})\BibitemShut {NoStop}%
\end{thebibliography}%

\end{document}